\documentclass[12pt]{article}

 \usepackage{graphicx}
 \usepackage{cite}

 \usepackage{mathtools} 

\long\def\begincomment#1\endcomment{}
\begin{document}


\begin{center}
{\LARGE\bf Fast and flexible implementations of Wilson,\\[2mm] Brillouin and Susskind fermions in lattice QCD}
\end{center}

\vspace{10pt}

\begin{center}
{\large\bf Stephan D\"urr$\,^{a,b}$}
\\[10pt]
${}^a${\sl Department of Physics, University of Wuppertal, 42119 Wuppertal, Germany}\\
${}^b${\sl J\"ulich Supercomputing Centre, Forschungszentrum J\"ulich, 52425 J\"ulich, Germany}
\end{center}

\vspace{10pt}

\begin{abstract}
\noindent
A modern Fortran implementation of three Dirac operators (Wilson, Brillouin, Susskind)
in lattice QCD is presented, based on OpenMP shared-memory parallelization and SIMD pragmas.
The main idea is to apply a Dirac operator to $N_v$ vectors simultaneously, to ease the memory bandwidth bottleneck.
All index computations are left to the compiler and maximum weight is given to portability and flexibility.
The lattice volume, $N_xN_yN_zN_t$, the number of colors, $N_c$, and the number of right-hand sides, $N_v$, are parameters defined at compile time.
Several memory layout options are compared.
The code performs well on modern many-core architectures (480\,Gflop/s, 880\,Gflop/s, and 780\,Gflop/s with $N_v=12$
for the three operators in single precision on a 72-core KNL processor, a $2\times24$-core Skylake node yields similar results).
Explicit run-time tests with CG/BiCGstab inverters confirm that the memory layout is relevant for the KNL, but less so for the Skylake architecture.
The ancillary code distribution contains all routines, including the single, double, and mixed precision Krylov space solvers, to render it self-contained and ready-to-use.
\end{abstract}

\vspace{10pt}

\newcommand{\pad}{\partial}
\newcommand{\hqu}{\hbar}
\newcommand{\til}{\tilde}
\newcommand{\pri}{^\prime}
\renewcommand{\dag}{^\dagger}
\newcommand{\<}{\langle}
\renewcommand{\>}{\rangle}
\newcommand{\gaf}{\gamma_5}
\newcommand{\nab}{\nabla}
\newcommand{\lap}{\triangle}
\newcommand{\dal}{{\sqcap\!\!\!\!\sqcup}}
\newcommand{\trc}{\mathrm{tr}}
\newcommand{\Trc}{\mathrm{Tr}}
\newcommand{\Mpi}{M_\pi}
\newcommand{\Fpi}{F_\pi}
\newcommand{\Mka}{M_K}
\newcommand{\Fka}{F_K}
\newcommand{\Met}{M_\et}
\newcommand{\Fet}{F_\et}
\newcommand{\Mss}{M_{\bar{s}s}}
\newcommand{\Fss}{F_{\bar{s}s}}
\newcommand{\Mcc}{M_{\bar{c}c}}
\newcommand{\Fcc}{F_{\bar{c}c}}

\newcommand{\al}{\alpha}
\newcommand{\be}{\beta}
\newcommand{\ga}{\gamma}
\newcommand{\de}{\delta}
\newcommand{\ep}{\epsilon}
\newcommand{\ve}{\varepsilon}
\newcommand{\ze}{\zeta}
\newcommand{\et}{\eta}
\renewcommand{\th}{\theta}
\newcommand{\vt}{\vartheta}
\newcommand{\io}{\iota}
\newcommand{\ka}{\kappa}
\newcommand{\la}{\lambda}
\newcommand{\rh}{\rho}
\newcommand{\vr}{\varrho}
\newcommand{\si}{\sigma}
\newcommand{\ta}{\tau}
\newcommand{\ph}{\phi}
\newcommand{\vp}{\varphi}
\newcommand{\ch}{\chi}
\newcommand{\ps}{\psi}
\newcommand{\om}{\omega}

\newcommand{\psb}{\bar{\psi}}
\newcommand{\etb}{\bar{\eta}}
\newcommand{\psd}{\psi^{\dagger}}
\newcommand{\etd}{\eta^{\dagger}}
\newcommand{\qh}{\hat{q}}
\newcommand{\kh}{\hat{k}}

\newcommand{\bdm}{\begin{displaymath}}
\newcommand{\edm}{\end{displaymath}}
\newcommand{\bea}{\begin{eqnarray}}
\newcommand{\eea}{\end{eqnarray}}
\newcommand{\beq}{\begin{equation}}
\newcommand{\eeq}{\end{equation}}

\newcommand{\mr}{\mathrm}
\newcommand{\mb}{\mathbf}
\newcommand{\ri}{\mr{i}}
\newcommand{\Nf}{N_{\!f}}
\newcommand{\Nc}{N_{ c }}
\newcommand{\Nt}{N_{ t }}
\newcommand{\Nv}{N_{ v }}
\newcommand{\Nthr}{N_\mr{thr}}
\newcommand{\Dst}{D^\mr{st}}
\newcommand{\DS}{D_\mr{S}}
\newcommand{\DW}{D_\mr{W}}
\newcommand{\DA}{D_\mr{A}}
\newcommand{\DB}{D_\mr{B}}
\newcommand{\MeV}{\,\mr{MeV}}
\newcommand{\GeV}{\,\mr{GeV}}
\newcommand{\fm}{\,\mr{fm}}
\newcommand{\MSbar}{\overline{\mr{MS}}}

\newcommand{\Nx}{N_x}
\newcommand{\Ny}{N_y}
\newcommand{\Nz}{N_z}


\hyphenation{topo-lo-gi-cal simu-la-tion theo-re-ti-cal mini-mum con-tinu-um}


\section{Introduction\label{sec:intro}}


An ideal lattice QCD code is short, easy to read (hence to enhance/modify), and compiles in a fully portable manner into a fast-performing executable.
Such codes are hard to find, as these requirements tend to be in conflict with each other.
But there are means and ways to mitigate the conflict, and this article reports on a specific effort in this direction.

Lattice QCD is the regulated theory of quarks and gluons \cite{Wilson:1974sk,Kogut:1974ag,Susskind:1976jm}.
While a significant amount of quantum field theory knowledge goes into its formulation, the actual computational problem is easy to describe.
It is about frequently solving large linear systems%
\footnote{Here ``frequently'' means $O(10^{5})$ times, ``large'' implies a $n\times n$ matrix with $n=402\,653\,184$ for a Wilson fermion on a $64^3\times128$ lattice,
and depending on the quark mass the condition number of $D\dag D$ is often in the range $10^{6}\ldots10^{8}$.
The factor $10^{5}$ reflects the production of an ensemble of $1000$ gauge configurations, separated by ten $\ta=1$ HMC trajectories, assuming that each of these requires $O(10)$ inversions.}
\beq
Du=b
\label{matrix_times_vector}
\eeq
for $u$, with a given vector $b$ and a given Dirac matrix $D$ (which is sparse and badly conditioned).
In this sense lattice QCD is a rather typical subfield of computational physics, except that $D$ is usually not available in a standard sparse format%
\footnote{Such as ``compressed sparse row'' (CSR) or ``compressed sparse column'' (CSC) format;
see e.g.\ the MATLAB documentation on {\tt sparse} for a quick introduction and a reference.},
but only implicitly, i.e.\ through a routine which implements the matrix-vector multiplication $v\leftarrow Du$.
One is forced to use Krylov-space solvers, such as CG \cite{Hestenes} or BiCGstab \cite{VanDerVorst}, whenever possible with some structure preserving preconditioning \cite{Saad}.
For a very nice overview of numerical issues in lattice QCD computations the reader is referred to Ref.~\cite{Luscher:2010ae}.
In short, this paper is about efficiently implementing the matrix-vector multiplication that is at the heart of solving Eq.~(\ref{matrix_times_vector}) by means of the
CG or BiCGstab algorithm, where $D$ is a sparse matrix that encodes the properties of a ``Wilson'', ``Brillouin'' or ``staggered/Susskind'' fermion, i.e.\ $D\in\{\DW,\DB,\DS\}$.

Lattice QCD production codes tend to use hybrid parallelism, i.e.\ there is a coarse-grained parallelism between nodes, and a fine-grained parallelism within each node.
The coarse-grained parallelism is usually implemented with MPI, the fine grained parallelism with OpenMP on CPUs (alternatively: OpenACC on GPUs).
The latter requires a symmetric multiprocessing (SMP) shared memory architecture within each node.
Typically, a geometric domain decomposition is used on the large scale, e.g.\ a $48^3\times96$ lattice might be distributed onto a $2\times2\times2\times4$ grid of nodes, each of which hosts a local $24^4$ lattice.
Explicit MPI commands are used to organize the ``halo exchange'' among the $32$ nodes, and OpenMP pragmas are used to organize the work sharing of the $\Nthr$ SMP-threads which handle a given local lattice.

To minimize execution time, the single-node performance must be maximized through OpenMP parallelization and vectorization pragmas,
while overlapping communication and computation must be organized through clever MPI calls.
These two issues tend to be logically separated, and this is why it makes sense%
\footnote{Unfortunately this is somewhat orthogonal to the way how CPU resources at large computational infra\-structure centers are allocated.
In the technical review the focus is on the scaling behavior versus the number of nodes. The worse the single-node performance, the easier it is to make the node-scaling graph look nice.}
to first upgrade a serial code into a fast OpenMP code (with vectorization pragmas), and to address MPI parallelization in the second step.
This paper reports on a dedicated effort to efficiently handle the first step, and establishes a ranking among various memory layout options.
With such results in hand, it will be a more straightforward endeavor to address the second step
in a forthcoming publication.

Two properties of the code presented in this article are essential for good balance between full portability and acceptable performance figures (see below).
The first feature is the use of multiple right-hand sides (RHS) in Eq.~(\ref{matrix_times_vector}), i.e.\ $b$ and $u$ are matrices of size $n\times\Nv$, where $\Nv$ is the number of RHS.
This is crucial on systems limited by memory bandwidth, since the Dirac matrix $D$ depends on a gauge field $U$ which is loaded from memory (besides $u$ and $v$, of course) in each call $v\leftarrow Du$.
The second feature is the decision to use the RHS-index as loop index for the single-instruction-multiple-data (SIMD) pipeline.
In QCD terminology one would say that -- besides the $\Nc$ color and $4$ spinor degrees of freedom -- an additional ``internal degree of freedom'' is introduced in the storage scheme.
The main difference is that its value is a matter of convenience, i.e.\ it has has no physics implication.
This avoids the reshuffle operation from a generic read-in format (where the site indices on $b$ or $v$ are slower than any internal degree of freedom) to a dedicated SIMD format
that is needed if, say, a subset of the space-time coordinates of the local lattice is used as SIMD index.

The philosophy behind the code presented in this article is to leave all index computations and all optimization work to the compiler.
Only a set of SIMD pragmas is used to tell the compiler that it should SIMD-ize loops over the RHS-index (which runs from $1$ to $\Nv$) with explicit unrolling of color and spinor operations inside the loop.
This turns out to be sufficient for reaching reasonable performance figures on many integrated core (MIC) architectures, such as the ``knights landing'' (KNL) chip by Intel which has up to $68$ physical cores.
A key feature of this processor (and more modern successors) is the ability to use the new intrinsic instruction set AVX-512.
Accordingly, excellent code performance hinges on the ability to organize fused multiply-add (FMA) operations on $512$ bit long%
\footnote{The $64$ bytes amount to SIMD pipelines handling 16\,real\_sp numbers (equivalently 8\,real\_dp, 8\,complex\_sp, or 4\,complex\_dp numbers) simultaneously.}
data sets.
Hence, the primary goal of this article is to explore whether this challenging task can be left to the compiler,
if all relevant information (e.g.\ the values $\Nc,\Nv$ and $N_x,N_y,N_z,N_t$) is given at compile time.

The code presented in this article handles three choices of the fermion discretization in lattice QCD.
The Wilson definition $\DW$ \cite{Wilson:1974sk}, and the staggered definition $\DS$ \cite{Kogut:1974ag,Susskind:1976jm} are well known,
with publicly available codes (see e.g.\ Refs.~\cite{Dasgupta:1996nj,DelDebbio:2006cn,DelDebbio:2007pz,Campos:2019kgw,Bach:2012iw,Bazavov:2014wgs,Boyle:2015tjk,Altenkort:2021fqk}).
The Brillouin definition $\DB$ \cite{Durr:2010ch,Durr:2012dw,Durr:2017wfi} is less popular -- in part since there is no publicly available implementation with full%
\footnote{At https://github.com/g-koutsou/qpb there is an undocumented C++ implementation by Giannis Koutsou.}
documentation.
In this article each of these formulations is implemented for several vector layouts (see below for a detailed specification), and the CG and BiCGstab inverter routines
(which are part of the ancillary code distribution) are written in a completely generic manner.
The author hopes that this will enable PhD students and young postdocs to write their own QCD code (e.g.\ for hadron spectroscopy) with manageable effort,
and give them a handle to study their field of interest with minimal human constraints and/or dependencies.

This brief exposition of the subject cannot do justice to the effort spent by other authors to maximize performance on a specific architecture for a given Dirac operator $D$.
Recent review talks on the interplay between algorithms and machines in lattice QCD include \cite{Majumdar:2016ixv,Boyle:2017wul,Rago:2017pyb,Lin:2018,Gregory:2019}.
In addition, there is a number of HPC projects in lattice QCD with similar objectives on several architectures
\cite{Arts:2015jia,Heybrock:2015kpy,Richtmann:2016kcq,Kobayashi:2016gog,Boku:2016dmw,DeTar:2016ndn,DeTar:2018pyj,Kanamori:2017urm,Kanamori:2018hwh,Alappat:2021icl,Ishikawa:2021iqw,Akahoshi:2021gvk}.
Preliminary accounts%
\footnote{The attentive reader will notice that performance figures increased quite a bit since these early accounts.}
of this work were given in \cite{Durr:2017clx,Durr:2018ayq}.
All performance measurements were done on three machines at J\"ulich Supercomputing Centre.
The KNL figures were obtained%
\footnote{In flat mode {\tt numactl --preferred 1 ./testknl\_main} is used, in cache mode {\tt ./testknl\_main} suffices.}
on DEEP-knl (booted in flat mode) and the JURECA-booster (booted in cache mode), but the performance difference was marginal.
The results for the Skylake architecture were obtained on DEEP-dam and on JUWELS.

The remainder of this article is organized as follows.
The coding guidelines, and the options for the internal degrees of freedom in the vectors $b,u,v$ are discussed in Sec.~\ref{sec:guidelines}.
A comparison of the vector layouts for the task of computing vector norms, dot-products, and vectorial multiply-adds (all employing SIMD pragmas) is found in Sec.~\ref{sec:norms}.
The coding of the clover term (which is used in conjunction with $\DW$ and $\DB$) is explained Sec.~\ref{sec:clov}.
The implementations of the Wilson Laplacian $\lap^\mr{std}$ and Dirac operator $\DW$ are specified in Sec.~\ref{sec:wils}.
The details for the Brillouin Laplacian $\lap^\mr{bri}$ and Dirac operator $\DB$ are arranged in Sec.~\ref{sec:bril}.
Analogous reasonings and timings for the staggered Dirac operator $\DS$ are found in Sec.~\ref{sec:stag}.
In addition, it is interesting to study the performance as a function of the compile-time parameters $\Nc$, $\Nv$, and $N_xN_yN_zN_t$; such results are assembled in Sec.~\ref{sec:parameters}.
The CG and BiCGstab inverters for all five operators (working with any ordering of the internal degrees of freedom) are presented Sec.~\ref{sec:inverters}.
Finally, in Sec.~\ref{sec:summary} a summary is attempted.
All technical issues (including a guide to the ancillary code distribution) are relegated to appendices \ref{app:A}-\ref{app:F}.


\section{Coding guidelines and vector layouts\label{sec:guidelines}}


The code to be presented is written in Fortran\,2008, which is an excellent choice for scientific problems with static data structures.
Lattice QCD is in this category, and after declaring the number of colors and the box size via the compile time parameters
\small
\begin{verbatim}
integer,parameter :: Nc= 3, Nv=Nc*4             !!! note: number of colors and rhs
integer,parameter :: Nx=34, Ny=Nx,Nz=Ny,Nt=2*Nz !!! note: box size for T=0 physics
\end{verbatim}
\normalsize
an object like the gauge field $U_\mu(n)$, with $\mu\in\{1,\ldots,4\}$ and $n=(x,y,z,t)$ the position in discrete four-dimensional space-time, is conveniently defined as a rank-seven array
\begin{verbatim}
complex(kind=sp),dimension(Nc,Nc,4,Nx,Ny,Nz,Nt) :: U
\end{verbatim}
\normalsize
with the intrinsic complex data type.
Objects defined in ``single precision'' (sp) use four bytes per real component, those in ``double precision'' (dp) use eight bytes per real component.

It is important to keep in mind that Fortran uses the ``column major'' convention for matrices and arrays;
in {\tt U} the first color index (from $1\ldots\Nc$) is the fastest moving index, while the fourth space-time coordinate (from $1\ldots N_t$) is the slowest moving index.
In a nested set of loops
\small
\begin{verbatim}
do t=1,Nt
do z=1,Nz
do y=1,Ny
do x=1,Nx
   do mu=1,4
      U(:,:,mu,x,y,z,t)=float(mu)*eye(:,:)
   end do
end do
end do
end do
end do
\end{verbatim}
\normalsize
the $t$-coordinate must thus be in the outermost loop, followed by the $z$, $y$, $x$ coordinates, the direction index $\mu$, the column and row indices in color space,
to ensure that the elements of $U$ are addressed in the order in which they lie in memory.
In this example we use the stride notation to access a contiguous patch of memory through two implied do loops (with correct ordering built-in);
the $\Nc^2$ complex unit long patch for the matrix $U_\mu(n)$ is overwritten by $\mu I$ where $I$ is the $\Nc\times\Nc$ identity matrix in color space, if {\tt eye(:,:)} was defined accordingly.

Furthermore, this example illustrates our decision to avoid explicit site-index computations in the code.
One might have used {\tt dimension(Nc,Nc,4,Nx*Ny*Nz*Nt)} in the definition of {\tt U}, and an extra line
{\tt n=(((t-1)*Nz+(z-1))*Ny+(y-1))*Nx+x} ahead of the loop over {\tt mu}, along with {\tt U(:,:,mu,n)=float(mu)*eye(:,:)}.
But this is potentially error prone (especially in view of a future MPI-spreading of up to four space-time dimensions over several nodes),
and it seems more elegant to leave all index computations to the compiler.

Given our goal of combining $\Nv$ column vectors into one object [for $D$ being the staggered operator $\DS$ the left-hand side of Eq.~(\ref{matrix_times_vector}) represents
an $n\times n$ matrix which acts on a $n\times\Nv$ matrix with $n\equiv\Nc\Nx\Ny\Nz\Nt$], it seems natural to define a staggered multi-RHS vector through
\small
\begin{verbatim}
complex(kind=sp),dimension(Nc,Nx,Ny,Nz,Nt,Nv) :: suv_sp
\end{verbatim}
\normalsize
but this would imply a detrimental loop-ordering in the staggered Dirac routine (cf.\ Sec.~\ref{sec:stag}).
For good performance it is crucial to promote the RHS-index to an \emph{internal degree of freedom} (like color) which is \emph{ahead} of all space-time indices.
This leaves us with the two options
\small
\begin{verbatim}
complex(kind=sp),dimension(Nc,Nv,Nx,Ny,Nz,Nt) :: suv_sp !!! [Nc,Nv]=layout1
complex(kind=sp),dimension(Nv,Nc,Nx,Ny,Nz,Nt) :: suv_sp !!! [Nv,Nc]=layout2
\end{verbatim}
\normalsize
for a ``staggered utility vector'', and we shall implement both of them and compare the respective timings.
For $\DW,\DB$ the vectors have yet another internal degree of freedom (called ``spinor'', ranging from $1$ to $4$), and this implies that we should consider the six options
\small
\begin{verbatim}
complex(kind=sp),dimension(Nc,04,Nv,Nx,Ny,Nz,Nt) :: vec_sp !!! [Nc,Ns,Nv]=layout1
complex(kind=sp),dimension(04,Nc,Nv,Nx,Ny,Nz,Nt) :: vec_sp !!! [Ns,Nc,Nv]=layout2
complex(kind=sp),dimension(Nc,Nv,04,Nx,Ny,Nz,Nt) :: vec_sp !!! [Nc,Nv,Ns]=layout3
complex(kind=sp),dimension(Nv,Nc,04,Nx,Ny,Nz,Nt) :: vec_sp !!! [Nv,Nc,Ns]=layout4
complex(kind=sp),dimension(04,Nv,Nc,Nx,Ny,Nz,Nt) :: vec_sp !!! [Ns,Nv,Nc]=layout5
complex(kind=sp),dimension(Nv,04,Nc,Nx,Ny,Nz,Nt) :: vec_sp !!! [Nv,Ns,Nc]=layout6
\end{verbatim}
\normalsize
for vectors to be used in conjunction with Wilson-type Dirac matrices (for $\DW$ see Sec.~\ref{sec:wils}, for $\DB$ see Sec.~\ref{sec:bril}), and compare the respective timings.

In Sec.~\ref{sec:inverters} it will be demonstrated that with $u,b$ in the class {\tt suv\_sp} or {\tt vec\_sp} a tolerance $\ep=10^{-12}$, where $\ep\equiv||Dx-b||/||b||$, cannot be reached.
Such relative residual norms tend to stagnate at $\ep\sim10^{-6}$, as is typical for attempts to solve Eq.~(\ref{matrix_times_vector}) in sp.
In practice, this means that, in order to reach a relative residual norm $\ep\ll10^{-6}$, in addition to $v_\mr{sp}\leftarrow D_\mr{sp}u_\mr{sp}$ also the
operation $v_\mr{dp}\leftarrow D_\mr{dp}u_\mr{dp}$ must be implemented, and a dp-solver must be used.
In other words, the code needs the ability to allocate objects of the classes {\tt suv\_dp} or {\tt vec\_dp}, with the same two or six options for the layout of the internal indices,
and to feed them to routines which code the left-multiplication with $\DS$, $\DW$, or $\DB$ in dp.

A peculiar feature of this lattice QCD code is that the gauge field $U$ is always in sp, in order to be conservative on using disk space and memory bandwidth.
Hence, even in the dp-version of the matrix-vector multiplication routine, the Dirac operator depends on a gauge field which is defined in sp, i.e.\ $v_\mr{dp}\leftarrow D_\mr{dp}[U_\mr{sp}]u_\mr{dp}$.
In Sec.~\ref{sec:inverters} it will be shown that this is perfectly sufficient%
\footnote{This should not come as a surprise; one might think of the routine $v_\mr{dp}\leftarrow D_\mr{dp}[U_\mr{dp}]u_\mr{dp}$ as a routine which
operates on a ``copy'' $U_\mr{dp}$ where all significant digits which are not present in $U_\mr{sp}$ are set to zero.}
to reach relative residual norms as small as $\ep=10^{-12}$.

Another feature is that the Dirac operators need not necessarily use the original (``thin link'') gauge field $U_\mu(n)$,
but may use a smeared (``fat link'') gauge background which is referred to as $V_\mu(n)$ in this article.
Specifically a stout-smearing routine \cite{Morningstar:2003gk} is included%
\footnote{An overview of the ancillary code distribution is given in App.~\ref{app:F}.}
in the module {\tt testknl\_util.f90}, with parameters $\rh_\mr{stout}=0.12,n_\mr{stout}=3$ as default%
\footnote{The user who prefers an unsmeared gauge field should set $n_\mr{stout}=0$; this will establish $V_\mu(n)=U_\mu(n)$.}.

In summary, the goal of this article is to find out whether one can use a high-level language such as Fortran\,2008 to write a lattice QCD code which performs well on modern many-core architectures.
By construction such a code is fully portable, but the decision to stay away from assembly-tuning and cache-line optimization means that all the burden is on the compiler.
Our strategy is to chose a suitable data layout, and to let the compiler know anything which eases the task of creating fast-performing code.
For a modern compiler it is ideal if the trip counts and the array extents are statically%
\footnote{With such knowledge in hand, more informed vectorization cost-model decisions (whether or not to peel, unroll factors, etc.) are possible.
Also compiler-based analysis of alignment for vectorization becomes more effective, and prefetch distances chosen by the compiler tend to be more adequate.
Finally, the compiler is able to do more efficient outer-loop optimizations,
for instance PRE [partial redundancy elimination] for address calculations, and PDSE [partial dead store elimination].}
known (via {\tt parameter} in Fortran or {\tt \#define} in C).
In the remainder of this paper we will explore whether this approach works in practice.


\section{Norms, dot-products and multiply-adds\label{sec:norms}}


A Krylov-space solver such as CG and BiCGstab is designed to solve Eq.~(\ref{matrix_times_vector}) via an iterative process \cite{Hestenes,VanDerVorst,Saad}.
The actual computer implementation (to be discussed in Sec.~\ref{sec:inverters} below) involves two ingredients.
On the one hand, a fast matrix-vector multiplication routine for the chosen operator $D\in\{\DW,\DB,\DS\}$ on a given (possibly smeared) gauge background $V_\mu(n)$ is required.
On the other hand, some linear algebra routines are needed, in particular one which determines the squared 2-norm $||v||_2^2$, and the dot-product%
\footnote{Throughout the article we use the physics convention where $\<u,v\>\in\mb{C}$ is linear in the \emph{second} argument.}
$\<u,v\>$ between two vectors.
The matrix-vector routines for the three Dirac operators will be presented in Secs.~\ref{sec:wils}, \ref{sec:bril} and \ref{sec:stag} (along with timings).
Here we provide similar information for the linear algebra routines mentioned.

For the first staggered layout (locally {\tt NcNv}) the norm-square routine takes the form
\small
\begin{verbatim}
function suv_normsqu_NcNv_sp(suv)
   implicit none
   complex(kind=sp),dimension(Nc,Nv,Nx,Ny,Nz,Nt),intent(in) :: suv
   real(kind=sp),dimension(Nv) :: suv_normsqu_NcNv_sp
   real(kind=dp),dimension(Nv) :: res !!! note: accumulation variable in "dp"
   integer :: x,y,z,t,rhs
   res(:)=0.0_dp
   !$OMP PARALLEL DO COLLAPSE(2) REDUCTION(+:res) SHARED(suv) ...
   do t=1,Nt
   do z=1,Nz
   do y=1,Ny
   do x=1,Nx
      !$OMP SIMD
      do rhs=1,Nv
         res(rhs)=res(rhs)+sum(myabssqu_sp(suv(:,rhs,x,y,z,t)))
      end do
   end do
   end do
   end do
   end do
   !$OMP END PARALLEL DO
   suv_normsqu_NcNv_sp=real(res,kind=sp)
end function suv_normsqu_NcNv_sp
\end{verbatim}
\normalsize
where the {\tt pure elemental} function {\tt myabssqu\_sp} returns $|z|^2$ in dp (for $z\in\mb{C}$ in sp), the relevant line being
{\tt myabssqu\_sp=dble(real(z)**2)+dble(aimag(z)**2)}.
This piece of code illustrates important principles of the ancillary code distribution.
Besides the (correctly ordered) loops over the space-time indices {\tt x,y,z,t} there is a sum over color,
and the loop over the RHS-index {\tt rhs} is equipped with a pragma that instructs the compiler to use it for filling the SIMD pipeline.
The construct {\tt !\$OMP PARALLEL DO} lets the compiler organize the work share among the $\Nthr$ threads which are spanned by the $z$ and $t$ loops, due to the clause {\tt COLLAPSE(2)}.
The variable {\tt suv} is shared, and the clause {\tt REDUCTION(+:res)} means that the results accumulated by individual threads are combined into the variable {\tt res}.
It is worth pointing out that all accumulation is done in dp, despite the input and output variables being in sp.

\begin{table}[tb]
\centering
\begin{tabular}{|c|cc|cc|}
\hline
{} & \multicolumn{2}{c|}{single precision} & \multicolumn{2}{c|}{double precision} \\
{} & $\Nthr=136$ & $\Nthr=272$ & $\Nthr=136$ & $\Nthr=272$ \\
\hline
{\tt NcNsNv} & 0.0090 & 0.0091 & 0.0174 & 0.0183 \\
{\tt NsNcNv} & 0.0088 & 0.0091 & 0.0173 & 0.0183 \\
{\tt NcNvNs} & 0.0089 & 0.0091 & 0.0174 & 0.0183 \\
{\tt NvNcNs} & 0.0109 & 0.0092 & 0.0173 & 0.0182 \\
{\tt NsNvNc} & 0.0188 & 0.0173 & 0.0177 & 0.0183 \\
{\tt NvNsNc} & 0.0104 & 0.0093 & 0.0174 & 0.0182 \\
\hline
{\tt NcNv}   &  0.0045 & 0.0040 & 0.0044 & 0.0046 \\
{\tt NvNc}   &  0.0028 & 0.0023 & 0.0044 & 0.0046 \\
\hline
\end{tabular}
\caption{\label{tab:vec_normsqu_knl}
Time in seconds to compute all $\Nv$ norms of a multi-RHS vector on the 68-core KNL architecture with the array allocated in MCDRAM.
The upper part is for Wilson-type vectors (spinor and color structure), the lower part for Susskind-type vectors (color structure only).
The lattice size is $34^3\times68$ with parameters $\Nc=3$, $\Nv=12$.
The rows refer to the array layout, the columns to the array precision and the number of OpenMP threads used.
Regardless of the vector precision, the accumulation variable of the $\Nv$ norms is in double precision.}
\bigskip
\begin{tabular}{|c|ccc|ccc|}
\hline
{} & \multicolumn{3}{c|}{single precision} & \multicolumn{3}{c|}{double precision} \\
{} & $\Nthr=24$ & $\Nthr=48$ & $\Nthr=96$ & $\Nthr=24$ & $\Nthr=48$ & $\Nthr=96$ \\
\hline
{\tt NcNsNv} & 0.0162 & 0.0134 & 0.0136 & 0.0278 & 0.0267 & 0.0270 \\
{\tt NsNcNv} & 0.0163 & 0.0134 & 0.0135 & 0.0278 & 0.0266 & 0.0270 \\
{\tt NcNvNs} & 0.0165 & 0.0135 & 0.0136 & 0.0277 & 0.0267 & 0.0270 \\
{\tt NvNcNs} & 0.0172 & 0.0135 & 0.0135 & 0.0279 & 0.0266 & 0.0269 \\
{\tt NsNvNc} & 0.0189 & 0.0136 & 0.0135 & 0.0285 & 0.0266 & 0.0269 \\
{\tt NvNsNc} & 0.0172 & 0.0135 & 0.0135 & 0.0280 & 0.0267 & 0.0269 \\
\hline
{\tt NcNv}   & 0.0045 & 0.0035 & 0.0033 & 0.0072 & 0.0067 & 0.0069 \\
{\tt NvNc}   & 0.0042 & 0.0033 & 0.0033 & 0.0071 & 0.0068 & 0.0069 \\
\hline
\end{tabular}
\caption{\label{tab:vec_normsqu_sky}
Same as Tab.~\ref{tab:vec_normsqu_knl}, but on the $2\times24$-core (dual socket) Skylake architecture.}
\end{table}

Such a routine needs to be written for each internal index ordering, i.e.\ two sp-routines for Susskind-type vectors and six for Wilson-type vectors.
In addition, a slightly modified version of these eight routines needs to be provided for input/output variables in dp (here the accumulation variable is still in dp, not in quadruple precision).
Overall, sixteen routines want to be tested, for various values of $\Nthr$.
Depending on the hyperthreading capabilities of the architecture, the values are $\Nthr\in\{2N_\mr{hw},4N_\mr{hw}\}$ for four threads per hardware core (e.g.\ the KNL chip), or $\Nthr\in\{N_\mr{hw},2N_\mr{hw}\}$ for two threads per hardware core (e.g.\ the Skylake chip).
The KNL data were obtained on a single-chip node with $N_\mr{hw}=68$ physical cores, the Skylake data on a dual-socket node with a total of $N_\mr{hw}=2\times24=48$ physical cores.

\begin{table}[tb]
\centering
\begin{tabular}{|c|cc|cc|}
\hline
{} & \multicolumn{2}{c|}{single precision} & \multicolumn{2}{c|}{double precision} \\
{} & $\Nthr=136$ & $\Nthr=272$ & $\Nthr=136$ & $\Nthr=272$ \\
\hline
{\tt NcNsNv} & 0.0392 & 0.0302 & 0.0575 & 0.0504 \\
{\tt NsNcNv} & 0.0268 & 0.0277 & 0.0448 & 0.0500 \\
{\tt NcNvNs} & 0.0296 & 0.0240 & 0.0426 & 0.0465 \\
{\tt NvNcNs} & 0.0214 & 0.0221 & 0.0424 & 0.0445 \\
{\tt NsNvNc} & 0.0214 & 0.0228 & 0.0433 & 0.0456 \\
{\tt NvNsNc} & 0.0215 & 0.0221 & 0.0425 & 0.0447 \\
\hline
{\tt NcNv}   & 0.0068 & 0.0055 & 0.0107 & 0.0111 \\
{\tt NvNc}   & 0.0054 & 0.0057 & 0.0107 & 0.0112 \\
\hline
\end{tabular}
\caption{\label{tab:vec_incr_knl}
Time in seconds to perform all $\Nv$ increment operations $u^{(i)}=u^{(i)}+v^{(i)}\al^{(i)}$ with $\al^{(i)}\in\mb{C}$ and $i=1,\ldots,\Nv$
on the 68-core KNL architecture with the multi-RHS vectors $u,v$ allocated in MCDRAM.
The lattice size is $34^3\times68$ with parameters $\Nc=3$, $\Nv=12$.}
\bigskip
\begin{tabular}{|c|ccc|ccc|}
\hline
{} & \multicolumn{3}{c|}{single precision} & \multicolumn{3}{c|}{double precision} \\
{} & $\Nthr=24$ & $\Nthr=48$ & $\Nthr=96$ & $\Nthr=24$ & $\Nthr=48$ & $\Nthr=96$ \\
\hline
{\tt NcNsNv} & 0.0584 & 0.0484 & 0.0475 & 0.1399 & 0.1087 & 0.1034 \\
{\tt NsNcNv} & 0.0507 & 0.0463 & 0.0479 & 0.1392 & 0.1091 & 0.1033 \\
{\tt NcNvNs} & 0.0468 & 0.0466 & 0.0450 & 0.0932 & 0.0902 & 0.0906 \\
{\tt NvNcNs} & 0.0471 & 0.0469 & 0.0451 & 0.0934 & 0.0925 & 0.0903 \\
{\tt NsNvNc} & 0.0463 & 0.0457 & 0.0452 & 0.0943 & 0.0901 & 0.0905 \\
{\tt NvNsNc} & 0.0470 & 0.0465 & 0.0451 & 0.0936 & 0.0930 & 0.0903 \\
\hline
{\tt NcNv}   & 0.0119 & 0.0115 & 0.0116 & 0.0238 & 0.0234 & 0.0233 \\
{\tt NvNc}   & 0.0118 & 0.0116 & 0.0115 & 0.0237 & 0.0231 & 0.0232 \\
\hline
\end{tabular}
\caption{\label{tab:vec_incr_sky}
Same as Tab.~\ref{tab:vec_incr_knl}, but on the $2\times24$-core (dual socket) Skylake architecture.}
\end{table}

The timings of the routines {\tt $\{$vec,suv$\}$\_normsqu\_$\{$sp,dp$\}$} are listed in Tabs.~\ref{tab:vec_normsqu_knl}, \ref{tab:vec_normsqu_sky} for the KNL and Skylake architectures, respectively.
The figures give the time needed to compute all $\Nv=12$ norm-squares with $\Nc=3$ colors on a lattice with $2\,672\,672$ sites.
A peculiarity of the KNL architecture is that the lattice fits into the MCDRAM, and the vector is initialized with the same number of threads.
This ``first touch'' policy is used in all subsequent timings.
On the KNL the MCDRAM high-bandwidth memory has an aggregate bandwidth%
\footnote{See e.g.\ {\tt https://colfaxresearch.com/knl-mcdram}.}
of about 450\,GB/s.
In sp the $\Nv$ (complex) staggered vectors occupy $8\Nc\Nv N_xN_yN_zN_t=769\,729\,536$ bytes in memory, hence transferring them to the registers takes 0.0017\,s (assuming zero latency).
In dp the bandwidth limit amounts to 0.0034\,s, and for Wilson-type vectors these figures are four-fold increased to 0.0068\,s and 0.0136\,s, respectively.
Comparing the actual entries in Tab.~\ref{tab:vec_normsqu_knl} to these lower bounds, we see that most of the timings are reasonably close to it,
only the staggered {\tt NcNv} layout in sp takes considerably longer.
Regarding two-fold versus four-fold hyperthreading, there is no universal law on the KNL architecture; sometimes one option is faster, sometimes the other.
The timings in Tab.~\ref{tab:vec_normsqu_sky} for the Skylake architecture are generally slower, but not dramatically so.
On this architecture the ordering of the internal indices seems irrelevant, and the norm-square operation does not benefit from two-fold hyperthreading.

\begin{table}[tb]
\centering
\begin{tabular}{|c|cc|cc|}
\hline
{} & \multicolumn{2}{c|}{single precision} & \multicolumn{2}{c|}{double precision} \\
{} & $\Nthr=136$ & $\Nthr=272$ & $\Nthr=136$ & $\Nthr=272$ \\
\hline
{\tt NcNsNv} & 0.0162 & 0.0174 & 0.0241 & 0.0253 \\
{\tt NsNcNv} & 0.0162 & 0.0176 & 0.0322 & 0.0346 \\
{\tt NcNvNs} & 0.0143 & 0.0114 & 0.0205& 0.0246 \\
{\tt NvNcNs} & 0.0101 & 0.0105 & 0.0200 & 0.0205 \\
{\tt NsNvNc} & 0.0162 & 0.0175 & 0.0322 & 0.0346 \\
{\tt NvNsNc} & 0.0112 & 0.0120 & 0.0219 & 0.0233 \\
\hline
{\tt NcNv}   & 0.0027 & 0.0030 & 0.0050 & 0.0051 \\
{\tt NvNc}   & 0.0026 & 0.0027 & 0.0048 & 0.0049 \\
\hline
\end{tabular}
\caption{\label{tab:vec_gamma_knl}
Time in seconds to apply a $\ga$-matrix to all $\Nv$ columns of a multi-RHS vector on the 68-core KNL architecture with the array allocated in MCDRAM.
The upper part is for $\gamma_5$ and Wilson-type vectors (spinor and color structure), the lower part for $\ep$ and Susskind-type vectors (color structure only).
The lattice size is $34^3\times68$ with parameters $\Nc=3$, $\Nv=12$.}
\bigskip
\begin{tabular}{|c|ccc|ccc|}
\hline
{} & \multicolumn{3}{c|}{single precision} & \multicolumn{3}{c|}{double precision} \\
{} & $\Nthr=24$ & $\Nthr=48$ & $\Nthr=96$ & $\Nthr=24$ & $\Nthr=48$ & $\Nthr=96$ \\
\hline
{\tt NcNsNv} & 0.0326 & 0.0328 & 0.0325 & 0.0525 & 0.0505 & 0.0514 \\
{\tt NsNcNv} & 0.0327 & 0.0330 & 0.0324 & 0.0658 & 0.0657 & 0.0651 \\
{\tt NcNvNs} & 0.0226 & 0.0208 & 0.0205 & 0.0455 & 0.0429 & 0.0416 \\
{\tt NvNcNs} & 0.0210 & 0.0205 & 0.0197 & 0.0454 & 0.0431 & 0.0414 \\
{\tt NsNvNc} & 0.0326 & 0.0330 & 0.0325 & 0.0658 & 0.0658 & 0.0652 \\
{\tt NvNsNc} & 0.0223 & 0.0217 & 0.0210 & 0.0470 & 0.0445 &  0.0435 \\
\hline
{\tt NcNv}   & 0.0061 & 0.0056 & 0.0055 & 0.0110 & 0.0104 & 0.0102 \\
{\tt NvNc}   & 0.0055 & 0.0055 & 0.0053 & 0.0109 & 0.0105 & 0.0102 \\
\hline
\end{tabular}
\caption{\label{tab:vec_gamma_sky}
Same as Tab.~\ref{tab:vec_gamma_knl}, but on the $2\times24$-core (dual socket) Skylake architecture.}
\end{table}

Changing the task from computing the squared norm $||v||_2^2$ to computing the dot-product $\<u,v\>$ will double both the memory traffic, and the flop count,
since $r=r+\mr{Re}^2(v)+\mr{Im}^2(v)$ takes four flops, while $r=r+\mr{Re}(u)\mr{Re}(v)-\mr{Im}(u)\mr{Im}(v), i=i+\mr{Re}(u)\mr{Im}(v)-\mr{Im}(u)\mr{Re}(v)$ takes eight flops.
We thus expect that all timings (regardless of precision, and architecture) will double.
A quick test reveals this is precisely what happens (tables not included).

Another important ingredient in an iterative solver is the vectorial multiply-add operation.
Instead of the generic $w^{(i)}\leftarrow u^{(i)}+v^{(i)}\al^{(i)}$ for $i\in\{1,\ldots,\Nv\}$, which is more demanding on memory bandwidth, we implement two multiply-add routines with overwrite
\bea
{\tt incr:} && u^{(i)}\leftarrow u^{(i)}+v^{(i)}\al^{(i)}\qquad(i=1,\ldots,\Nv)
\label{def_incr}
\\
{\tt anti:} && v^{(i)}\leftarrow u^{(i)}+v^{(i)}\al^{(i)}\qquad(i=1,\ldots,\Nv)
\label{def_anti}
\eea
for $\al^{(i)}$ in $\mb{R}$ or $\mb{C}$.
For the precision of the (complex) vectors there are three possibilities:
($i$) $u,v$ both in sp, ($ii$) $u,v$ both in dp, and ($iii$) $u$ in sp and $v$ in dp (relevant to the mixed-precision solvers mentioned in Sec.~\ref{sec:inverters}).
Furthermore, each of the $\Nv$ vectors may have spinor degrees of freedom (length $4\Nc\Nx\Nx\Nz\Nt$) or not (length $\Nc\Nx\Nx\Nz\Nt$).
Hence, the names of these routines are {\tt $\{$vec,suv$\}$\_$\{$incr,anti$\}$\_$\{$r,c$\}$$\{$sp,dp$\}$}, and the ``{\tt r}'' or ``{\tt c}'' indicates whether the $\al^{(i)}$ are real or complex.
The respective timings on the KNL architecture are summarized in Tab.~\ref{tab:vec_incr_knl}.
In sp the layouts with the color degree of freedom first (i.e.\ {\tt NcNsNv} and {\tt NcNvNs}) take a little longer than the remaining four,
and for the latter ones four-fold hyperthreading does not seem to bring any advantage over two-fold hyperthreading.
Quite generally, there is factor two difference between sp-timings and dp-timings of (\ref{def_incr}) and (\ref{def_anti}).
Analogous timing results on the Skylake architecture are listed in Tab.~\ref{tab:vec_incr_sky}.
In this case using one socket ($N_\mr{thr}=24$) or both sockets without ($N_\mr{thr}=48$) or with ($N_\mr{thr}=96$) hyperthreading yields similar figures.

Another routine that proves relevant below is the left-multiplication with $\ga_5$ for Wilson-type vectors and $\ep\equiv(-1)^{x_1+x_2+x_3+x_4}\doteq\ga_5\otimes\xi_5$ for Susskind-type vectors.
Such results are collected in Tab.~\ref{tab:vec_gamma_knl} for the KNL, and in Tab.~\ref{tab:vec_gamma_sky} for the Skylake architecture.
Once more, we find that the vector layout has a mild effect on the actual timing on the KNL, and essentially no effect on the Skylake node.
And the effect of hyperthreading is negligible on both architectures.

Overall, the chosen vector layout (i.e.\ the order of the internal indices color/spinor/RHS) affects the timings of the linear algebra routines by just a few percent.
As we shall see below, for the Wilson and Susskind operators the matrix-vector operation is about an order of magnitude more expensive than the linear algebra operations.
And the Brillouin operator is almost two orders of magnitude more time consuming.
In view of these forthcoming results, it is fair to say that the linear algebra routines have been optimized to the point where
their CPU share is a subdominant part of the overall solver time (cf.\ Sec.~\ref{sec:inverters} below).


\section{Clover routine\label{sec:clov}}


The clover routine is a matrix-vector routine which is applied%
\footnote{With $c_\mr{SW}=0$ both $\DW$ and $\DB$ induce cut-off effects $O(a)$.
With the tree-level value $c_\mr{SW}=1$ the latter are mitigated to $O(\al a)$, and with the one-loop value to $O(\al^2 a)$, where $\al=g_\mr{ren}^2/(4\pi)$ is the strong coupling constant.
With a non-perturbatively determined value $c_\mr{SW}^\mr{NP}$ cut-off effects can be lifted to $O(a^2)$ \cite{Sheikholeslami:1985ij,Aoki:2003sj,Luscher:1996sc}.}
in addition to the Wilson or Brillouin Dirac operator $D$.
The operation is $v\leftarrow(D+C)u$ with $D\in\{\DW,\DB\}$ and
\beq
C(n,m)=-\frac{c_\mr{SW}}{2}\sum_{\mu<\nu}\si_{\mu\nu}F_{\mu\nu}(n)\de_{n,m}
\label{def_clov}
\eeq
where $n$ and $m$ are positions in the four-dimensional lattice.
It acts on the vector $v$ non-trivially in color and spinor space, but as an identity in RHS space and position space.
In other words, it acts locally in space-time, so $v(n)$ depends on $u(m)$ only for $n=m$.
Depending on the layout (i.e.\ the ordering of the color/spinor/RHS indices) Eq.~(\ref{def_clov}) is thus a shorthand for
\beq
C(n,m)=-\frac{c_\mr{SW}}{2}\sum_{\mu<\nu}\de_{n,m} \otimes
\left\{
\begin{array}{ccc}
I \otimes \si_{\mu\nu} \otimes F_{\mu\nu}(n) & \mbox{for} & {\tt NcNsNv} \\
I \otimes F_{\mu\nu}(n) \otimes \si_{\mu\nu} & \mbox{for} & {\tt NsNcNv} \\
\si_{\mu\nu} \otimes I \otimes F_{\mu\nu}(n) & \mbox{for} & {\tt NcNvNs} \\
\si_{\mu\nu} \otimes F_{\mu\nu}(n) \otimes I & \mbox{for} & {\tt NvNcNs} \\
F_{\mu\nu}(n) \otimes I \otimes \si_{\mu\nu} & \mbox{for} & {\tt NsNvNc} \\
F_{\mu\nu}(n) \otimes \si_{\mu\nu} \otimes I & \mbox{for} & {\tt NvNsNc}
\end{array}
\right.
\label{options_clov}
\eeq
where the sum%
\footnote{Since $\si_{\mu\nu}$ and $F_{\mu\nu}$ are both anti-symmetric in $\mu\leftrightarrow\nu$, the sum may be written without the constraint among $\mu,\nu$, but with a prefactor $c_\mr{SW}/4$ instead of $c_\mr{SW}/2$.}
is over the six plaquette orientations with $\mu<\nu$.
For each orientation the $4\times4$ matrix $\si_{\mu\nu}\equiv\frac{\ri}{2}[\ga_\mu,\ga_\nu]$ is hermitean in spinor space,
and the clover-leaf field-strength operator $F_{\mu\nu}(n)$ is hermitean in color space.
In consequence the clover term (\ref{def_clov}) is a hermitean contribution to the combined matrix-vector operation $v\leftarrow(D+C)u$.

The field-strength operator $F_{\mu\nu}$ is based on the smeared (``fat-link'') gauge field $V_\mu(n)$ which derives from the original (``thin-link'') gauge field $U_\mu(n)$.
Here, any type of smearing may be used; the code uses stout-smearing \cite{Morningstar:2003gk} which produces differentiable links (though this point is not relevant to this article).
The field-strength is precomputed in a routine which takes the (possibly smeared) gauge field {\tt V} as input; the result is stored in the rank-seven array
\small
\begin{verbatim}
complex(kind=sp),dimension(Nc,Nc,6,Nx,Ny,Nz,Nt) :: F
\end{verbatim}
\normalsize
since a solver requests dozens to millions of operations $v\leftarrow(D+C)u$ in which $U$ (and thus $V$) stays unchanged.
Since $C$ depends on $U$ only via $F$, it pays%
\footnote{With HPC architectures becoming increasingly memory bandwidth limited, this may change in the future.}
to compute $F_{\mu\nu}$ only once.
The six orientations stand for $(\mu,\nu)\in\{(1,2),(1,3),(1,4),(2,3),(2,4),(3,4)\}$.

\begin{table}[tb]
\centering
\begin{tabular}{|c|cc|cc|}
\hline
{} & \multicolumn{2}{c|}{single precision} & \multicolumn{2}{c|}{double precision} \\
{} & $\Nthr=136$ & $\Nthr=272$ & $\Nthr=136$ & $\Nthr=272$ \\
\hline
{\tt NcNsNv} & 0.0426 & 0.0375 & 0.0737 & 0.0642 \\
{\tt NsNcNv} & 0.0424 & 0.0375 & 0.0728 & 0.0635 \\
{\tt NcNvNs} & 0.0461 & 0.0447 & 0.0640 & 0.0595 \\
{\tt NvNcNs} & 0.0349 & 0.0296 & 0.0617 & 0.0570 \\
{\tt NsNvNc} & 0.0435 & 0.0370 & 0.0634 & 0.0589 \\
{\tt NvNsNc} & 0.0340 & 0.0291 & 0.0547 & 0.0559 \\
\hline
\end{tabular}
\caption{\label{tab:clov_knl}
Time in seconds to apply the clover term (\ref{def_clov}) to all $\Nv$ columns of a multi-RHS vector on the 68-core KNL architecture with the vector and the precomputed field-strength allocated in MCDRAM.
The lattice size is $34^3\times68$ with parameters $\Nc=3$, $\Nv=12$.
The best timings correspond to 2500\,Gflop/s in sp, and 1300\,Gflop/s in dp -- see App.~\ref{app:E} for details.}
\bigskip
\begin{tabular}{|c|ccc|ccc|}
\hline
{} & \multicolumn{3}{c|}{single precision} & \multicolumn{3}{c|}{double precision} \\
{} & $\Nthr=24$ & $\Nthr=48$ & $\Nthr=96$ & $\Nthr=24$ & $\Nthr=48$ & $\Nthr=96$ \\
\hline
{\tt NcNsNv} & 0.0606 & 0.0519 & 0.0519 & 0.1222 & 0.1004 & 0.0979 \\
{\tt NsNcNv} & 0.0612 & 0.0519 & 0.0518 & 0.1212 & 0.0994 & 0.0982 \\
{\tt NcNvNs} & 0.0621 & 0.0523 & 0.0519 & 0.1137 & 0.0977 & 0.0973 \\
{\tt NvNcNs} & 0.0560 & 0.0516 & 0.0520 & 0.1148 & 0.0986 & 0.0975 \\
{\tt NsNvNc} & 0.0605 & 0.0520 & 0.0521 & 0.1140 & 0.0971 & 0.0974 \\
{\tt NvNsNc} & 0.0560 & 0.0513 & 0.0520 & 0.1168 & 0.0991 & 0.0976 \\
\hline
\end{tabular}
\caption{\label{tab:clov_sky}
Same as Tab.~\ref{tab:clov_knl}, but on the $2\times24$-core (dual socket) Skylake architecture.
The best timings correspond to 1450\,Gflop/s in sp, and 750\,Gflop/s in dp.}
\end{table}

\begin{figure}[p]
\vspace*{-8mm}
\includegraphics[width=0.99\textwidth]{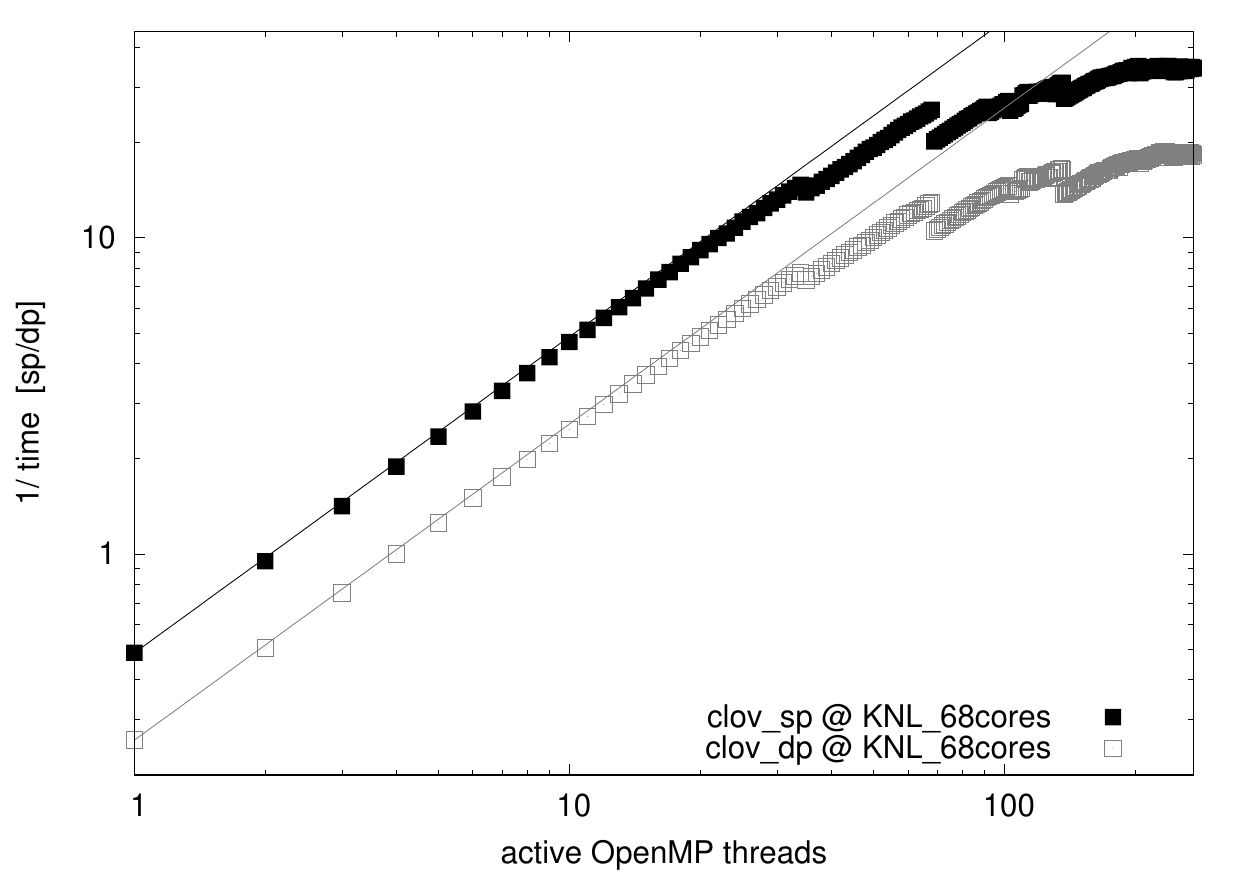}\\
\includegraphics[width=0.99\textwidth]{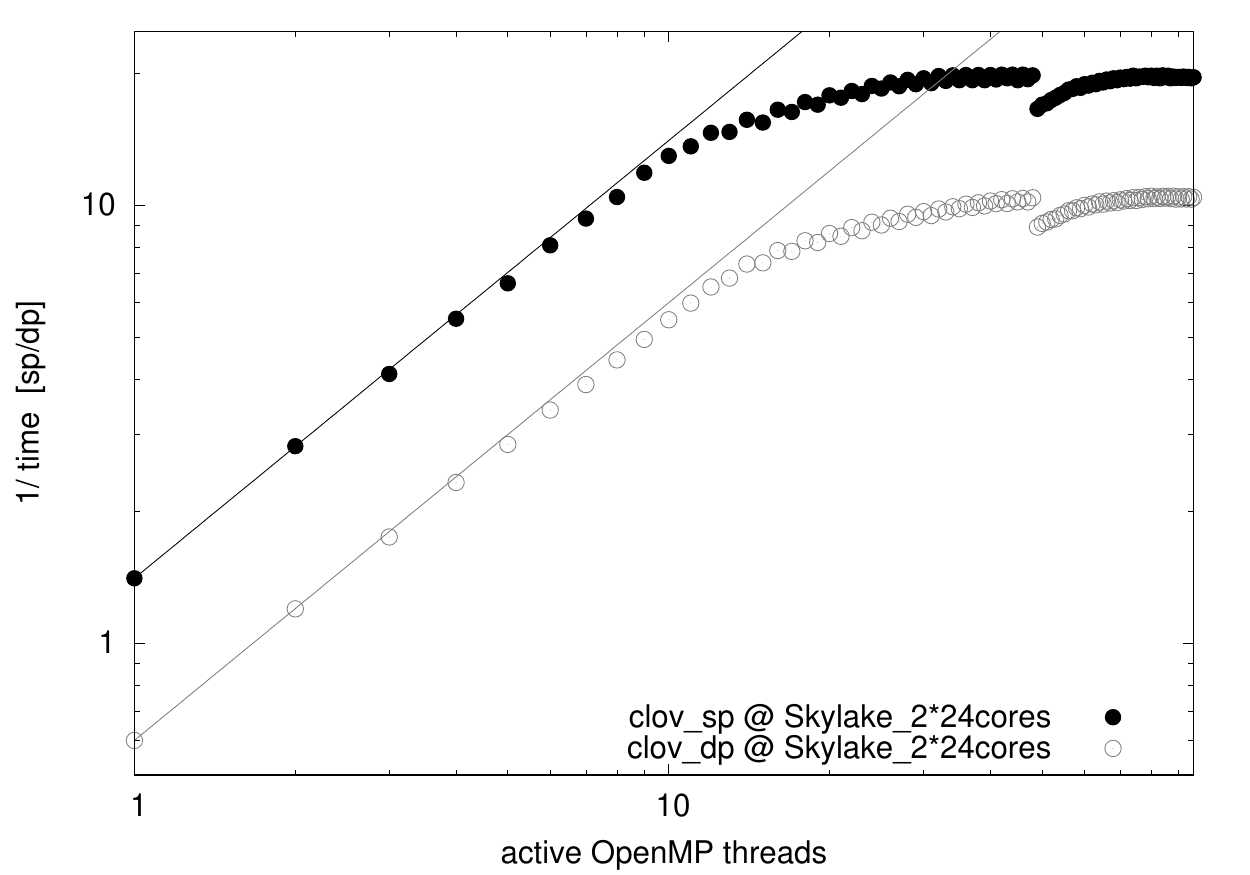}
\vspace*{-3mm}
\caption{\label{fig:clov}%
Clover term performance versus the number of active threads in sp and dp, for the KNL and the dual-socket Skylake architectures (same parameters as in Tab.~\ref{tab:clov_knl}).}
\end{figure}

The clover routine takes the field-strength array {\tt F} as input, as well as the vector {\tt old}, denoted $u$ in the line above (\ref{def_clov}).
The vector {\tt new} at position $n$, i.e.\ $v(n)$ above (\ref{def_clov}), is incremented by $-\frac{1}{2}c_\mr{SW}\sum_{\mu<\nu}\si_{\mu\nu}F_{\mu\nu}(n)$ applied to $u(n)$.
The operation is thus \emph{site-diagonal}, and this means that the OpenMP thread-parallelization is easily achieved by declaring {\tt SHARED(old,new,F)} in the {\tt !\$OMP PARALLEL DO} construct.
A nice side-effect is that any read-collision or write-collision among the threads is excluded by construction, since each thread reads from (and writes to) a specified segment in memory,
with no overlap among the segments (there is an implied barrier at the end of the {\tt !\$OMP PARALLEL DO} construct).
The structure of the routine is thus similar to the function {\tt suv\_normsqu\_sp} displayed in Sec.~\ref{sec:norms}, with four nested do-loops to cover all space-time points.
Since all site-index computations are handled by the compiler no large lookup-table is needed.
The reader is reminded that the coding must be done for sp and dp vectors separately.
Within sp or dp, also for each vector layout a dedicated routine is needed to achieve good performance
(with a wrapper routine which presents them as a single routine to the outside world).

The timings of the clover routine (which is specific to Wilson-type vectors) are listed in Tabs.~\ref{tab:clov_knl}, \ref{tab:clov_sky} for the KNL and Skylake architectures, respectively.
On the KNL, the layouts {\tt NvNcNs} and {\tt NvNsNc} are faster than the remaining four.
This is unsurprising, since the SIMD operation, which is over the RHS-index {\tt 1:Nv}, should affect the fastest (in Fortran: first) index of the array.
Obviously, this performance hierarchy among the six possible layouts will persist in the Wilson-type Dirac operators.
In the sections on $\DW,\DB$ we shall thus restrict ourselves to these two layouts, plus the most naive {\tt NcNsNv} for comparison.
On the Skylake architecture, the performance of the clover term is essentially the same for all layout options.
On the latter architecture the memory bandwidth is the limiting factor; so the AVX-512 instruction set is pointless for this routine (though it exists on the Skylake architecture).

The scaling of the clover routine (in sp and dp, for the {\tt NvNsNc} layout, with $34^3\times68$ volume, and $\Nc=3$, $\Nv=12$) as a function of the number of active threads is shown in Fig.~\ref{fig:clov}.
On the KNL architecture the number of threads ranges from $1$ to $4N_\mr{cpu}=272$, where $N_\mr{cpu}=68$ is the number of physical cores.
On the Skylake architecture the number of threads ranges from $1$ to $2N_\mr{cpu}=96$, where $N_\mr{cpu}=48$ is the number of physical cores in the full dual-socket node.
On the KNL chip we find essentially perfect scaling behavior until every physical core hosts one thread.
There is a reduced slope associated with the second thread on a physical core (from $69$ to $136$), and modest improvement with the third (from $137$ to $204$) and fourth (from $205$ to $272$) thread on a given core.
On the Skylake architecture saturation effects set in at $O(10)$ threads; this is a clear sign of the process being limited by memory bandwidth.

A time of $30\,\mr{ms}$ for $\Nv=12$ in Tab.~\ref{tab:clov_knl} amounts to the clover operator being applied to a single vector with $3\cdot4\cdot34^3\cdot68=32\,072\,064$ complex elements within $2.5\,\mr{ms}$.
As we shall see in Secs.~\ref{sec:wils} and \ref{sec:bril}, this is about one third of the time needed to apply the Wilson Dirac operator (\ref{def_wils}),
and the Brillouin Dirac operator (\ref{def_bril}) takes an order of magnitude longer.
In short, the clover routine has been optimized to the point where further optimization would speed up the solver routines (to be discussed in Sec.~\ref{sec:inverters} below) only marginally.


\section{Wilson Laplace and Dirac routines\label{sec:wils}}


For a given $V_\mu(n)$ the Wilson Dirac operator with optional clover term (\ref{def_clov}) is defined as~\cite{Wilson:1974sk}
\beq
D_\mr{W}(n,m)=\sum_\mu \ga_\mu \nab_\mu^\mr{std}(n,m)
-\frac{ar}{2}\lap^\mr{std}(n,m)+m_0\de_{n,m}
+aC(n,m)
\label{def_wils}
\eeq
where $a$ is the lattice spacing, $\nab_\mu^\mr{std}$ is the 2-point discretization of the covariant derivative
\beq
a\nab_\mu^\mr{std}(n,m)=\frac{1}{2}\,\big[V_{\mu}(n)\de_{n+\hat\mu,m}-V_\mu\dag(n-\hat\mu)\de_{n-\hat\mu,m}\big]
\label{def_sder}
\eeq
and $\lap^\mr{std}$ is the 9-point discretization of the covariant Laplacian
\beq
a^2\lap^\mr{std}(n,m)=-\,8\,\de_{n,m}+\sum_\mu\big[V_\mu(n)\de_{n+\hat\mu,m}+V_\mu\dag(n-\hat\mu)\de_{n-\hat\mu,m}\big]
\;.
\label{def_wlap}
\eeq
The sums in (\ref{def_wils}, \ref{def_wlap}) extend over the positive Euclidean directions $\mu\in\{1,\ldots,4\}$, and the bare quark mass $m_0$ undergoes both additive and multiplicative renormalization.
How the $V$-links in $\nab_\mu^\mr{std},\lap^\mr{std}$ and $C$ relate to the original $U$-links has been explained in Sec.~\ref{sec:clov}.
Note that the ``standard derivative'' (\ref{def_sder}) is anti-hermitean, while the ``standard/Wilson Laplacian'' \ref{def_wlap} and the clover term are hermitean operators.
The species-lifting parameter is typically set to $r=1$, and the operator~(\ref{def_wils}) is HPC friendly, since its stencil contains at most 1-hop terms.

The action of the Wilson operator~(\ref{def_wils}) at $r=1,c_\mr{SW}=0$ on a Dirac vector $\ps$ (spinor$\otimes$color internal degrees of freedom) with periodic boundary conditions in all directions is given by
\beq
(D_\mr{W}\ps)(n)=
\frac{1}{2}\sum_\mu
\big\{
[(\ga_\mu-1)\otimes V_\mu(n)]\ps(n+\hat\mu)-[(\ga_\mu+1)\otimes V_\mu\dag(n-\hat\mu)]\ps(n-\hat\mu)
\big\}
+(4+m_0)\ps(n)
\label{app_wils}
\eeq
and our task is to implement a routine which performs this operation efficiently.
The action of the embedded Laplace operator (\ref{def_wlap}) alone (which we implement for comparison) is
\beq
\Big(-\frac{1}{2}\lap^\mr{std}+\frac{m_0^2}{2}\Big)\ps(n)=
\frac{1}{2}\sum_\mu
\big[
-V_\mu(n)\ps(n+\hat\mu)-V_\mu\dag(n-\hat\mu)\ps(n-\hat\mu)
\big]
+(4+\frac{m_0^2}{2})\ps(n)\quad
\label{app_wlap}
\eeq
with the mass parameter $m_0$ replaced by $m_0^2/2$, in line with the standard boson propagator in quantum field theory.
In (\ref{app_wils}) and (\ref{app_wlap}) we have taken the liberty to set the lattice spacing $a=1$.

\begin{table}[tb]
\centering
\begin{tabular}{|c|cc|cc|}
\hline
{} & \multicolumn{2}{c|}{single precision} & \multicolumn{2}{c|}{double precision} \\
{} & $\Nthr=136$ & $\Nthr=272$ & $\Nthr=136$ & $\Nthr=272$ \\
\hline
{\tt NcNsNv} & 0.1387 & 0.1375 & 0.2094 & 0.2162 \\
{\tt NvNcNs} & 0.1031 & 0.0973 & 0.1808 & 0.1727 \\
{\tt NvNsNc} & 0.0749 & 0.0745 & 0.1485 & 0.1602 \\
\hline
\end{tabular}
\caption{\label{tab:wlap_knl}
Time in seconds per matrix times multi-RHS vector operation for the Wilson Laplace operator (\ref{def_wlap}) on the 68-core KNL architecture, with all variables allocated in MCDRAM.
The lattice size is $34^3\times68$ with parameters $\Nc=3$, $\Nv=12$.
The best timings correspond to 1000\,Gflop/s in sp, and 500\,Gflop/s in dp -- see App.~\ref{app:E} for details.}
\bigskip
\begin{tabular}{|c|ccc|ccc|}
\hline
{} & \multicolumn{3}{c|}{single precision} & \multicolumn{3}{c|}{double precision} \\
{} & $\Nthr=24$ & $\Nthr=48$ & $\Nthr=96$ & $\Nthr=24$ & $\Nthr=48$ & $\Nthr=96$ \\
\hline
{\tt NcNsNv} & 0.1366 & 0.1148 & 0.1197 & 0.2581 & 0.2215 & 0.2450 \\
{\tt NvNcNs} & 0.1262 & 0.1141 & 0.1190 & 0.2579 & 0.2216 & 0.2451 \\
{\tt NvNsNc} & 0.1221 & 0.1141 & 0.1192 & 0.2390 & 0.2200 & 0.2443 \\
\hline
\end{tabular}
\caption{\label{tab:wlap_sky}
Same as Tab.~\ref{tab:wlap_knl}, but on the $2\times24$-core (dual socket) Skylake architecture.
The best timings correspond to 650\,Gflop/s in sp, and 340\,Gflop/s in dp.}
\end{table}

\begin{figure}[p]
\vspace*{-8mm}
\includegraphics[width=0.99\textwidth]{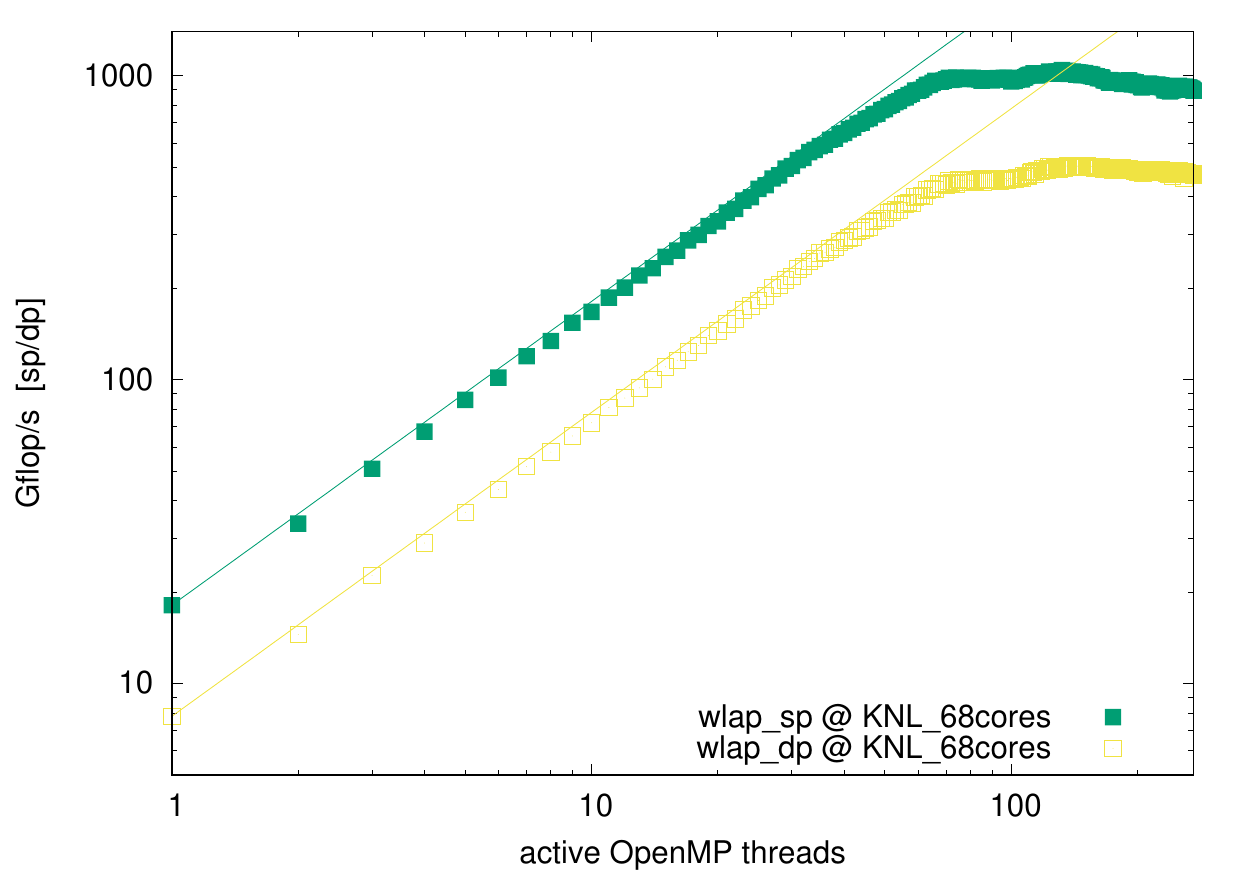}\\
\includegraphics[width=0.99\textwidth]{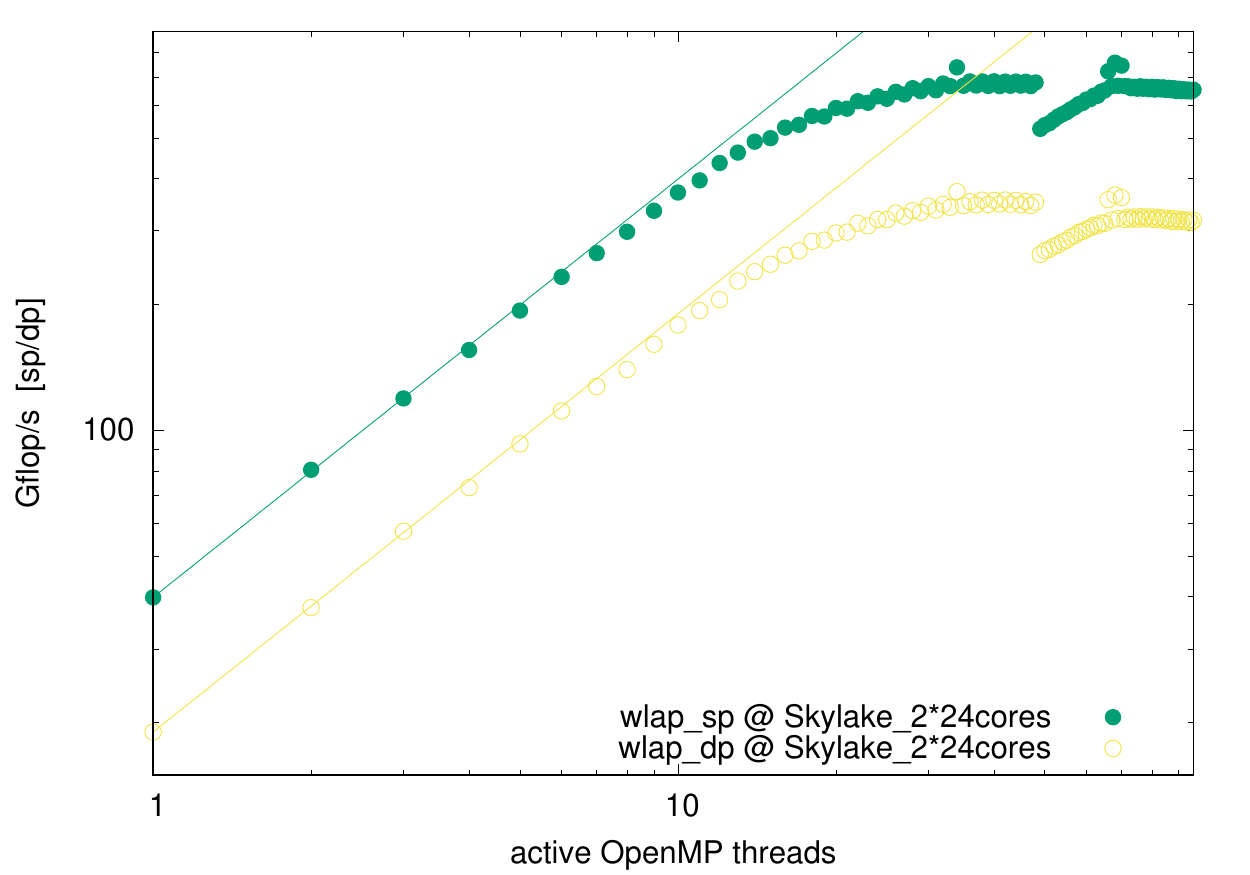}
\vspace*{-3mm}
\caption{\label{fig:wlap}%
Wilson Laplace operator performance versus the number of threads in sp and dp, for the KNL and the dual-socket Skylake architectures (same parameters as in Tab.~\ref{tab:wlap_knl}).}
\end{figure}

From a coding viewpoint it is clear that the Wilson Laplace operator (\ref{app_wlap}) is easier to implement than the Wilson Dirac operator (\ref{app_wils}), since it acts trivially%
\footnote{Hence one might let $\lap^\mr{std}$ act on a Susskind-type vector (color structure only), and the ancillary code distribution
contains such a multiplication routine under the label {\tt app\_wsuv\_$\{$sp,dp$\}$}, see App.~\ref{app:F} for details.\label{foot:wsuv}}
in spinor space, and we proceed with implementation details of this routine.
The main difference to the clover routine discussed in the previous section is that the Wilson Laplacian (\ref{app_wlap}) is \emph{not} site diagonal;
it has a 9-point stencil, since $\ps_\mr{new}(n)$ depends, besides $\ps_\mr{old}(n)$, on the eight points $\ps_\mr{old}(n\pm\hat\mu)$ with $\mu\in\{1,2,3,4\}$.
From the perspective of a given thread, the problem with SMP shared-memory parallelization is that these data may lay in the patch of memory which is associated%
\footnote{No inconsistency may arise, but read/write-collisions among threads are detrimental to the performance.}
with another thread.
Hence read-collisions (or even worse write-collisions) may arise, if different threads attempt to read from (or write to) one portion of memory at the same time.
The good news is that there is --~within the SMP-paradigm~-- a simple and effective strategy which bans write-collisions fully, and makes read-collisions very unlikely \cite{Durr:2017clx,Durr:2018ayq}.
One accumulates, for each $n$, the nine contributions (from the central point and from the eight nearest neighbors) in a thread-private variable, e.g.\ {\tt site(1:Nv,1:4,1:Nc)}, and writes it \emph{once} to $\ps_\mr{new}(n)$.
The design of the routine is thus governed by a set of four nested loops (over the $x,y,z,t$-directions, respectively) that generate the space-time point $n$ of the out-vector $\ps_\mr{new}$,
with a {\tt SHARED(old,new,V)} clause in the {\tt !\$OMP PARALLEL DO} construct.
In the accumulation process one uses {\tt !\$OMP SIMD} pragmas to vectorize the summation over the RHS-index, with explicit unrolling of color/spinor indices inside.
The usual comments regarding separate implementations of the sp/dp-versions (and the various vector layouts) apply; for details see App.~\ref{app:B}.

The timings of the Wilson Laplace routine are listed in Tabs.~\ref{tab:wlap_knl}, \ref{tab:wlap_sky} for the KNL and Skylake architectures, respectively.
On the KNL chip the vector layout is important; for good performance the SIMD index {\tt rhs} must be first, and the option {\tt NvNsNc} wins the contest.
On the Skylake node all vector layouts deliver comparable speed.
Most notably, already $24$ threads yield almost maximal performance; hence for this routine the second socket
(which is populated by threads $25-48$ and $73-96$ under full load) is essentially pointless.
For the Wilson Laplacian the Skylake timing surplus (relative to KNL) is about a factor two.

The scaling of the Wilson Laplace routine (in sp and dp, for the {\tt NvNsNc} layout) as a function of the number of active threads is shown in Fig.~\ref{fig:wlap}.
The parameters, and the range over which the number of threads is varied, are the same as in Sec.~\ref{sec:clov}.
On the KNL architecture we find nearly perfect scaling behavior until every physical core hosts one thread.
A second thread per core brings a tiny improvement, while a third and fourth thread tend to deteriorate performance.
On the Skylake architecture the bottleneck in memory bandwidth is effective from $O(20)$ threads; and there is a local maximum at $68$ threads.


\begin{table}[tb]
\centering
\begin{tabular}{|c|cc|cc|}
\hline
{} & \multicolumn{2}{c|}{single precision} & \multicolumn{2}{c|}{double precision} \\
{} & $\Nthr=136$ & $\Nthr=272$ & $\Nthr=136$ & $\Nthr=272$ \\
\hline
{\tt NcNsNv} & 0.1269 & 0.1306 & 0.1953 & 0.2246 \\
{\tt NvNcNs} & 0.1025 & 0.1013 & 0.2032 & 0.1992 \\
{\tt NvNsNc} & 0.0938 & 0.0913 & 0.1654 & 0.1633 \\
\hline
\end{tabular}
\caption{\label{tab:wils_knl}
Time in seconds per matrix times multi-RHS vector operation for the Wilson Dirac operator (\ref{def_wils}) on the 68-core KNL architecture, with all variables allocated in MCDRAM.
The lattice size is $34^3\times68$ with parameters $\Nc=3$, $\Nv=12$, $c_\mr{SW}=0$.
The best timings correspond to 480\,Gflop/s in sp, and 260\,Gflop/s in dp -- see App.~\ref{app:E} for details.}
\bigskip
\begin{tabular}{|c|ccc|ccc|}
\hline
{} & \multicolumn{3}{c|}{single precision} & \multicolumn{3}{c|}{double precision} \\
{} & $\Nthr=24$ & $\Nthr=48$ & $\Nthr=96$ & $\Nthr=24$ & $\Nthr=48$ & $\Nthr=96$ \\
\hline
{\tt NcNsNv} & 0.1472 & 0.1148 & 0.1183 & 0.2768 & 0.2204 & 0.2460 \\
{\tt NvNcNs} & 0.1274 & 0.1145 & 0.1194 & 0.3009 & 0.2378 & 0.2485 \\
{\tt NvNsNc} & 0.1233 & 0.1138 & 0.1179 & 0.2466 & 0.2218 & 0.2448 \\
\hline
\end{tabular}
\caption{\label{tab:wils_sky}
Same as Tab.~\ref{tab:wils_knl} but for the $2\times24$-core (dual socket) Skylake architecture.
The best timings correspond to 350\,Gflop/s in sp, and 180\,Gflop/s in dp.}
\end{table}

\begin{figure}[p]
\vspace*{-8mm}
\includegraphics[width=0.99\textwidth]{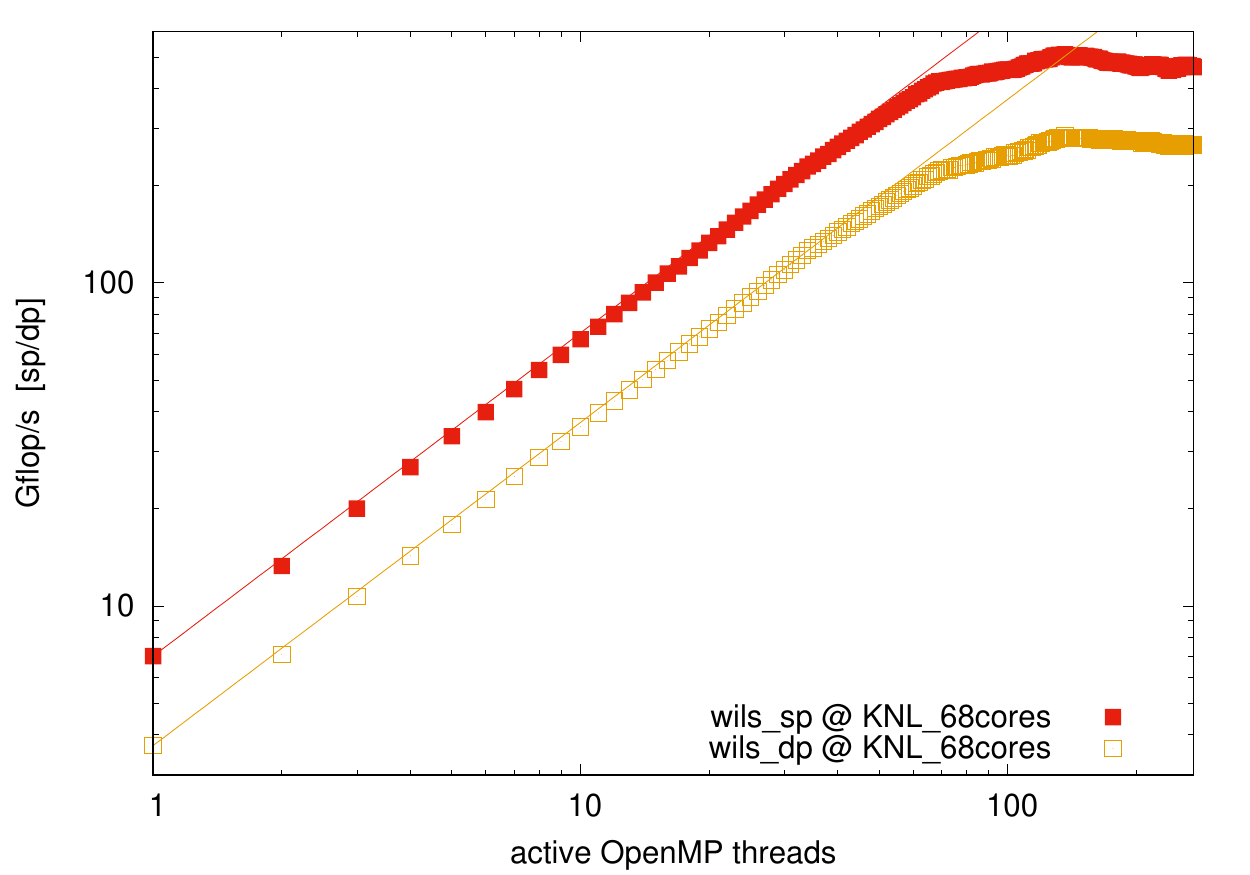}\\
\includegraphics[width=0.99\textwidth]{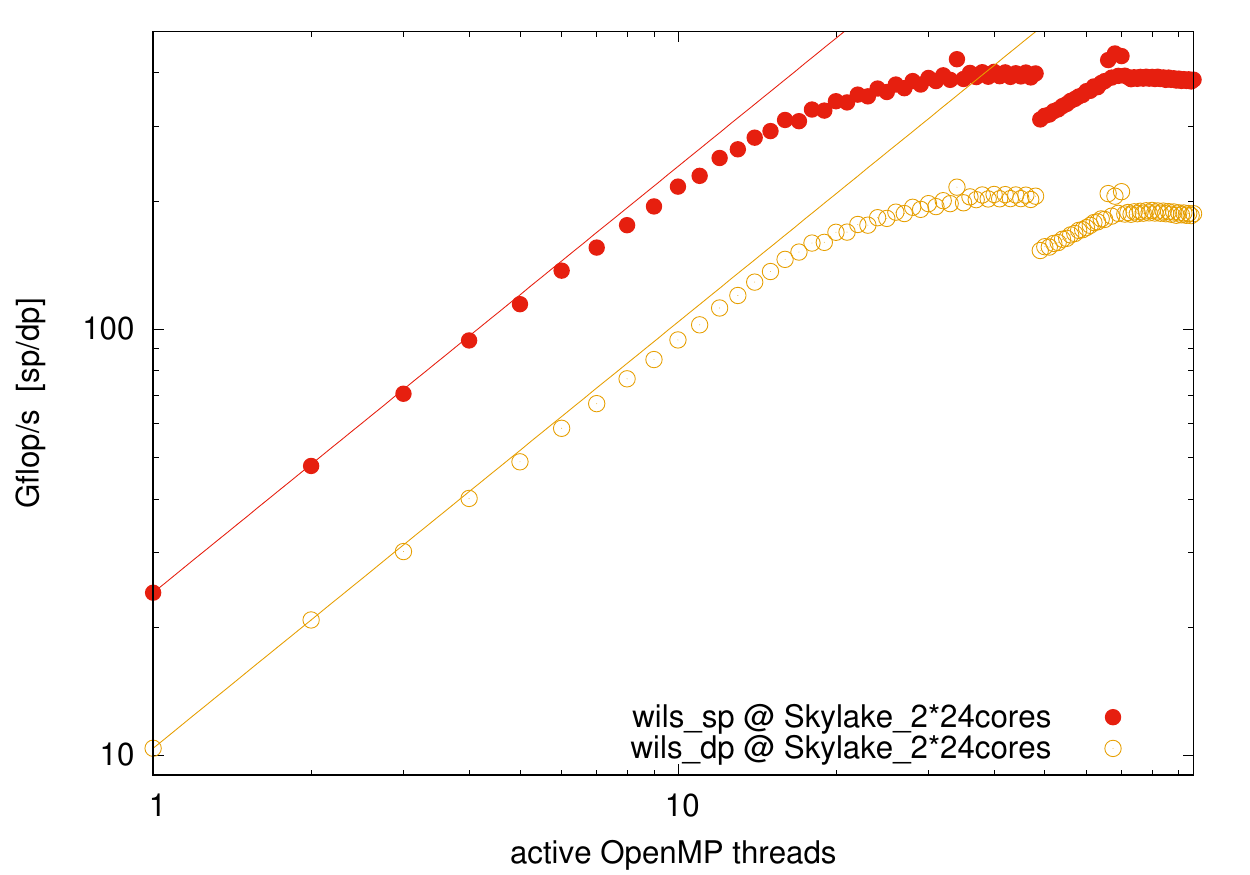}
\vspace*{-3mm}
\caption{\label{fig:wils}%
Wilson Dirac operator performance versus the number of threads in sp and dp, for the KNL and the dual-socket Skylake architectures (same parameters as in Tab.~\ref{tab:wils_knl}).}
\end{figure}

The coding of the Wilson Dirac operator is similar to the Laplacian, except that extra operations with $\pm\ga_\mu$ are involved.
Hence, one would naively expect that the matrix-vector operation (\ref{app_wils}) takes twice as long as (\ref{app_wlap}).
Fortunately, actual timings are in the same ball-park (see below).
The reason is that for every $\mu$ the operator $\frac{1}{2}(\ga_\mu\mp1)$ in (\ref{app_wils}) is a projector.
As a result, for each $\mu$ the multiplication with $V_\mu(n)$ or $V_\mu\dag(n-\hat{\mu})$ in color space can be limited to an object which is \emph{half in size}, see App.~\ref{app:A} for details.
Otherwise the implementation follows the example of the Laplacian sibling routine, with an overall {\tt !\$OMP PARALLEL DO} construct on the set of nested space-time loops,
and SIMD vectorization over the RHS-index in the part which increments the thread-private variable {\tt site(1:Nv,1:4,1:Nc)}.
This accumulation variable is eventually written into the memory block of $\ps_\mr{new}(n)$ as specified in App.~\ref{app:B}.

The timings of the Wilson Dirac routine at $c_\mr{SW}=0$ are listed in Tabs.~\ref{tab:wils_knl}, \ref{tab:wils_sky} for the KNL and Skylake architectures, respectively.
On the KNL chip good performance is achieved whenever the SIMD index {\tt rhs} is in front, and the option {\tt NvNsNc} wins the contest.
On the Skylake node all memory layouts fare in the same league, and $48$ threads (i.e.\ one per physical core on either socket) reach maximum performance.

The scaling of the Wilson Dirac routine (in sp and dp, for the {\tt NvNsNc} layout) as a function of the number of active threads is shown in Fig.~\ref{fig:wils}.
On the KNL architecture we find nearly perfect scaling behavior until every physical core hosts one thread, with mild improvement by a second thread per core, and flat or decreasing behavior after this point.
On the Skylake architecture maximum performance is already reached with one thread per physical core on one socket.
Again, there is a local maximum at $68$ threads, and the figure looks like a carbon copy of Fig.~\ref{fig:wlap}.
The computational intensity (flops per load from memory, see App.~\ref{app:E}) of the Wilson Dirac operator is so low that the second Skylake socket hardly boosts performance.
This will be different with the Brillouin Laplace and Dirac operators.


\section{Brillouin Laplace and Dirac routines\label{sec:bril}}


For a given $V_\mu(n)$ the Brillouin Dirac operator with optional clover term (\ref{def_clov}) is defined as~\cite{Durr:2010ch,Durr:2012dw}
\beq
D_\mr{B}(n,m)=\sum_\mu \ga_\mu \nab_\mu^\mr{iso}(n,m)
-\frac{ar}{2}\lap^\mr{bri}(n,m)+m_0\de_{n,m}
+aC(n,m)
\label{def_bril}
\eeq
where the isotropic derivative $\nab_\mu^\mr{iso}$ is the 54-point discretization of the covariant derivative
\bea
a\nab_\mu^\mr{iso}(n,m)&=&\rh_1\,\big[W_{\mu}(n)\de_{n+\hat\mu,m}-W_{-\mu}(n)\de_{n-\hat\mu,m}\big]
\nonumber\\
&+&\rh_2\sum\nolimits_{\neq(\nu;\mu)}\big[W_{\mu\nu}(n)\de_{n+\hat\mu+\hat\nu,m}-(\mu\to-\mu)\big]
\nonumber\\
&+&\rh_3\sum\nolimits_{\neq(\nu,\rh;\mu)}\big[W_{\mu\nu\rh}(n)\de_{n+\hat\mu+\hat\nu+\hat\rh,m}-(\mu\to-\mu)\big]
\nonumber\\
&+&\rh_4\sum\nolimits_{\neq(\nu,\rh,\si;\mu)}\big[W_{\mu\nu\rh\si}(n)\de_{n+\hat\mu+\hat\nu+\hat\rh+\hat\si,m}-(\mu\to-\mu)\big]
\label{def_ider}
\eea
and the Brillouin Laplacian $\lap^\mr{bri}$ is the 81-point discretization of the covariant Laplacian
\bea
a^2\lap^\mr{bri}(n,m)=\la_0\,\de_{n,m}
&+&\la_1\sum\nolimits_{\mu}W_{\mu}(n)\de_{n+\hat\mu,m}
\nonumber\\
&+&\la_2\sum\nolimits_{\neq(\mu,\nu)}W_{\mu\nu}(n)\de_{n+\hat\mu+\hat\nu,m}
\nonumber\\
&+&\la_3\sum\nolimits_{\neq(\mu,\nu,\rh)}W_{\mu\nu\rh}(n)\de_{n+\hat\mu+\hat\nu+\hat\rh,m}
\nonumber\\
&+&\la_4\sum\nolimits_{\neq(\mu,\nu,\rh,\si)}W_{\mu\nu\rh\si}(n)\de_{n+\hat\mu+\hat\nu+\hat\rh+\hat\si,m}
\label{def_blap}
\eea
with $(\rh_1,\rh_2,\rh_3,\rh_4)\equiv(64,16,4,1)/432$ and $(\la_0,\la_1,\la_2,\la_3,\la_4)\equiv(-240,8,4,2,1)/64$.
The sum in (\ref{def_bril}) extends over the positive Euclidean directions, i.e.\ $\mu\in\{1,\ldots,4\}$, and the bare quark mass $m_0$ undergoes both additive and multiplicative renormalization.
In Eq.~(\ref{def_ider}) the last sum extends over (positive and negative) indices $(\nu,\rh,\si)$ whose absolute values are pairwise unequal and different from $\mu$ (which is $>0$).
In Eq.~(\ref{def_blap}) the last sum extends over indices $(\mu,\nu,\rh,\si)$ whose absolute values are pairwise unequal.
Here $W_\mr{dir}(n)$ denotes a link in direction ``dir'' which may be on-axis (dir=$\mu$) or off-axis with Euclidean length $\sqrt{2}$ (dir=$\mu\nu$), $\sqrt{3}$ (dir=$\mu\nu\rh$), $\sqrt{4}$ (dir=$\mu\nu\rh\si$).
This $W_\mr{dir}(n)$ is defined as the average of all chains of $V$-links that connect $n$ and $n+\mr{dir}$ with the minimum number of hops.
How the $V$-links (contained in $W$ and $C$) relate to the original $U$-links has been explained in Sec.~\ref{sec:clov}.
As a result, $W_\mr{dir}(n)$ is a legitimate parallel transporter from $n+\mr{dir}$ to $n$, see Tab.~\ref{tab:off-axis} for details.
More details on the physics motivation and the free-field behavior of this operator are given in Refs.~\cite{Durr:2010ch,Durr:2017wfi}.

From the viewpoint of computational expedience, it pays%
\footnote{This holds on current CPU architectures, including the KNL and Skylake chips.
With the anticipated increase of the compute-to-bandwidth capacity ratio, this may change at some point in the future.}
to precompute the $W$-links, and to feed the routine that eventually implements (\ref{def_bril}) with $W_\mr{dir}(n)$ [and possibly $F_{\mu\nu}(n)$, if $c_\mr{SW}>0$].
Similar as with the Wilson Dirac operator the ``isotropic derivative'' (\ref{def_ider}) is anti-hermitean, while the ``Brillouin Laplacian'' (\ref{def_blap}) and the clover term are hermitean operators.
The species-lifting parameter is typically set to $r=1$, and from a HPC viewpoint the challenge with the operator~(\ref{def_bril}) is that its stencil contains up to 4-hop terms.

\begin{table}[tb]
\centering
\begin{tabular}{|cccl|}
\hline
hops & terms & paths & formula \\
\hline
1 &  8 &  1 & $W_{\mu}(n)=V_{\mu}(n)$ using smeared link with $\mu\in\{\pm1,\pm2,\pm3,\pm4\}$\\
2 & 24 &  2 & $W_{\mu\nu}(n)=\frac{1}{2}[V_\mu(n)V_\nu(n\!+\!\hat\mu)+\mr{perm}]$\\
3 & 32 &  6 & $W_{\mu\nu\rh}(n)=\frac{1}{6}[V_\mu(n)V_\nu(n\!+\!\hat\mu)V_\rh(n\!+\!\hat\mu\!+\!\hat\nu)+\mr{perms}]$\\
4 & 16 & 24 & $W_{\mu\nu\rh\si}(n)=\frac{1}{24}[V_\mu(n)V_\nu(n\!+\!\hat\mu)V_\rh(n\!+\!\hat\mu\!+\!\hat\nu)V_\si(n\!+\!\hat\mu\!+\!\hat\nu\!+\!\hat\si)+\mr{perms}]$\\
\hline
\end{tabular}
\caption{\label{tab:off-axis}
Summary of the on/off-axis links $W_\mr{dir}(n)$, with lengths ranging from 1 to 4 hops.
The number of paths contributing to a given term matches the (total) number of permutations.
Starting at site $n$, there are $1+8+24+32+16=81$ directions, but one is trivial (``no hop''), and the remaining 80 can be reduced to 40, based on $W_\mr{-dir}^{}(n)=W_\mr{dir}\dag(n-\mr{dir})$.
In the code $W$ is precomputed and stored in the array {\tt W(Nc,Nc,40,Nx,Ny,Nz,Nt)}.
Note that for 36 of the 40 directions the entry is not special unitary; here gauge compression is not possible.
Alternatively, the prefactors $\frac{1}{2},\frac{1}{6},\frac{1}{24}$ might be replaced by $P_{SU(\Nc)}$; in this case gauge compression is possible.}
\end{table}

The action of the Brillouin operator~(\ref{def_bril}) at $r=1,c_\mr{SW}=0$ on a Dirac vector $\ps$ (spinor$\otimes$color internal degrees of freedom) is given by
[with $\ga_{-\mu}\equiv-\ga_\mu$ for $\mu>0$]
\bea
(D_\mr{B}\ps)(n)&=&
(m_0-\frac{\la_0}{2})\ps(n)
+\sum_{\mu}\Big[(\rh_1\ga_\mu-\frac{\la_1}{2})\otimes W_\mu(n)\Big]\ps(n+\hat\mu)
\nonumber
\\
&+&\sum_{\neq(\mu,\nu)}\Big[(\rh_2\ga_\mu-\frac{\la_2}{2})\otimes W_{\mu\nu}(n)\Big]\ps(n+\hat\mu+\hat\nu)
\nonumber
\\
&+&\sum_{\neq(\mu,\nu,\rh)}\Big[(\rh_3\ga_\mu-\frac{\la_3}{2})\otimes W_{\mu\nu\rh}(n)\Big]\ps(n+\hat\mu+\hat\nu+\hat\rh)
\nonumber
\\
&+&\sum_{\neq(\mu,\nu,\rh,\si)}\Big[(\rh_4\ga_\mu-\frac{\la_4}{2})\otimes W_{\mu\nu\rh\si}(n)\Big]\ps(n+\hat\mu+\hat\nu+\hat\rh+\hat\si)
\label{app_bril}
\eea
where now even $\mu$ admits negative values, and our task is to implement a routine which performs this operation efficiently.
The action of the embedded Laplace operator (\ref{def_blap}) is
\bea
\Big(-\frac{1}{2}\lap^\mr{bri}+\frac{m_0^2}{2}\Big)\ps(n)&=&
(\frac{m_0^2}{2}-\frac{\la_0}{2})\ps(n)
-\frac{\la_1}{2}\sum_{\mu}W_\mu(n)\ps(n+\hat\mu)
\nonumber
\\
&-&\frac{\la_2}{2}\sum_{\neq(\mu,\nu)}W_{\mu\nu}(n)\ps(n+\hat\mu+\hat\nu)
\nonumber
\\
&-&\frac{\la_3}{2}\sum_{\neq(\mu,\nu,\rh)}W_{\mu\nu\rh}(n)\ps(n+\hat\mu+\hat\nu+\hat\rh)
\nonumber
\\
&-&\frac{\la_4}{2}\sum_{\neq(\mu,\nu,\rh,\si)}W_{\mu\nu\rh\si}(n)\ps(n+\hat\mu+\hat\nu+\hat\rh+\hat\si)
\label{app_blap}
\eea
where the same comment on $\mu$ applies, and the mass parameter reflects the usual choice for the Euclidean boson propagator.
Hence, in both (\ref{app_bril}) and (\ref{app_blap}) the first sum is over $8$ directions, the second one over $8\cdot6/2=24$ directions, the third sum is over $8\cdot6\cdot4/6=32$ directions,
and the last one over $8\cdot6\cdot4\cdot2/24=16$ directions (cf.\ Tab.~\ref{tab:off-axis}).
All together, we have $80$ non-trivial directions, and this is exactly what we expect if the total sum extends over a $3^4$ cube centered around the space-time point $n=(x,y,z,t)$.
In (\ref{app_bril}) and (\ref{app_blap}) we have taken the liberty to set the lattice spacing $a=1$.

\begin{table}[tb]
\centering
\begin{tabular}{|c|cc|cc|}
\hline
{} & \multicolumn{2}{c|}{single precision} & \multicolumn{2}{c|}{double precision} \\
{} & $\Nthr=136$ & $\Nthr=272$ & $\Nthr=136$ & $\Nthr=272$ \\
\hline
{\tt NcNsNv} & 0.9814 & 1.0363 & 1.6139 & 2.3746  \\
{\tt NvNcNs} & 0.8146 & 0.8685 & 1.6750 & 2.4589 \\
{\tt NvNsNc} & 0.6574 & 0.7611 & 1.4889 & 2.1980 \\
\hline
\end{tabular}
\caption{\label{tab:blap_knl}
Time in seconds per matrix times multi-RHS vector operation for the Brillouin Laplace operator (\ref{def_blap}) on the 68-core KNL architecture, with variables allocated in MCDRAM.
The lattice size is $34^3\times68$ with parameters $\Nc=3$, $\Nv=12$.
The best timings correspond to 1220\,Gflop/s in sp, and 540\,Gflop/s in dp -- see App.~\ref{app:E} for details.}
\bigskip
\begin{tabular}{|c|ccc|ccc|}
\hline
{} & \multicolumn{3}{c|}{single precision} & \multicolumn{3}{c|}{double precision} \\
{} & $\Nthr=24$ & $\Nthr=48$ & $\Nthr=96$ & $\Nthr=24$ & $\Nthr=48$ & $\Nthr=96$ \\
\hline
{\tt NcNsNv} & 1.2716 & 0.6758 & 0.5580 & 2.0792 & 1.1825 & 1.0767 \\
{\tt NvNcNs} & 1.1534 & 0.6266 & 0.5324 & 1.8679 & 1.0877 & 1.0182 \\
{\tt NvNsNc} & 1.1746 & 0.6288 & 0.5355 & 1.9440 & 1.1489 & 1.0844 \\
\hline
\end{tabular}
\caption{\label{tab:blap_sky}
Same as Tab.~\ref{tab:blap_knl} but for the $2\times24$-core (dual socket) Skylake architecture.
The best timings correspond to 1500\,Gflop/s in sp, and 790\,Gflop/s in dp.}
\end{table}

\begin{figure}[p]
\vspace*{-8mm}
\includegraphics[width=0.99\textwidth]{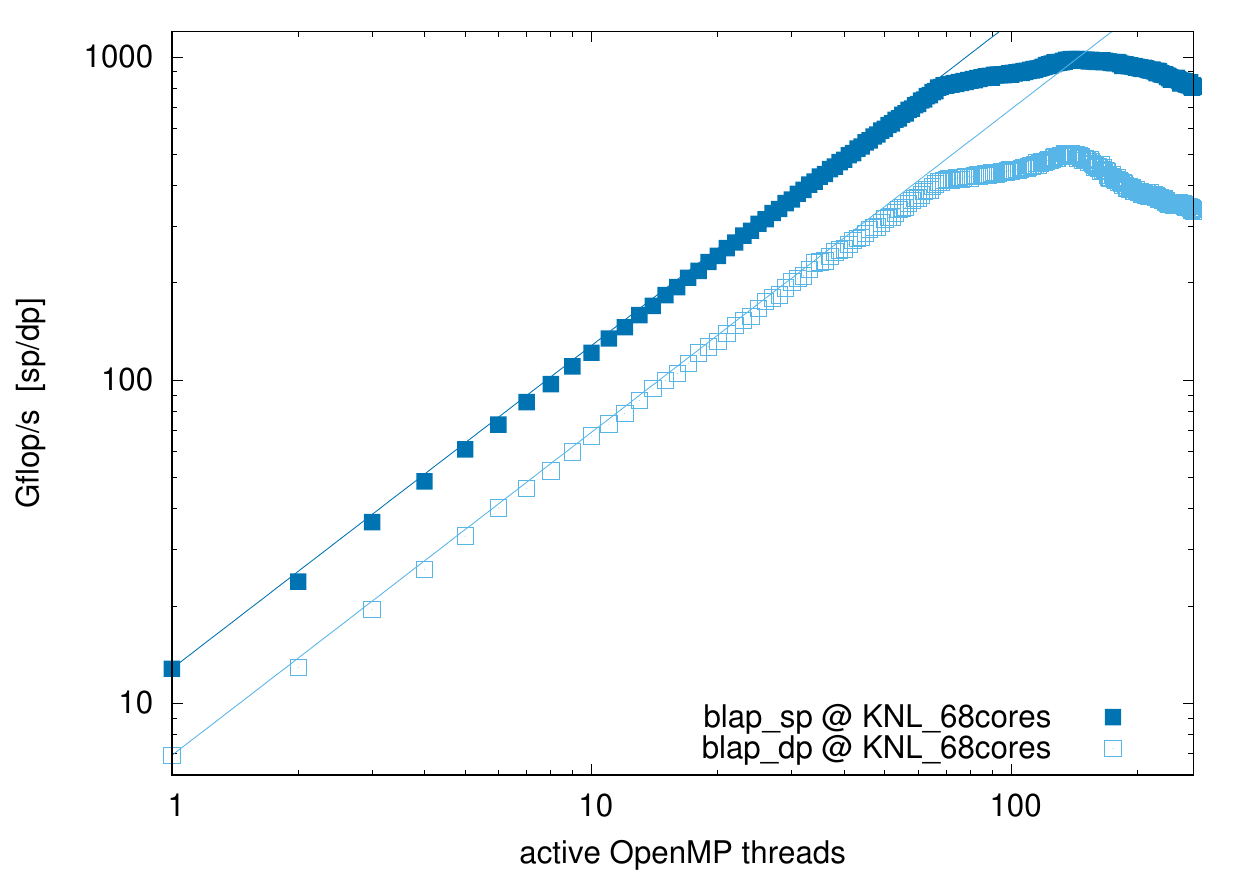}\\
\includegraphics[width=0.99\textwidth]{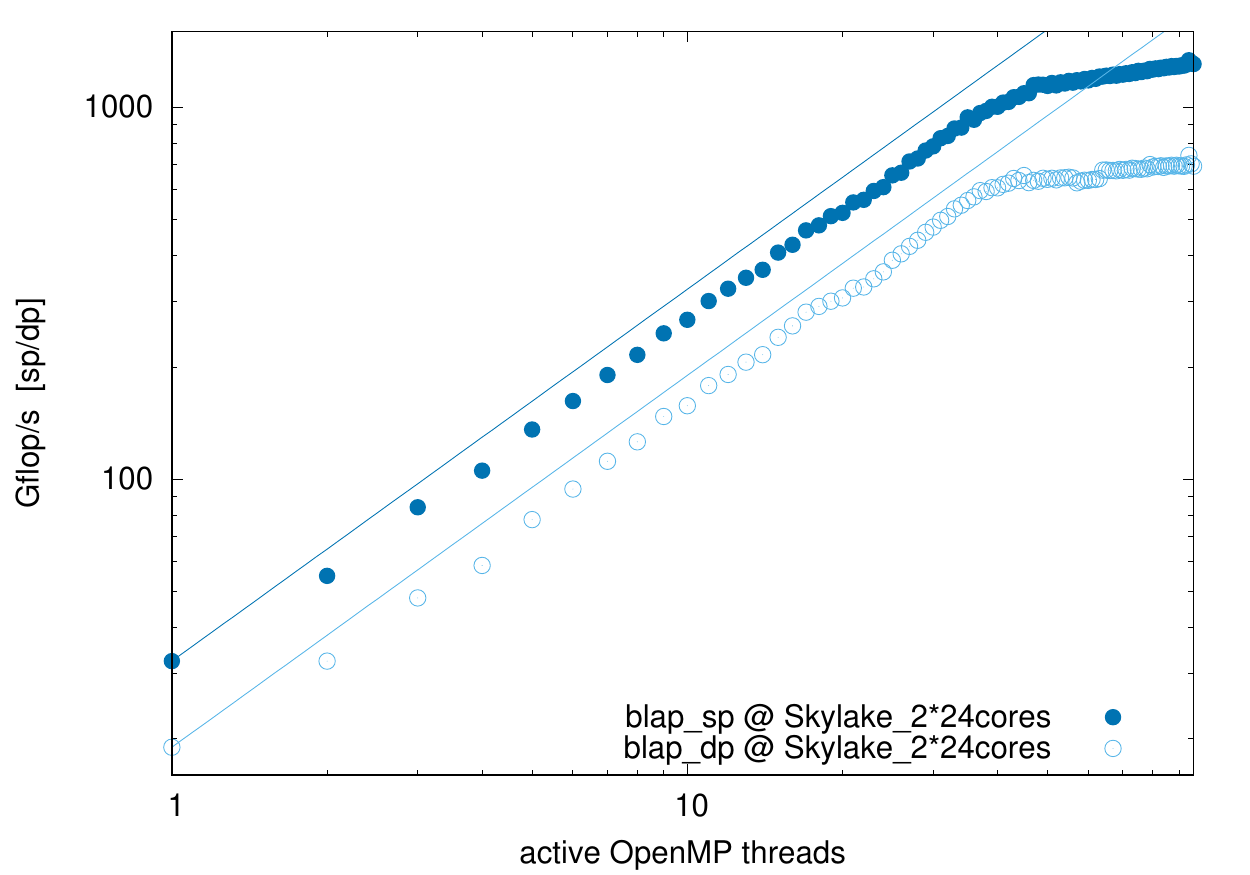}
\vspace*{-3mm}
\caption{\label{fig:blap}%
Brillouin Laplace operator performance versus the number of threads in sp and dp, for the KNL and the dual-socket Skylake architectures (same parameters as in Tab.~\ref{tab:blap_knl}).}
\end{figure}

Obviously the Brillouin Laplace operator (\ref{app_blap}) is easier to implement than the Brillouin Dirac operator (\ref{app_bril}), since it acts trivially%
\footnote{Hence one might let $\lap^\mr{bri}$ act on a Susskind-type vector (color structure only), and the ancillary code distribution
contains such a multiplication routine under the label {\tt app\_bsuv\_$\{$sp,dp$\}$}, see App.~\ref{app:F} for details.\label{foot:bsuv}}
in spinor space, so we consider this routine first.
At first sight (\ref{app_blap}) looks complicated, but the five terms can be combined in a convenient manner.
Put differently, there is no need to organize the matrix-vector operation separately for 0-hop, 1-hop, $\ldots$, 4-hop terms.
Given that the off-axis links $W_\mu(n)$, $W_{\mu\nu}(n)$, $W_{\mu\nu\rh}(n)$, $W_{\mu\nu\rh\si}(n)$ are contained in the single array {\tt W(Nc,Nc,40,Nx,Ny,Nz,Nt)},
it is more convenient (and faster) to organize the routine through a nested set of four additional loops (the variables {\tt go\_x,go\_y,go\_z,go\_t} take the values $-1,0,+1$)
which address the $81$ positions $m$ of $\ps_\mr{old}(m)$ in the $3^4$ hypercube centered around $n$.
To pick the right prefactor from the set $\{\frac{1}{2}m_0^2-\frac{1}{2}\la_0,-\frac{1}{2}\la_1,\ldots,-\frac{1}{2}\la_4\}$,
it suffices to know which distance (in ``taxi-driver metric'') the vector {\tt [go\_x,go\_y,go\_z,go\_t]} would bridge.
This is achieved through {\tt count([go\_x,go\_y,go\_z,go\_t].ne.0)} or via a look-up table which uses {\tt min(dir,82-dir)},
where {\tt dir} is the direction count (from $1$ to $81$ in the nested {\tt [go\_x,go\_y,go\_z,go\_t]} loop).
Again, all contributions are accumulated in the thread-private variable {\tt site(1:Nv,1:4,1:Nc)}, which is eventually written into the memory block of $\ps_\mr{new}(n)$.
More details are provided in App.~\ref{app:C}.

The timings of the Brillouin Laplace routine are listed in Tabs.~\ref{tab:blap_knl}, \ref{tab:blap_sky} for the KNL and Skylake architectures, respectively.
On the KNL chip the vector layout is important; the layout {\tt NvNsNc} wins the contest.
On the Skylake architecture the layouts {\tt NvNcNs}, {\tt NvNsNc} (with the SIMD index {\tt rhs} in front) are just slightly better than {\tt NcNsNv}.
Unlike with the Wilson Laplace routine (see Tab.~\ref{tab:wlap_sky}) more threads yield higher performance; the table culminates in a whopping 1500\,Gflops (for sp vectors) with $96$ threads.
In other words, this is the first operator for which the dual-socket Skylake node delivers higher performance than a single KNL chip.

The scaling of the Brillouin Laplace routine (in sp and dp, for the {\tt NvNsNc} layout) as a function of the number of active threads is shown in Fig.~\ref{fig:blap}.
The parameters, and the range over which the number of threads is varied, are the same as in Secs.~\ref{sec:clov}, \ref{sec:wils}.
On the KNL architecture we find nearly perfect scaling behavior until every physical core hosts one thread.
A second thread per core brings modest improvement, while a third and fourth thread tend to deteriorate performance.
By contrast, on the Skylake architecture performance increases (both in sp and dp) until the global maximum is reached near $96$ threads.


\begin{table}[tb]
\centering
\begin{tabular}{|c|cc|cc|}
\hline
{} & \multicolumn{2}{c|}{single precision} & \multicolumn{2}{c|}{double precision} \\
{} & $\Nthr=136$ & $\Nthr=272$ & $\Nthr=136$ & $\Nthr=272$ \\
\hline
{\tt NcNsNv} & 1.3264 & 1.2675 & 2.0097 & 2.6167 \\
{\tt NvNcNs} & 1.1326 & 1.0912 & 2.0275 & 2.6740 \\
{\tt NvNsNc} & 1.1265 & 1.1102 & 2.0228 & 2.5207 \\
\hline
\end{tabular}
\caption{\label{tab:bril_knl}
Time in seconds per matrix times multi-RHS vector operation for the Brillouin Dirac operator~(\ref{def_bril}) on the 68-core KNL architecture, with variables allocated in MCDRAM.
The lattice size is $34^3\times68$ with parameters $\Nc=3$, $\Nv=12$, and $c_\mr{SW}=0$.
The best timings correspond to 880\,Gflop/s in sp, and 480\,Gflop/s in dp -- see App.~\ref{app:E} for details.}
\bigskip
\begin{tabular}{|c|ccc|ccc|}
\hline
{} & \multicolumn{3}{c|}{single precision} & \multicolumn{3}{c|}{double precision} \\
{} & $\Nthr=24$ & $\Nthr=48$ & $\Nthr=96$ & $\Nthr=24$ & $\Nthr=48$ & $\Nthr=96$ \\
\hline
{\tt NcNsNv} & 1.6864 & 0.8700 & 0.7433 & 2.8651 & 1.5404 & 1.4205 \\
{\tt NvNcNs} & 1.5488 & 0.8161 & 0.7262 & 2.7298 & 1.4811 & 1.3666 \\
{\tt NvNsNc} & 1.7860 & 0.9200 & 0.7426 & 2.7952 & 1.5275 & 1.4288 \\
\hline
\end{tabular}
\caption{\label{tab:bril_sky}
Same as Tab.~\ref{tab:bril_knl} but for the $2\times24$-core (dual socket) Skylake architecture.
The best timings correspond to 1320\,Gflop/s in sp, and 700\,Gflop/s in dp.}
\end{table}

\begin{figure}[p]
\vspace*{-8mm}
\includegraphics[width=0.99\textwidth]{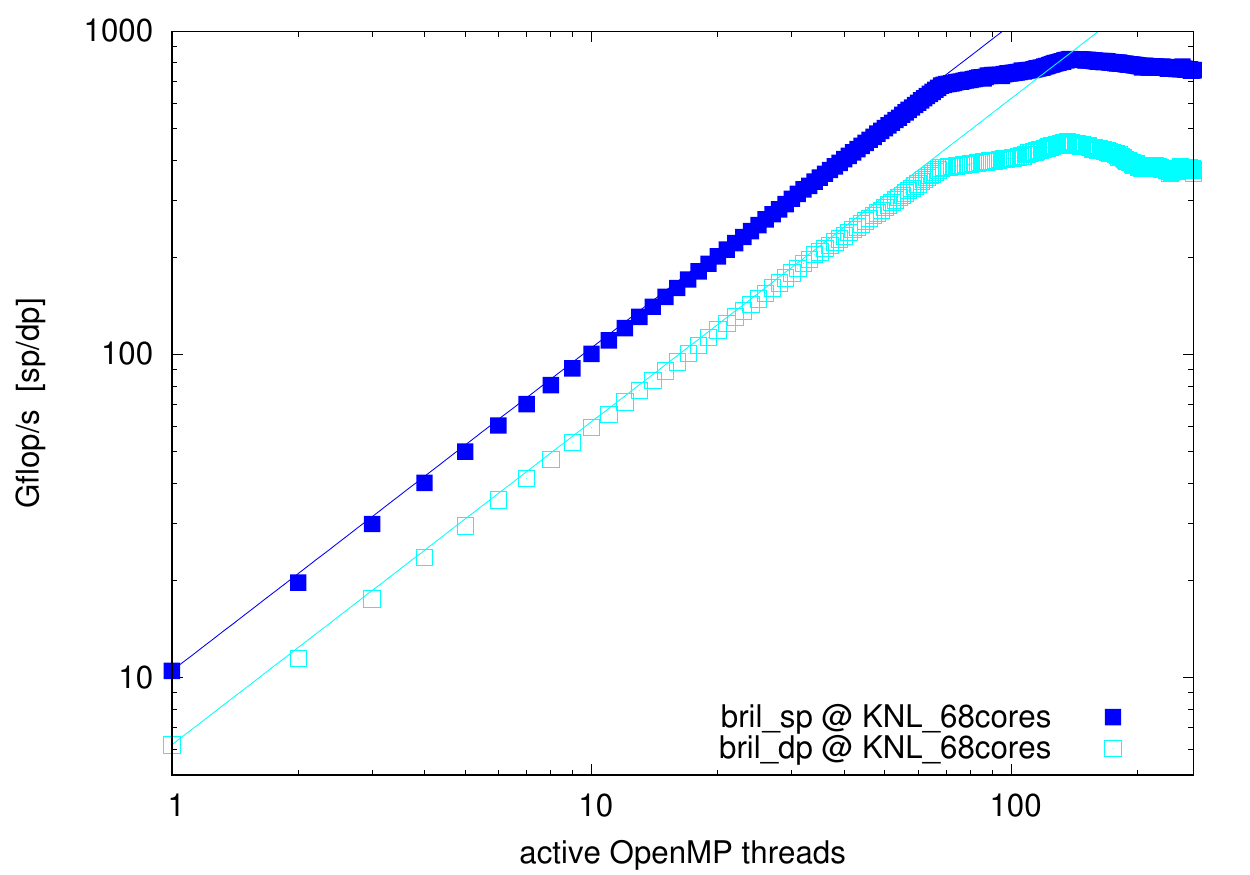}\\
\includegraphics[width=0.99\textwidth]{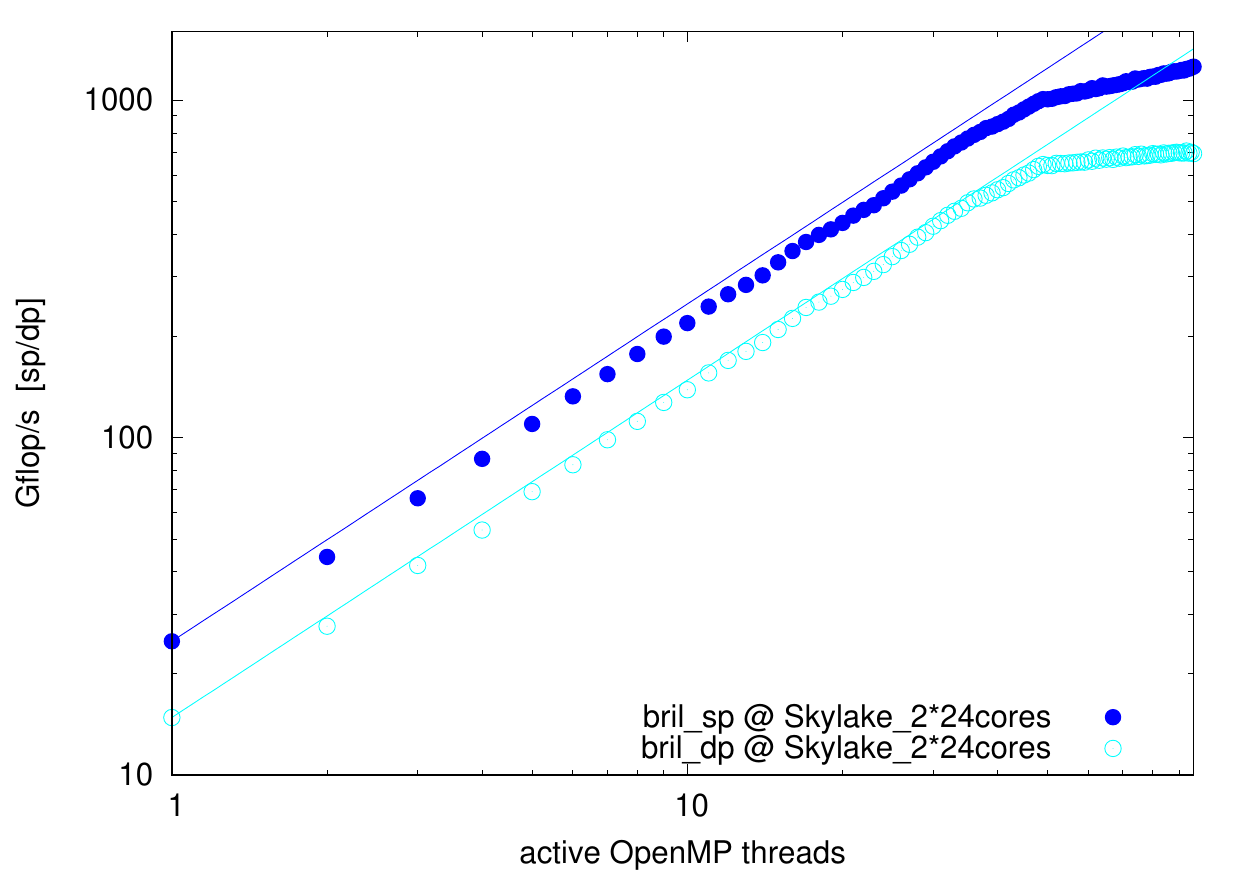}
\vspace*{-3mm}
\caption{\label{fig:bril}%
Brillouin Dirac operator performance versus the number of threads in sp and dp, for the KNL and the dual-socket Skylake architectures (same parameters as in Tab.~\ref{tab:bril_knl}).}
\end{figure}

The coding of the Brillouin Dirac operator (\ref{app_bril}) proceeds analogous to (\ref{app_blap}), except that it acts non-trivially in spinor space, too.
These extra terms involve the ``isotropic derivative'' (\ref{def_ider}) which again leads to a $81$-point stencil (with the contraction, for each $\mu$ it is fewer points).
To maximize performance it pays to combine both gauge-field dependent terms in the Brillouin Dirac operator, so that each $W_\mr{dir}(n)$ is loaded once.
Similar to the Brillouin Laplace sibling routine, one uses a set of four nested loops (with variables {\tt go\_x,go\_y,go\_z,go\_t} taking the values $-1,0,+1$ each)
to visit the $81$ points $m$ of $\ps_\mr{old}(m)$ in the hypercube around $n$.
Unlike the Laplace term (which requires only knowledge abut the taxi-driver distance between $m$ and $n$) the derivative term combines the details of $m-n$ with the detailed choice of the Dirac matrix
(in the code the chiral representation specified in App.~\ref{app:A} is used).
All contributions are accumulated in the thread-private variable {\tt site(1:Nv,1:4,1:Nc)}, which is eventually written into the memory block of $\ps_\mr{new}(n)$.
More details are provided in App.~\ref{app:C}.

The timings of the Brillouin Dirac routine at $c_\mr{SW}=0$ are listed in Tabs.~\ref{tab:bril_knl}, \ref{tab:bril_sky} for the KNL and Skylake architectures, respectively.
On the KNL chip the vector layouts {\tt NvNcNs}, {\tt NvNsNc} (with the SIMD index {\tt rhs} in front) yield best performance, with $N_\mr{thr}=136$ and $N_\mr{thr}=272$ virtually on par.
On the Skylake architecture the layout {\tt NvNcNs} seems slightly better than the other two.
The peak is at $96$ threads, i.e.\ with all available threads on the dual-socket node.
Also for this operator the maximum performance available on the Skylake dual-socket node (1320\,Gflops) exceeds the best figure on the KNL node (880\,Gflops) by $50\%$.

The scaling of the Brillouin Dirac routine (in sp and dp, for the {\tt NvNsNc} layout) as a function of the number of active threads is shown in Fig.~\ref{fig:bril}.
On the KNL we find (again) nearly perfect scaling behavior until every physical core hosts one thread.
The second thread still yields some improvement, while the third and fourth threads per physical core deteriorate performance.
By contrast, on the Skylake architecture performance increases (both in sp and dp) until the global maximum is reached at $96$ threads.
Apart from an overall vertical shift, the entire figure looks like a carbon copy of Fig.~\ref{fig:blap} (a phenomenon encountered in Sec.~\ref{sec:wils}, too).

The main lesson from this section is that the Brillouin (Laplace and Dirac) operators have a higher computational intensity than the Wilson (Laplace and Dirac) operators (see App.~\ref{app:E}).
This lets the Brillouin operators ($81$-point stencil) benefit from the compute power of the second socket in the Skylake node, while the Wilson operators ($9$-point stencil) barely do so.


\section{Susskind ``staggered'' Dirac routine\label{sec:stag}}


For a given $V_\mu(n)$ the Susskind (``staggered'') Dirac operator is defined as~\cite{Kogut:1974ag,Susskind:1976jm}
\beq
D_\mr{S}(n,m)=\sum_{\mu} \et_\mu(n)\,\frac{1}{2}\,[V_{\mu}(n)\de_{n+\hat\mu,m}-V_{\mu}\dag(n-\hat\mu)\de_{n-\hat\mu,m}] + m_0\de_{n,m}
\label{def_stag}
\eeq
with $\et_1(n)=1$, $\et_2(n)=(-1)^{x}$, $\et_3(n)=(-1)^{x+y}$, $\et_4(n)=(-1)^{x+y+z}$ and $n=(x,y,z,t)$.
Here $V_\mu(n)$ represents a smeared version of the (original) gauge link $U_\mu(n)$, i.e.\ a gauge-covariant parallel transporter from $n+\hat\mu$ to $n$.
Its main purpose is to reduce taste-symmetry breaking \cite{DeGrand:1999gp,Orginos:1999cr}, but there are more sophisticated alternatives with a larger stencil \cite{Lepage:1998vj}.

The main physics difference between the Susskind (``staggered'') action and previously discussed fermion actions is that the operator (\ref{def_stag}) yields four species in the continuum, not just one.
Furthermore, the bare quark mass $m_0$ gets multiplicatively renormalized only (for Wilson and Brillouin fermions there is also an additive shift).

The action of the operator~(\ref{def_stag}) on a Susskind vector $\ph$ (internal color structure only) is
\beq
(D_\mr{S}\ph)(n)=\sum_\mu
\et_\mu(n)\,\frac{1}{2}\,[V_\mu(n)\ph(n+\hat\mu)-V_\mu\dag(n-\hat\mu)\ph(n-\hat\mu)] + m_0\ph(n)
\eeq
and our task is to implement a routine which performs this operation efficiently.

\begin{table}[tb]
\centering
\begin{tabular}{|c|cc|cc|}
\hline
{} & \multicolumn{2}{c|}{single precision} & \multicolumn{2}{c|}{double precision} \\
{} & $\Nthr=136$ & $\Nthr=272$ & $\Nthr=136$ & $\Nthr=272$ \\
\hline
{\tt NcNv} & 0.0331 & 0.0306 & 0.0558 & 0.0506 \\
{\tt NvNc} & 0.0245 & 0.0237 & 0.0455 & 0.0429 \\
\hline
\end{tabular}
\caption{\label{tab:stag_knl}
Time in seconds per matrix times multi-RHS vector operation for the staggered Dirac operator~(\ref{def_stag}) on the 68-core KNL architecture, with all variables allocated in MCDRAM.
The lattice size is $34^3\times68$ with parameters $\Nc=3$, $\Nv=12$.
The best timings correspond to 780\,Gflop/s in sp, and 440\,Gflop/s in dp -- see App.~\ref{app:E} for details.}
\bigskip
\begin{tabular}{|c|ccc|ccc|}
\hline
{} & \multicolumn{3}{c|}{single precision} & \multicolumn{3}{c|}{double precision} \\
{} & $\Nthr=24$ & $\Nthr=48$ & $\Nthr=96$ & $\Nthr=24$ & $\Nthr=48$ & $\Nthr=96$ \\
\hline
{\tt NcNv} & 0.0396 & 0.0342 & 0.0366 & 0.0676 & 0.0613 & 0.0633 \\
{\tt NvNc} & 0.0342 & 0.0340 & 0.0363 & 0.0649 & 0.0612 & 0.0633 \\
\hline
\end{tabular}
\caption{\label{tab:stag_sky}
Same as Tab.~\ref{tab:stag_knl} but for the $2\times24$-core (dual socket) Skylake architecture.
The best timings correspond to 550\,Gflop/s in sp, and 300\,Gflop/s in dp.}
\end{table}

\begin{figure}[p]
\vspace*{-8mm}
\includegraphics[width=0.99\textwidth]{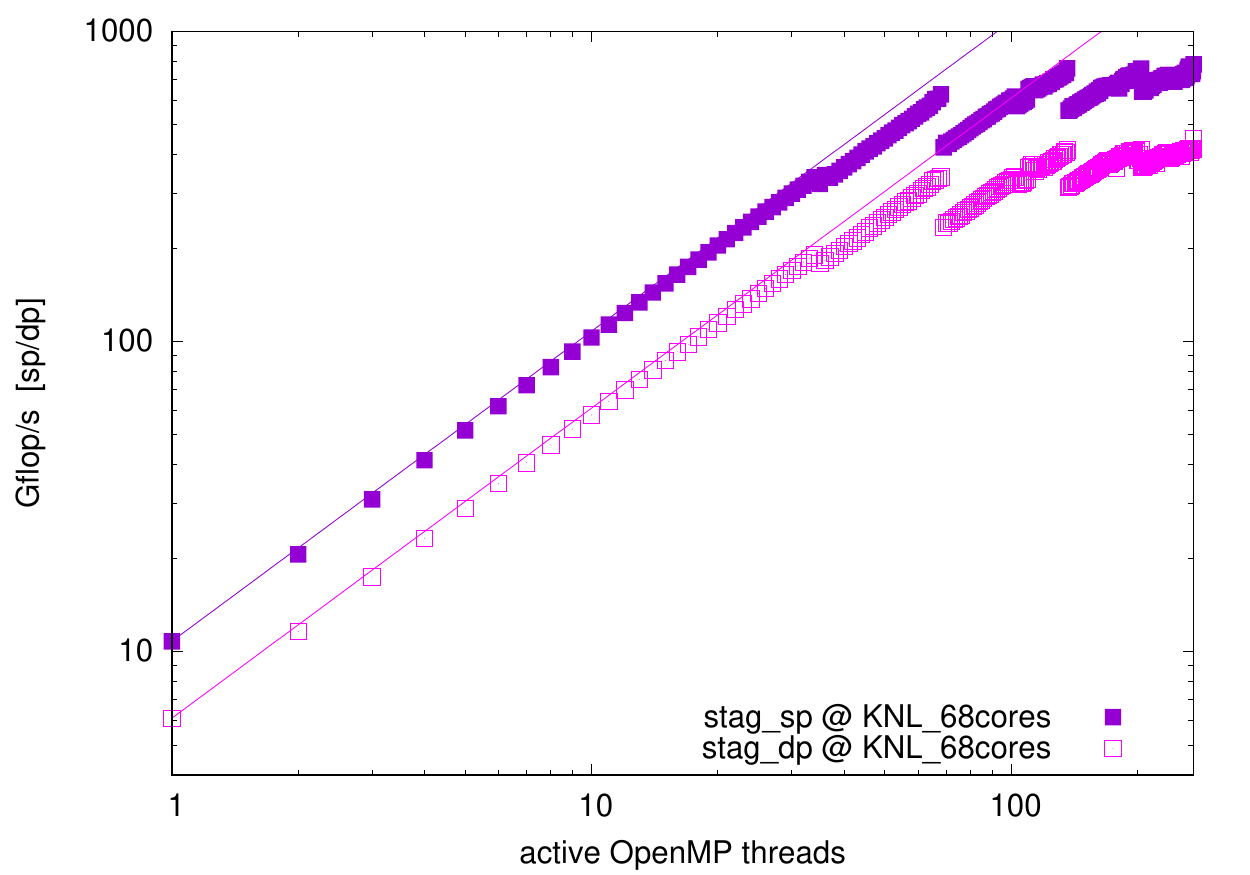}\\
\includegraphics[width=0.99\textwidth]{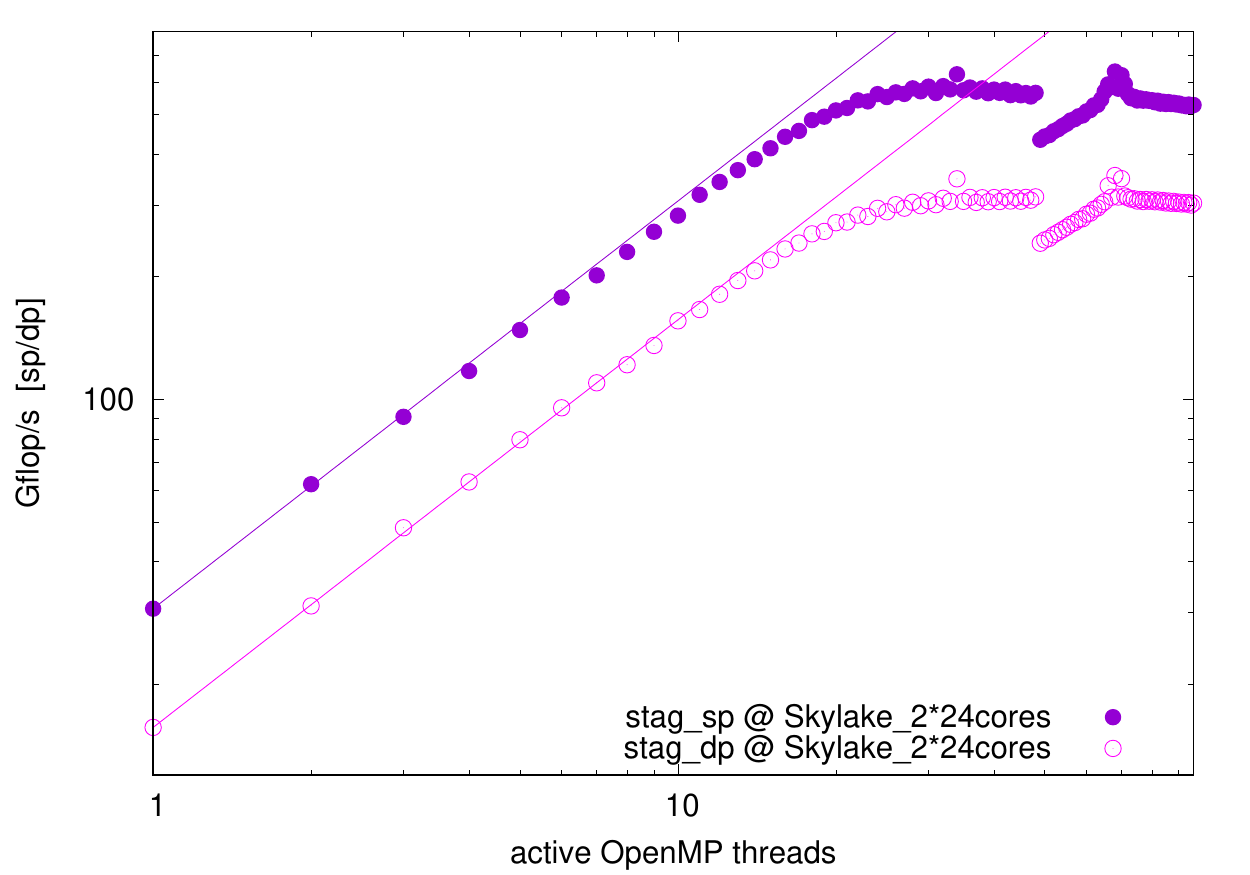}
\vspace*{-3mm}
\caption{\label{fig:stag}%
Staggered Dirac operator performance versus the number of threads in sp and dp, for the KNL and the dual-socket Skylake architectures (same parameters as in Tab.~\ref{tab:stag_knl}).}
\end{figure}

To prepare for his task, it helps to ``downgrade'' the Wilson Laplace routine (as discussed in Sec.~\ref{sec:wils}) to the ``staggered utility vector'', i.e.\ to remove the spinorial degrees of freedom
(which, in this case, were trivially acted on), and the resulting routine {\tt app\_wsuv\_$\{$sp,dp$\}$} was mentioned in footnote \ref{foot:wsuv}.
In a similar vein, the Brillouin Laplace operator (as discussed in Sec.~\ref{sec:bril}) can be ``downgraded'' to act on a ``staggered utility vector'',
and the resulting routine {\tt app\_bsuv\_$\{$sp,dp$\}$} was mentioned in footnote \ref{foot:bsuv}.
In either case, the six layout options (among the internal degrees of freedom {\tt rhs,spi,col}) collapse into two layout options (among {\tt rhs,col}).
Starting from {\tt app\_wsuv\_$\{$sp,dp$\}$}, it is easy to get the staggered routine; one just inserts%
\footnote{In some legacy lattice QCD codes, the gauge field $V_\mu(n)$ is dressed with these factors.
On modern architectures, e.g.\ the KNL and Skylake processors used in this work, this does not buy an advantage any more.}
the extra phase factors $\et_2,\et_3,\et_4$, where appropriate ($\et_1=1$ is a constant).

The timings of the staggered Dirac routine are listed in Tabs.~\ref{tab:stag_knl}, \ref{tab:stag_sky} for the KNL and Skylake architectures, respectively.
On the KNL chip the layout {\tt NvNc} outpaces the reverse ordering, with a peak at $136$ threads (both in sp and dp).
On the Skylake node the difference between the two layouts is marginal.
Here $24$ threads (one thread per physical core on one socket) yield better performance than $48$ or $96$ threads.
The relative strength of a single KNL versus a dual-socket Skylake node is opposite to what we have seen with the Brillouin operator;
for the staggered operator the KNL chip outpaces the Skylake performance by about $40\%$.

The scaling of the staggered Dirac routine (in sp and dp, for the {\tt NvNc} layout) as a function of the number of active threads is shown in Fig.~\ref{fig:stag}.
On the KNL we find (again) nearly perfect scaling behavior until every physical core hosts one thread.
After a tiny dip, the second thread yields minor improvement, whereas the third and fourth thread barely warrant the extra scheduling cost
(the dips in this figure reflect the OpenMP scheduling option {\tt static}).
By contrast, on the Skylake architecture a local extremum is reached at $68$ threads (i.e.\ one thread per timeslice)).
The flat structure after saturation at $O(20)$ threads suggests the memory bandwidth is the bottleneck on this architecture (similar to the Wilson case discussed in Sec.~\ref{sec:wils}).


\section{Dependence on compile-time parameters\label{sec:parameters}}


The question behind this article is whether it is possible to write in a high-level language a piece of code
that yields decent performance for an arbitrary number of colors, $\Nc$, number of RHS, $\Nv$, and lattice volume $\Nx\times\Ny\times\Nz\times\Nt$.
In the following, we shall test whether this design goal has been met.
We restrict ourselves to the three Dirac operators $\DW$, $\DB$ (both at $c_\mr{SW}=0$) and $\DS$, in sp and with the best-performing layout on the KNL architecture, i.e.\ {\tt NvNsNc} and {\tt NvNc}, respectively.
The number of active threads is $\Nthr=136$ or $\Nthr=272$.

\begin{table}
\centering
\begin{tabular}{|c|cc|cc|cc|}
\hline
{} & \multicolumn{2}{c|}{Wilson} & \multicolumn{2}{c|}{Brillouin} & \multicolumn{2}{c|}{Susskind} \\
{} & $\Nthr=136$ & $\Nthr=272$ & $\Nthr=136$ & $\Nthr=272$ & $\Nthr=136$ & $\Nthr=272$ \\
\hline
$L=16$ & 460 & 440 & 800 & 710 & 630 & 620 \\
$L=20$ & 480 & 480 & 820 & 740 & 760 & 830 \\
$L=24$ & 490 & 470 & 850 & 770 & 790 & 780 \\
$L=32$ & 410 & 440 & 760 & 660 & 730 & 690 \\
$L=40$ & 470 & 470 & 830 & 810 & 730 & 710 \\
$L=48$ & 380 & 400 & 790 & 740 & 670 & 670 \\
\hline
\end{tabular}
\caption{\label{tab:dependence_vol}
Performance in Gflop/s of the Wilson, Brillouin and Susskind routines in sp on the KNL chip as a function of the volume $L^3\times T$ with $T=2L$, at fixed $\Nc=3$, $\Nv=12$.}
\bigskip
\begin{tabular}{|c|cc|cc|cc|}
\hline
{} & \multicolumn{2}{c|}{Wilson} & \multicolumn{2}{c|}{Brillouin} & \multicolumn{2}{c|}{Susskind} \\
{} & $\Nthr=136$ & $\Nthr=272$ & $\Nthr=136$ & $\Nthr=272$ & $\Nthr=136$ & $\Nthr=272$ \\
\hline
$\Nv= 1\Nc$ & 240 & 270 &  190 & 280 &  210 & 240 \\
$\Nv= 2\Nc$ & 430 & 450 &  530 & 670 &  380 & 460 \\
$\Nv= 4\Nc$ & 480 & 450 &  840 & 720 &  770 & 780 \\
$\Nv= 8\Nc$ & 530 & 500 & 1050 & 710 & 1000 & 850 \\
$\Nv=16\Nc$ & 450 & 420 &  630 & 644 &  990 & 860 \\
$\Nv=32\Nc$ & 390 & 390 &  580 & 570 &  960 & 890 \\
\hline
\end{tabular}
\caption{\label{tab:dependence_nvec}
Performance in Gflop/s of the Wilson, Brillouin and Susskind routines in sp on the KNL chip as a function of $\Nv$, at fixed volume $24^4\times48$ and $\Nc=3$.}
\bigskip
\begin{tabular}{|c|cc|cc|cc|}
\hline
{} & \multicolumn{2}{c|}{Wilson} & \multicolumn{2}{c|}{Brillouin} & \multicolumn{2}{c|}{Susskind} \\
{} & $\Nthr=136$ & $\Nthr=272$ & $\Nthr=136$ & $\Nthr=272$ & $\Nthr=136$ & $\Nthr=272$ \\
\hline
$\Nc= 2$ & 410 & 390 &  850 &  630 &  770 &  670 \\
$\Nc= 3$ & 520 & 500 & 1040 &  710 &  990 &  860 \\
$\Nc= 4$ & 540 & 530 &  910 &  830 & 1100 & 1010 \\
$\Nc= 6$ & 630 & 530 & 1030 & 1050 & 1160 & 1080 \\
$\Nc= 8$ & 610 & 400 & 1230 & 1210 & 1260 & 1090 \\
$\Nc=12$ & 520 & 350 & 1280 &  910 &  720 &  870 \\
\hline
\end{tabular}
\caption{\label{tab:dependence_ncol}
Performance in Gflop/s of the Wilson, Brillouin and Susskind routines in sp on the KNL chip as a function of $\Nc$, at fixed volume $24^4\times48$ and $\Nv=24$.}
\bigskip
\begin{tabular}{|c|cc|cc|cc|}
\hline
{} & \multicolumn{2}{c|}{Wilson} & \multicolumn{2}{c|}{Brillouin} & \multicolumn{2}{c|}{Susskind} \\
{} & $\Nthr=136$ & $\Nthr=272$ & $\Nthr=136$ & $\Nthr=272$ & $\Nthr=136$ & $\Nthr=272$ \\
\hline
$\Nc= 2$ & 460 & 400 &  840 &  700 &  640 &  550 \\
$\Nc= 3$ & 470 & 470 &  840 &  720 &  770 &  780 \\
$\Nc= 4$ & 480 & 500 &  920 &  830 &  960 &  830 \\
$\Nc= 6$ & 630 & 530 & 1030 & 1060 & 1180 & 1080 \\
$\Nc= 8$ & 570 & 400 & 1210 &  960 & 1260 & 1210 \\
$\Nc=12$ & 410 & 310 &  880 &  780 & 1090 & 1130 \\
\hline
\end{tabular}
\caption{\label{tab:dependence_ncolcol}
Performance in Gflop/s of the Wilson, Brillouin and Susskind routines in sp on the KNL chip as a function of $\Nc$, at fixed volume $24^4\times48$ with proportionality $\Nv=4\Nc$.}
\end{table}

How the performance depends on the volume in lattice units, $\Nx\times\Ny\times\Nz\times\Nt$, is summarized in Tab.~\ref{tab:dependence_vol}.
For all three operators the performance seems largely independent of the volume, apart from a few minor dips (which may be influenced by some OS jitter).

In Tab.~\ref{tab:dependence_nvec} the dependence on the number of RHS, $\Nv$, is shown.
The volume $24^3\times48$ is fixed, and $\Nv$ is an integer multiple of $\Nc=3$.
For each Dirac operator the performance grows initially with $\Nv$, reaching a maximum at $\Nv=24$.
Beyond this point performance degrades%
\footnote{Recall that on the KNL chip the first 16\,GB is allocated in MCDRAM, the remainder in DDR4 memory.}
for $\DW$ and $\DB$, while $\DS$ stays more-or-less constant.

The dependence on $\Nc$ may be discussed in two settings (both of which keep the volume $24^3\times48$ fixed).
In Tab.~\ref{tab:dependence_ncol} $\Nv=24$ is kept fixed, and $\Nc$ is taken to be an integer divisor of $24$.
In Tab.~\ref{tab:dependence_ncolcol} $\Nv=4\Nc$ scales with $\Nc$.
In both cases%
\footnote{The attentive reader may notice the $\Nc=6$ rows of Tabs.~\ref{tab:dependence_ncol} and \ref{tab:dependence_ncolcol} refer to the same situation (though from different runs).
Comparing them one finds that rounding results to multiples of $10\,\mr{Gflop/s}$ is reasonable.}
performance grows with $\Nc$ (apart from a few minor dips) up to $\Nc=6$, where it reaches a maximum for Wilson quarks, or $\Nc=8$, where the maximum for Brillouin and staggered fermions is found.

The main finding is that no ``odd corners'' with dramatically reduced performance are detected.
Just limiting the number of RHS to small numbers (say $\Nv\leq4$) seems inadvisable; this backs the arguments presented in Sec.~\ref{sec:intro} for pursuing a multi-RHS strategy.
In summary, the code seems fairly robust against changes of the volume, the number of RHS, and the number of colors.
It appears to be a useful tool for studying QCD at large $\Nc$.


\section{Krylov space inverters\label{sec:inverters}}


We have all ingredients needed to compare the various vector layout options in an attempt to tackle Eq.~(\ref{matrix_times_vector}).
Krylov-space solvers are iterative procedures to solve the system $Au=b$, for a given RHS $b$, to a predefined tolerance $\ep$.
The solver is stopped, as soon as the norm of the residual $r\equiv b-Au$ satisfies $||r|| < \ep ||b||$.
In this article $||.||$ is taken to be the 2-norm, and the stringent tolerance $\ep=10^{-12}$ (which can be reached in dp but not in sp) is used.

We aim to compare the {\tt vec}-layouts for the Conjugate Gradient (CG) algorithm with one of the hermitean positive definite (HPD) operators
$A=-\frac{1}{2}\lap^\mr{std}+\frac{1}{2}m_0^2$, $A=\DW\dag\DW$, $A=-\frac{1}{2}\lap^\mr{bri}+\frac{1}{2}m_0^2$ or $A=\DB\dag\DB$,
as well as the BiCGstab algorithm with $A=\DW$ or $A=\DB$ (which are neither hermitean nor anti-hermitean but $\gaf$-hermitean).
And we compare the {\tt suv}-layouts for the CG algorithm with the HPD operator $A=\DS\dag\DS$.
The bare mass $am_0=0.01$ is used for all operators; for $\DW$ and $\DB$ it is combined with $c_\mr{SW}=1$.
We use a quenched $24^3\times48$ configuration with $a\simeq0.9\fm$, and $\Nv=12$.
The gauge field $V_\mu(n)$ [from which $F_{\mu\nu}(n)$ derives] is constructed from $U_\mu(n)$ via three steps of $\rho=0.12$ stout smearing \cite{Morningstar:2003gk}.

The solvers (CG and BiCGstab) are written in a generic way, i.e.\ they operate on vectors with any of the six (two) layout options for Wilson-type (Susskind-type) Dirac operators.
For the matrix times vector operation they call a ``wrapper'' routine which eventually calls the optimized routine for the specific layout (a similar statement holds w.r.t.\ the linear algebra routines).
The residual is ``recomputed'' (i.e.\ $r=b-Au$ explicitly formed with $u$ the current approximation) rather than ``updated'' (via a cheaper vector-only operation), if one of the following conditions is met:
$(i)$   the iteration count is an integer multiple of $20$ in sp ($50$ in dp),
$(ii)$  the updated residual $r$ suggests that $||r||/||b||<\ep$ might hold for each RHS,
$(iii)$ the maximum iteration count occurs.
The routine exits if either $||r||/||b||<\ep$ for each RHS (based on the recomputed $r=b-Au$), or the discrepancy between the updated and actual residual norm exceeds 1\%, or the maximum iteration count is reached.
To avoid excessively long log files the relative norm is printed (and thus plotted) every ten iterations only.
In this section half of the maximum number of OpenMP threads is used on either architecture, i.e.\ $\Nthr=136$ on the KNL chip and $\Nthr=48$ on the Skylake node.

\begin{figure}[p]
\vspace*{-8mm}
\includegraphics[width=0.99\textwidth]{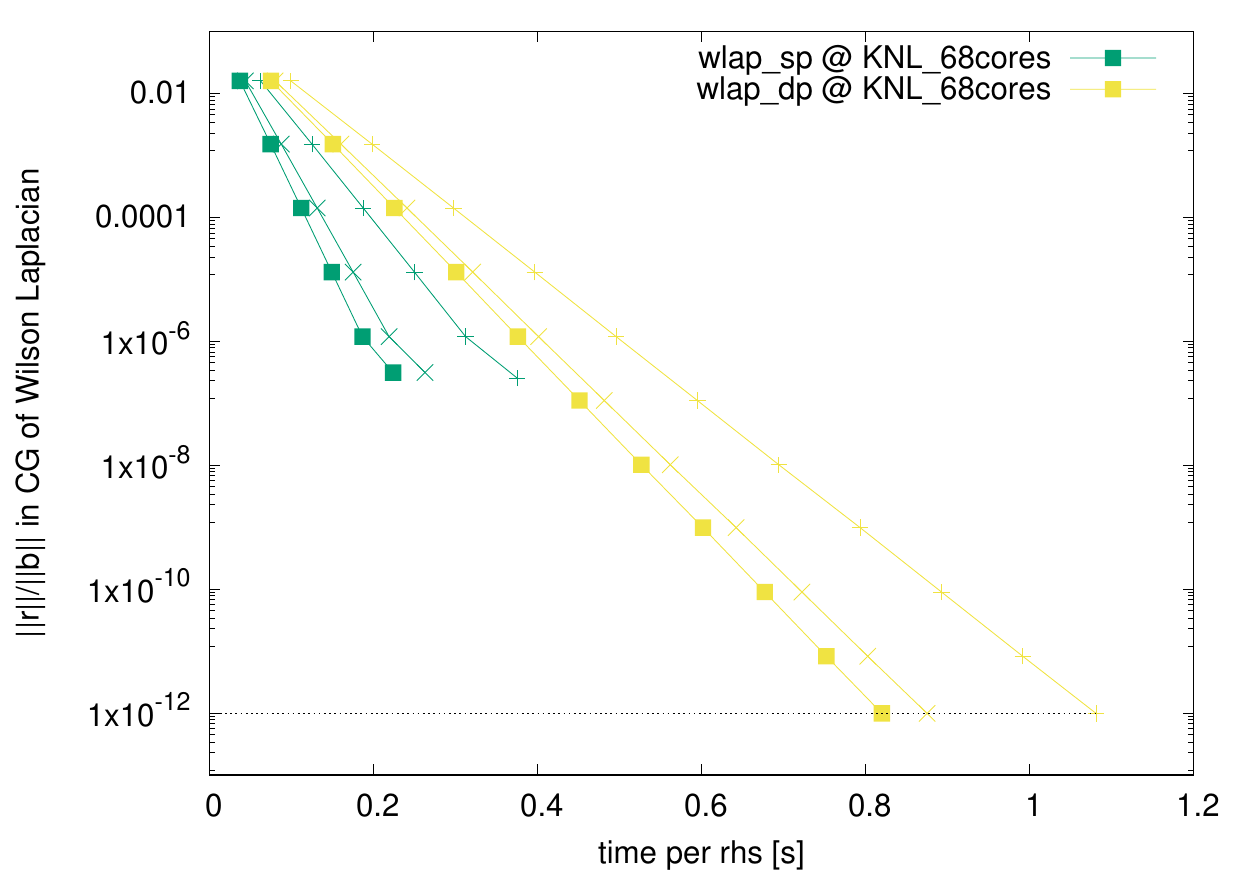}\\[-2mm]
\includegraphics[width=0.99\textwidth]{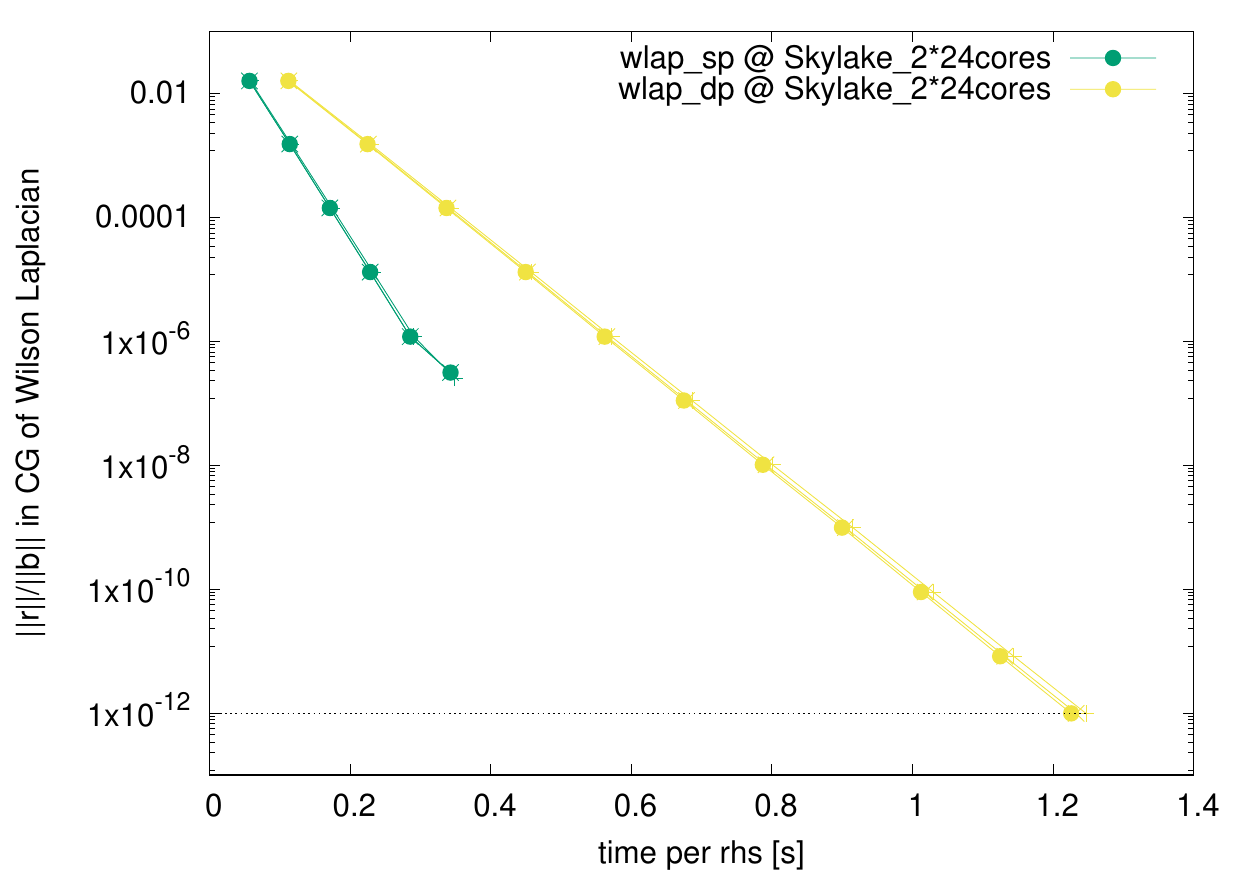}
\vspace*{-3mm}
\caption{\label{fig:convergence_wlap}%
Relative residual norm versus time of the CG solver for the Wilson Laplacian at $am_0=0.01$ on a $24^3\times48$ lattice in sp/dp (layout {\tt NvNsNc} filled, other $+,\times$) on the KNL and Skylake node.
In all cases the solver exits after 60 (109) iterations in sp (dp).}
\vspace*{-3mm}
\end{figure}

The CG history of $A=-\frac{1}{2}\lap^\mr{std}+\frac{1}{2}m_0^2$ is shown in Fig.\,\ref{fig:convergence_wlap}.
On the KNL chip the layout {\tt NcNsNv} is slowest ($+$ symbol), {\tt NvNcNs} is better ($\times$ symbol), and {\tt NvNsNc} is best (filled boxes), both in sp and dp.
It is worth mentioning that the physics content of the RHS vector $b$ is the same for the three vector layouts, i.e.\ the memory content of $b$ in {\tt NvNsNc} is just a permuted version of {\tt NcNsNv}.
The $x$-axis shows the time per RHS, the $y$-axis the \emph{worst} of the $\Nv$ relative residual norms.
In dp the target precision $\ep=10^{-12}$ is met after $109$ iterations (for each layout option), while in sp after $60$ iterations the algorithm notices that
the updated residual norm $||r||/||b||=1.11\times10^{-7}$ differs from the recomputed residual norm $||r||/||b||=3.12\times10^{-7}$ by more than 1\%, and thus stops%
\footnote{If one were to continue beyond this point, the recomputed residual norm would stagnate or grow.}.
On the Skylake architecture the difference among the layouts is gone, but the best overall time increases quite a bit (from 0.8\, to 1.2\,s in dp).

\begin{figure}[p]
\vspace*{-8mm}
\includegraphics[width=0.99\textwidth]{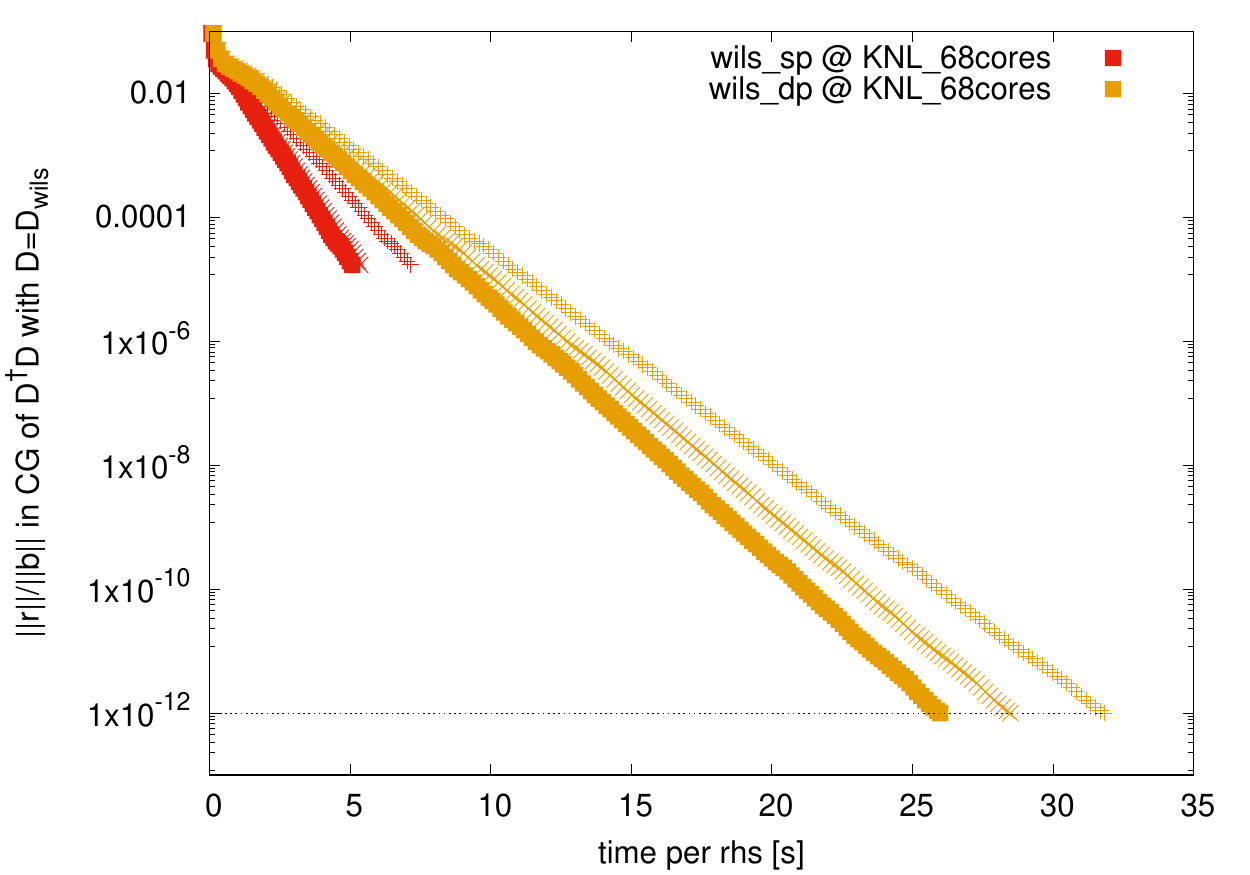}\\[-2mm]
\includegraphics[width=0.99\textwidth]{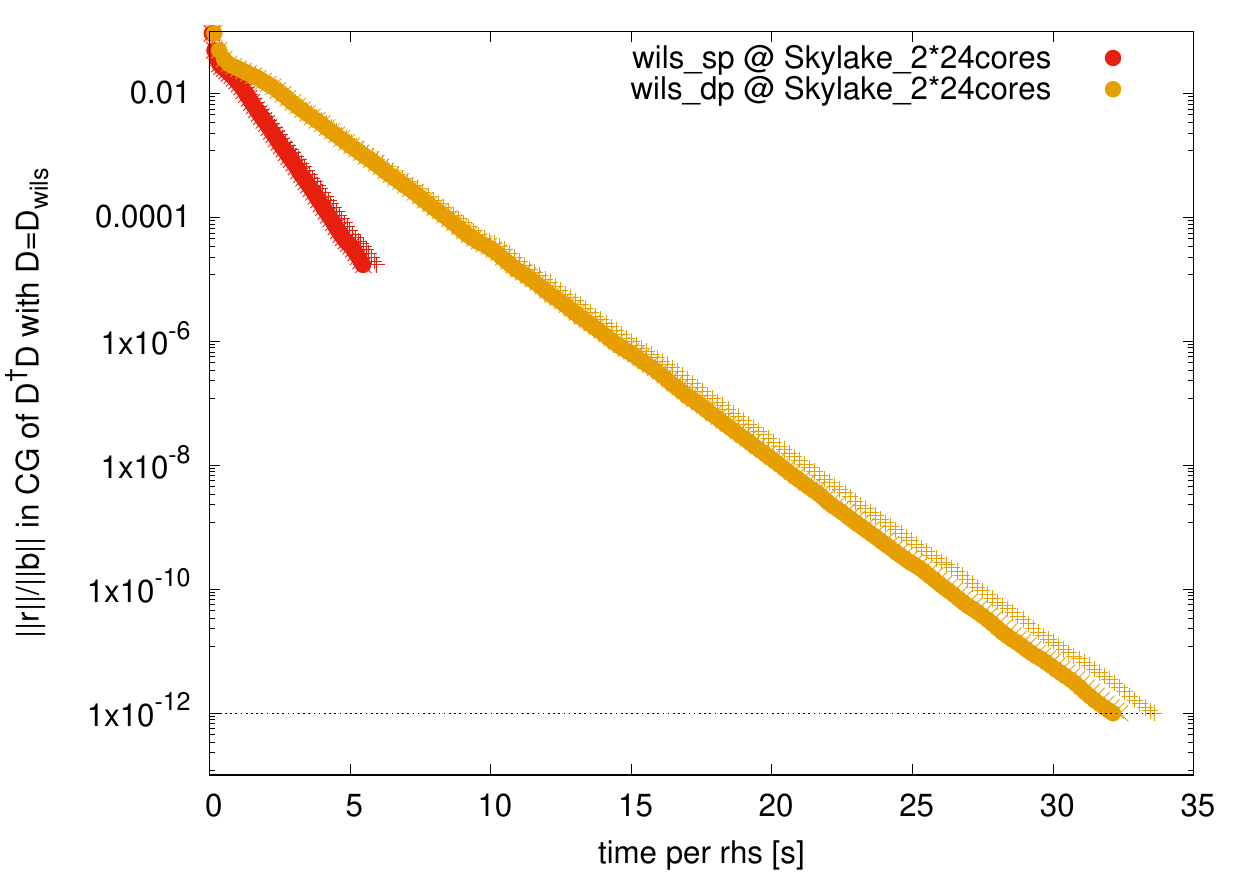}
\vspace*{-3mm}
\caption{\label{fig:convergence_wils}%
Relative residual norm versus time of the CG solver for the Wilson $\DW\dag\DW$ at $am_0=0.01$, $c_\mr{SW}=1$ on a $24^3\times48$ lattice in sp/dp (layout {\tt NvNsNc} filled, other $+,\times$) on the KNL and Skylake node.
In all cases the solver exits after 620 (1858) iterations in sp (dp).}
\vspace*{-3mm}
\end{figure}

The CG history of $A=\DW\dag\DW$ is shown in Fig.\,\ref{fig:convergence_wils}.
On the KNL chip the layout {\tt NvNsNc} is the fasted one (both in sp and dp, plotted with filled symbols).
In dp the target precision $\ep=10^{-12}$ is met after $1858$ iterations (for each layout option), while in sp after $620$ iterations the algorithm notices that
the updated residual norm $||r||/||b||=1.69\times10^{-5}$ differs from the recomputed residual norm $||r||/||b||=1.71\times10^{-5}$ by more than 1\%, and thus stops%
\footnote{The attentive reader may notice these figures are significantly larger than those mentioned in the discussion of the CG histories of $-\frac{1}{2}\lap^\mr{std}+\frac{1}{2}m_0^2$.
This suggests there is no universal tolerance that can be reached in sp; the minimum $||r||/||b||$ depends on the type of operator used and $am_0$.}.
On the Skylake architecture the difference among the layouts is almost gone, but the best overall execution time increases a bit (from 26\,s to 32\,s in dp).

\begin{figure}[p]
\vspace*{-8mm}
\includegraphics[width=0.99\textwidth]{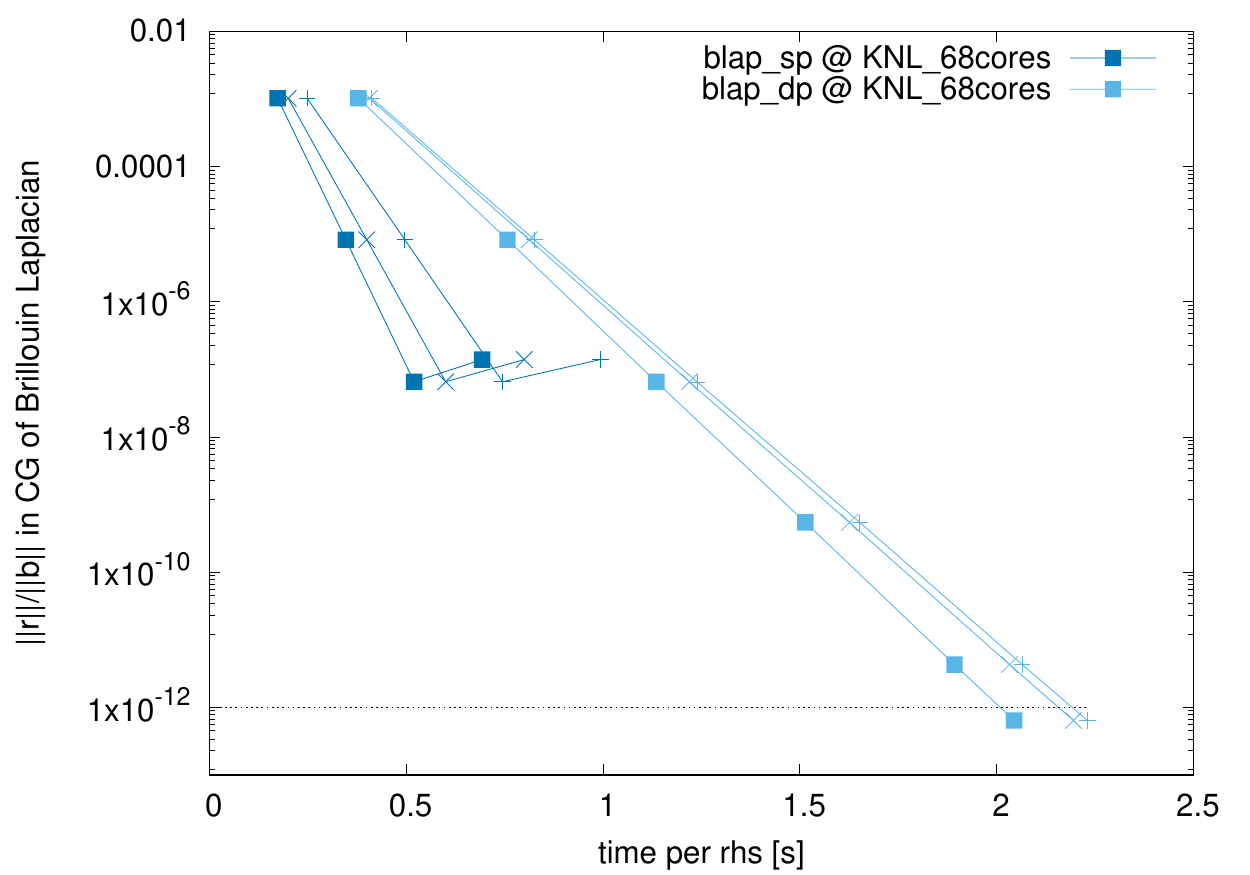}\\[-2mm]
\includegraphics[width=0.99\textwidth]{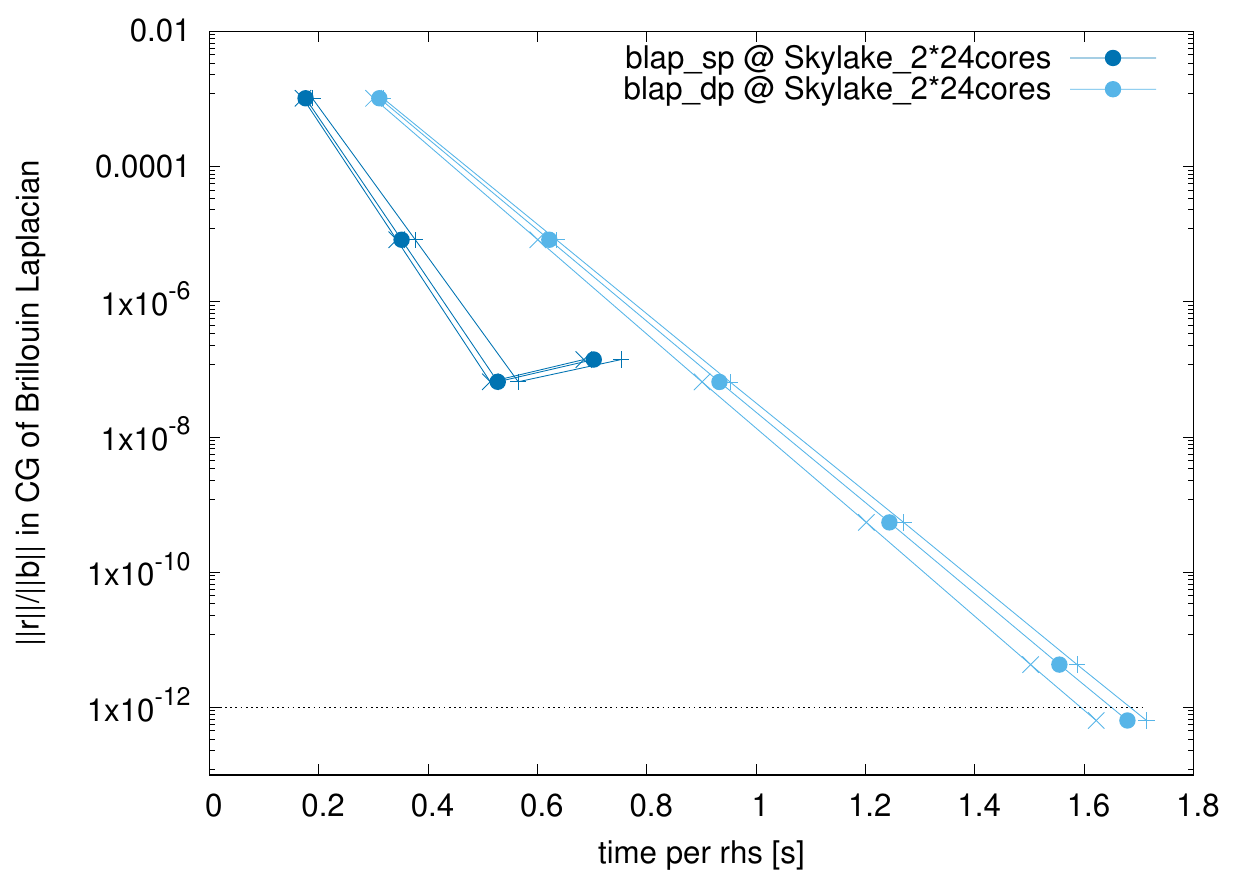}
\vspace*{-3mm}
\caption{\label{fig:convergence_blap}%
Relative residual norm versus time of the CG solver for the Brillouin Laplacian at $am_0=0.01$ on a $24^3\times48$ lattice in sp/dp (layout {\tt NvNsNc} filled, other $+,\times$) on the KNL and Skylake node.
In all cases the solver exits after 40 (54) iterations in sp (dp); cf.\ footnote \ref{foot:updated_recomputed}.}
\vspace*{-3mm}
\end{figure}

The CG history of $A=-\frac{1}{2}\lap^\mr{bri}+\frac{1}{2}m_0^2$ is shown in Fig.\,\ref{fig:convergence_blap}.
On the KNL chip the layout {\tt NvNsNc} is again the fastest one (both in sp and dp, plotted with filled symbols).
In dp the target precision $\ep=10^{-12}$ is met after $54$ iterations (for each layout option), while in sp after $40$ iterations the algorithm notices that
the updated residual norm $||r||/||b||=5.46\times10^{-10}$ differs from the recomputed residual norm $||r||/||b||=1.40\times10^{-7}$ by more than 1\%, and thus stops%
\footnote{In sp the third (fourth) box shows the updated (recomputed) relative residual norm at iteration 30 (40).\label{foot:updated_recomputed}}.
No signs of numerical instability are seen; the six symbols at a given precision and iteration count are just horizontally displaced from each other.
On the Skylake architecture the difference among the layouts is gone, while the best overall execution time is a bit shorter than on the KNL (from 2.0\,s to 1.7\,s in dp).

\begin{figure}[p]
\vspace*{-8mm}
\includegraphics[width=0.99\textwidth]{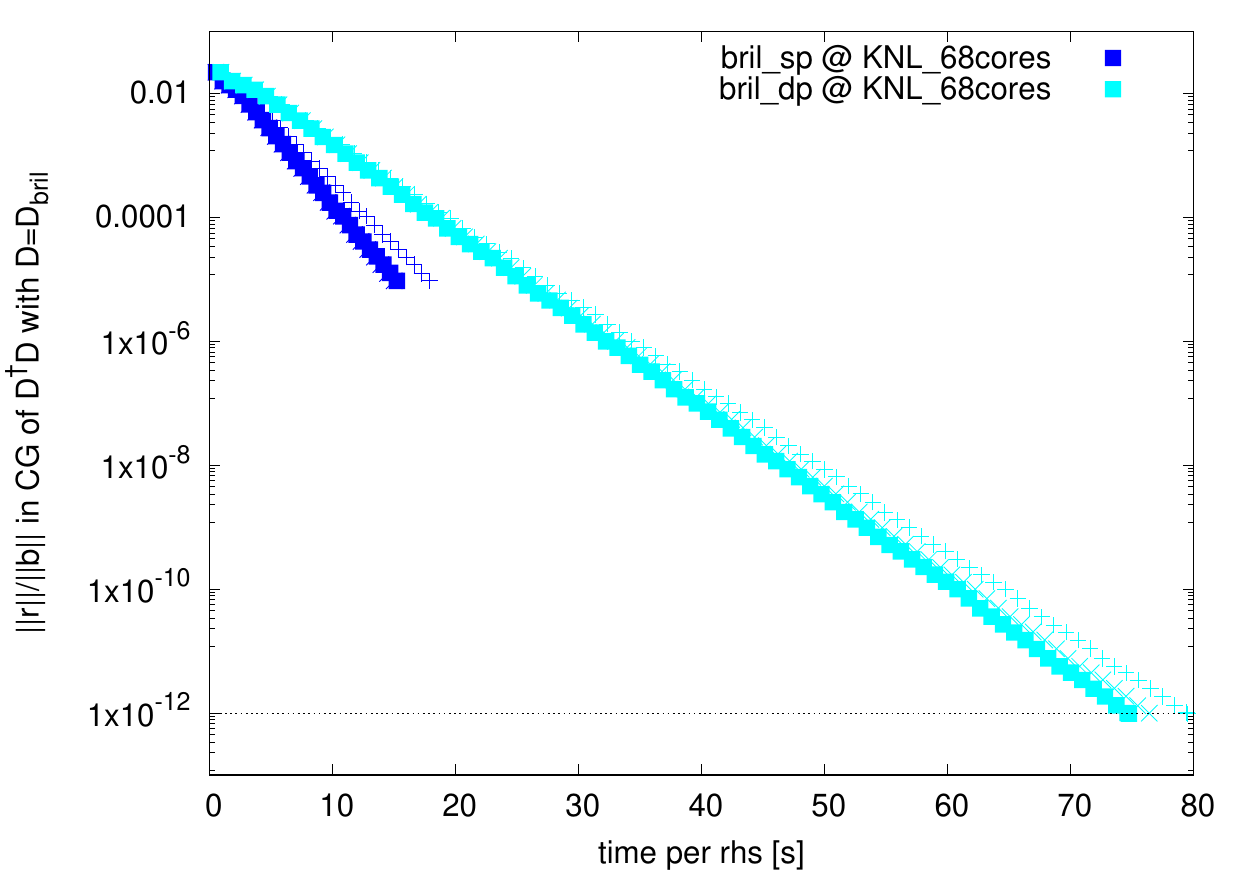}\\[-2mm]
\includegraphics[width=0.99\textwidth]{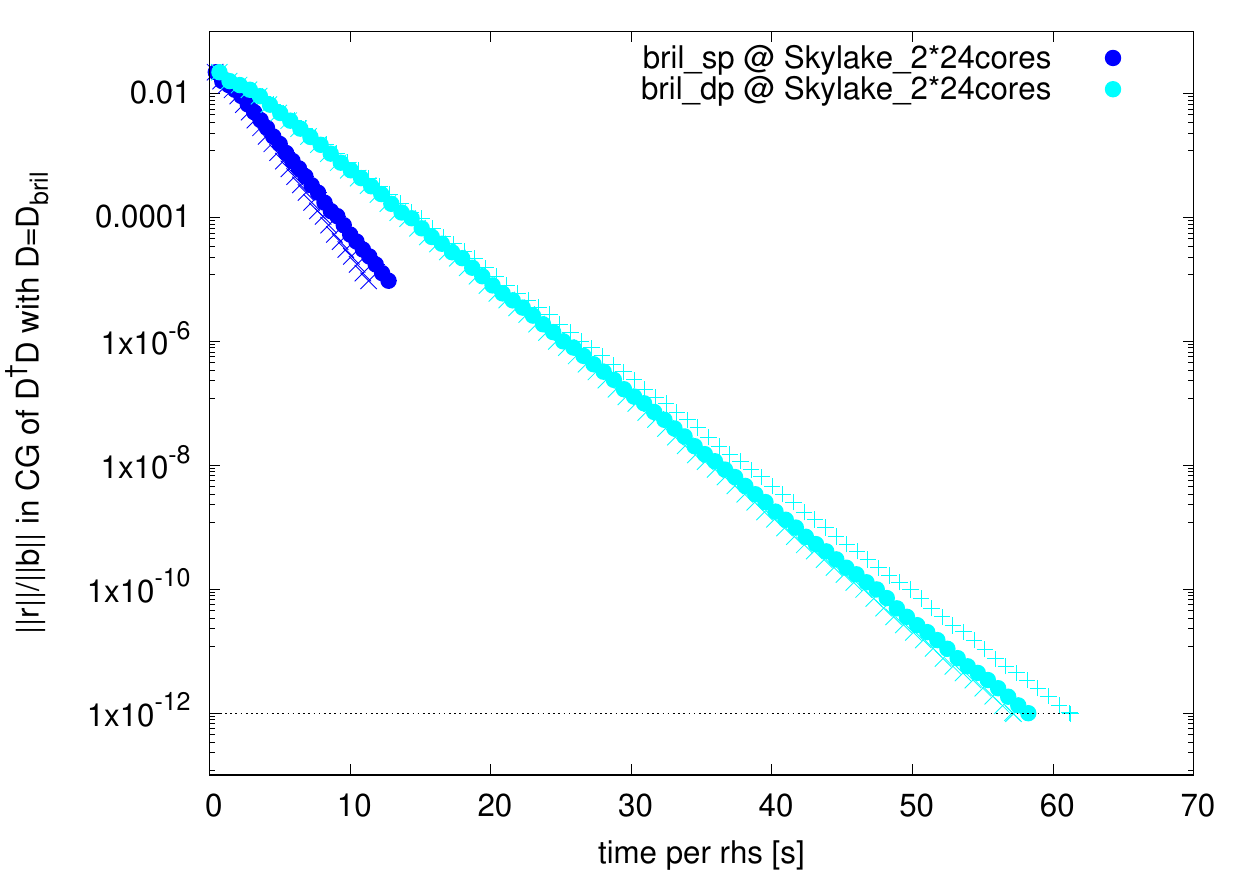}
\vspace*{-3mm}
\caption{\label{fig:convergence_bril}%
Residual residual norm versus time of the CG solver for the Brillouin $\DB\dag\DB$ at $am_0=0.01$, $c_\mr{SW}=1$ on a $24^3\times48$ lattice in sp/dp (layout {\tt NvNsNc} filled, other $+,\times$) on the KNL and Skylake node.
In all cases the solver exits after 280 (811) iterations in sp (dp).}
\vspace*{-3mm}
\end{figure}

The CG history of $A=\DB\dag\DB$ is shown in Fig.\,\ref{fig:convergence_bril}.
In dp the target precision $\ep=10^{-12}$ is met after $811$ iterations (for each layout option), while in sp after $280$ iterations the algorithm notices that
the updated residual norm $||r||/||b||=9.25\times10^{-6}$ differs from the recomputed residual norm $||r||/||b||=9.44\times10^{-6}$ by more than 1\%, and thus stops.
Here the difference among the three layout options is mild, and out of the KNL and Skylake architectures the latter one fares significantly better (75\,s versus 58\,s in dp).

In Figs.\,\ref{fig:convergence_wlap}--\ref{fig:convergence_bril} no sign of numerical imprecision is seen; the three symbols at a given iteration count (for either sp or dp) are just horizontally displaced.
A second issue is worth mentioning.
On the Skylake architecture the Brillouin operator \emph{converges in about twice the time} of the Wilson operator.
The additive mass shift of the two Dirac operators is roughly in the same%
\footnote{Preliminary spectroscopy on a handful of configurations suggests $\Mpi^\mr{wils}\simeq760\MeV$ and $\Mpi^\mr{bril}\simeq670\MeV$.} 
ballpark.
Thus the timings of Sec.~\ref{sec:wils} and Sec.~\ref{sec:bril} (where $\DB$ seemed about an order of magnitude more expensive than $\DW$) do not represent the last word on the relative cost of these two Dirac operators.
The reason is the more compact eigenvalue spectrum of $\DB$ [reaching up to $\mr{Re}(z)=2+am_0$ in the free field case] in comparison to $\DW$ [which extends to $\mr{Re}(z)=8+am_0$].
Hence at fixed pion mass, the matrix-vector cost explosion (in trading $\DW$ for $\DB$) is mitigated by a reduced condition number (see also the discussion in Refs.~\cite{Durr:2010ch,Durr:2012dw,Durr:2017wfi}).

\begin{figure}[p]
\vspace*{-8mm}
\includegraphics[width=0.99\textwidth]{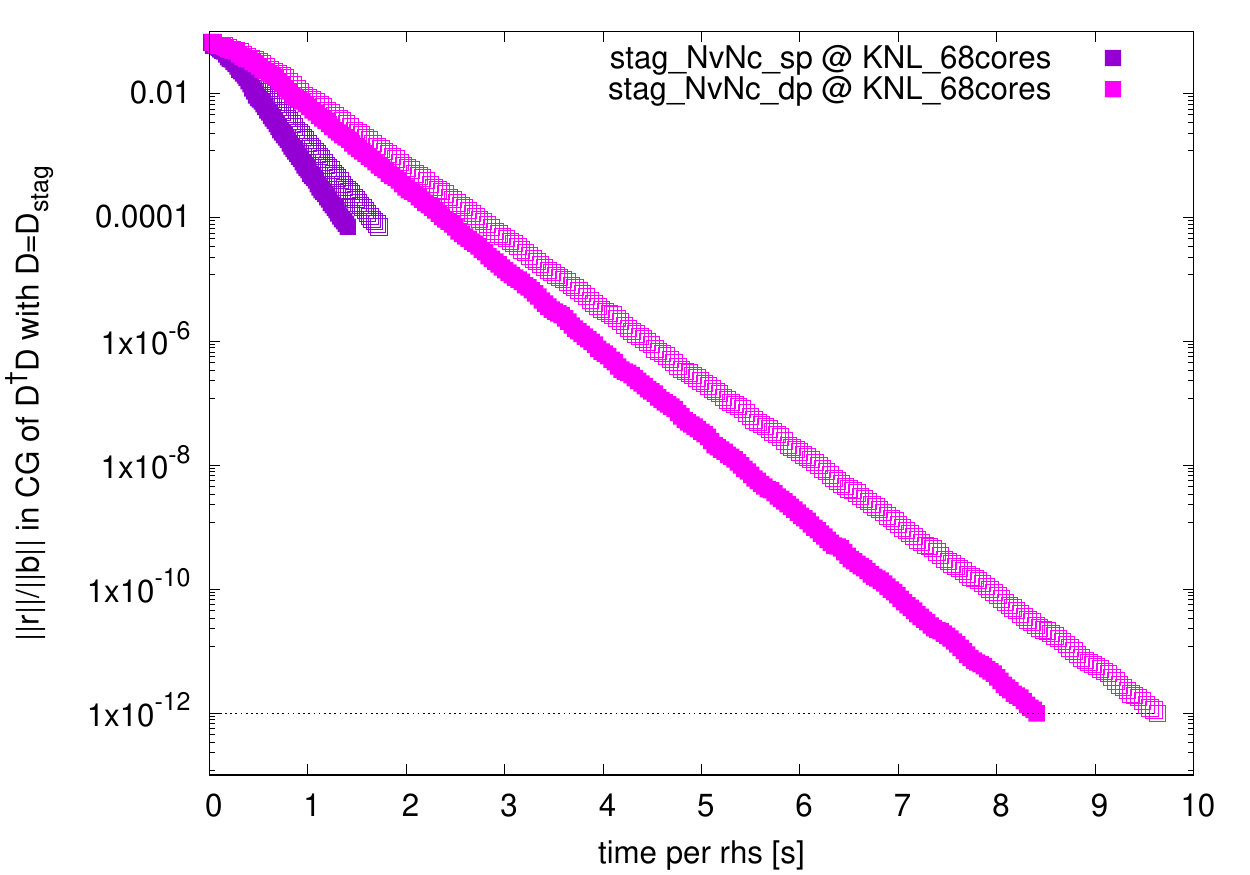}\\[-2mm]
\includegraphics[width=0.99\textwidth]{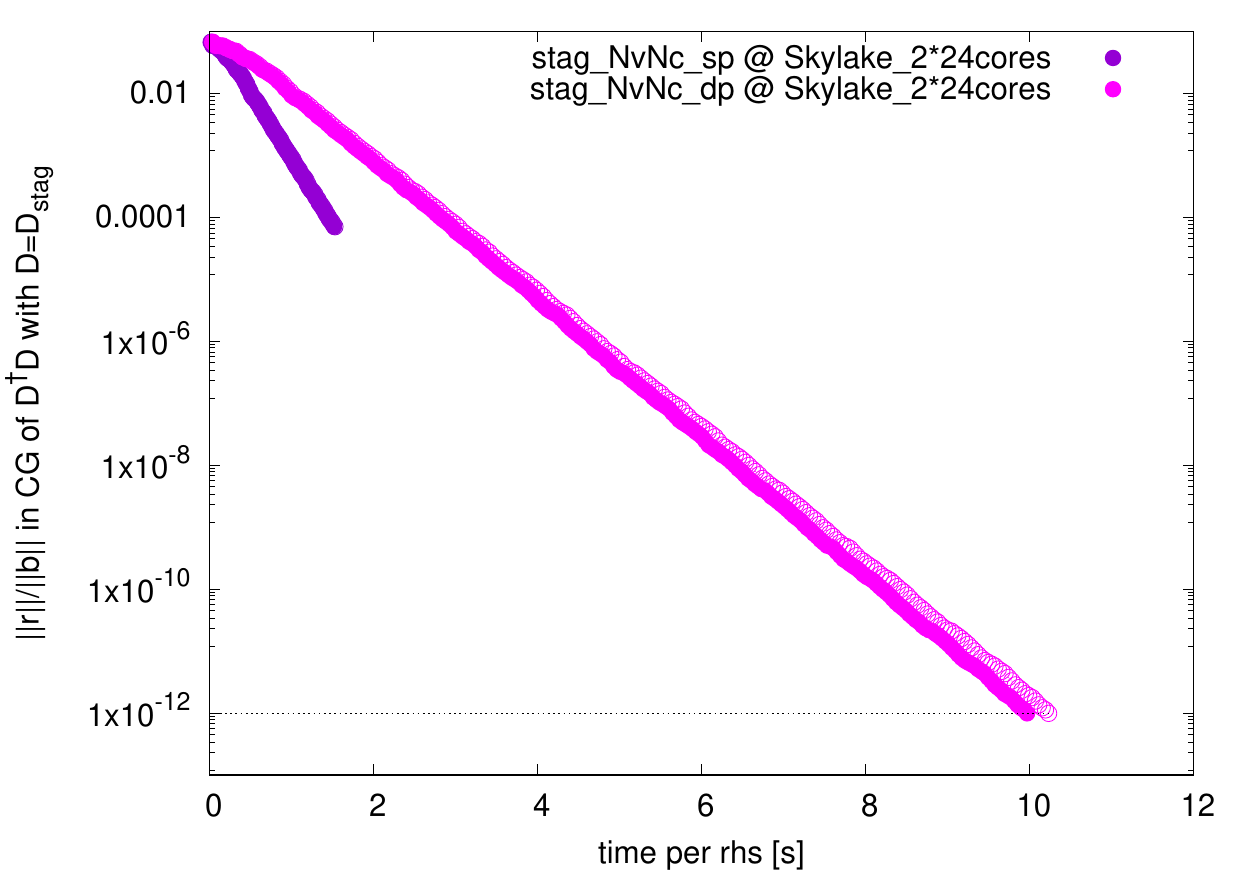}
\vspace*{-3mm}
\caption{\label{fig:convergence_stag}%
Relative residual norm versus time of the CG solver for the staggered $\DS\dag\DS$ at $am_0=0.01$ on a $24^3\times48$ lattice in sp/dp (layout {\tt NvNc} filled, other open) on the KNL and Skylake node.
In all cases the solver exits after 740 (2489) iterations in sp (dp).}
\vspace*{-3mm}
\end{figure}

The CG history of $A=\DS\dag\DS$ is shown%
\footnote{Preliminary spectroscopy on a handful of configurations suggests $\Mpi^\mr{stag}\simeq340\MeV.$}                                   
in Fig.\,\ref{fig:convergence_stag} (as usual adjacent boxes are separated by ten iterations).
On the KNL chip the layout {\tt NvNc} (filled symbols) is faster than {\tt NcNv} (open symbols); on the Skylake architecture the difference is marginal.
In dp the target precision $\ep=10^{-12}$ is met after $2489$ iterations (for either layout), while in sp after $740$ iterations the algorithm notices that
the updated residual norm $||r||/||b||=6.95\times10^{-5}$ differs from the recomputed residual norm $||r||/||b||=7.06\times10^{-5}$ by more than 1\%, and thus stops.

\begin{figure}[p]
\vspace*{-8mm}
\includegraphics[width=0.99\textwidth]{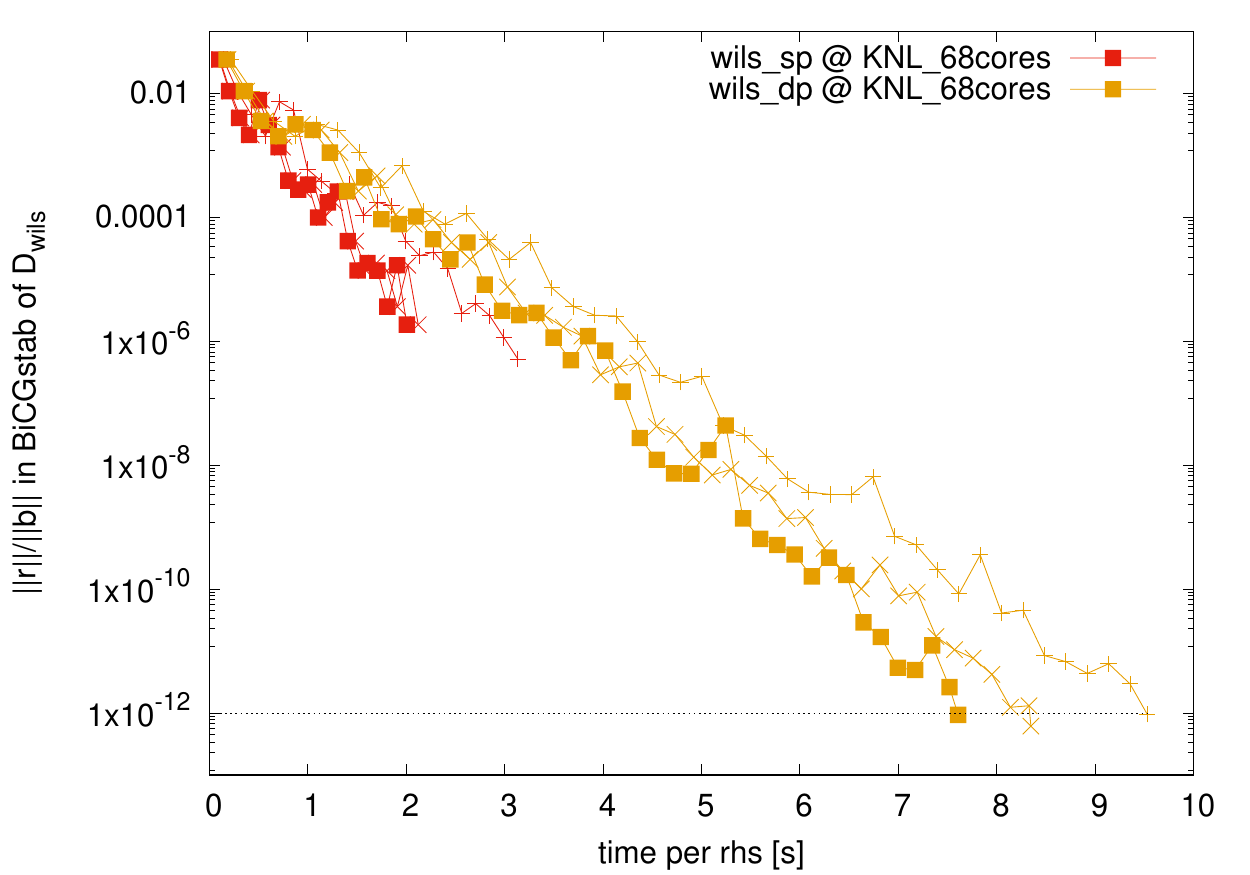}\\[-2mm]
\includegraphics[width=0.99\textwidth]{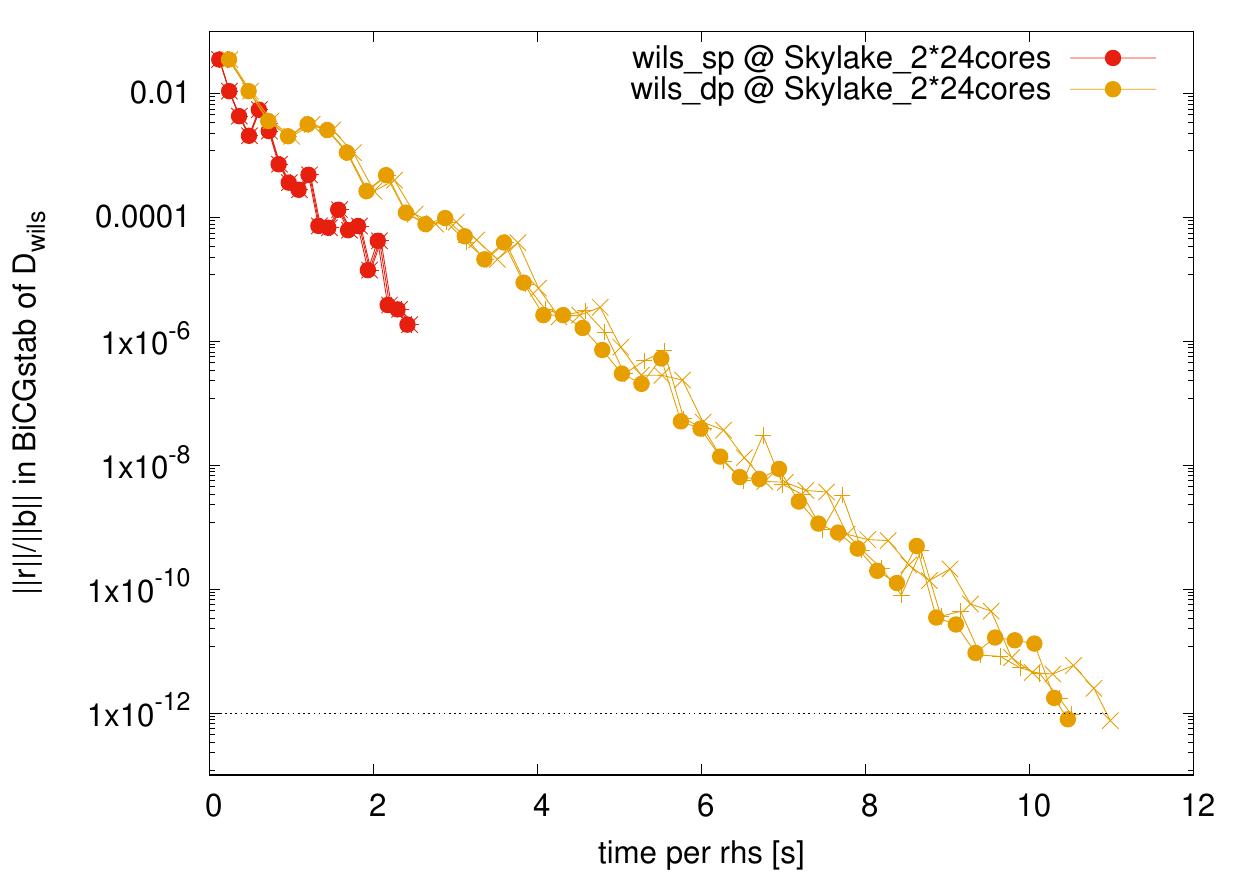}
\vspace*{-3mm}
\caption{\label{fig:conbicgstab_wils}%
Relative residual norm versus time of the BiCGstab solver for $\DW$ at $am_0=0.01$, $c_\mr{SW}=1$ on a $24^3\times48$ lattice in sp/dp (layout {\tt NvNsNc} filled, other $+,\times$) on the KNL and Skylake node.
In all cases the solver exits after 200 ($\sim435$) iterations in sp (dp).}
\vspace*{-3mm}
\end{figure}

Fig.\,\ref{fig:conbicgstab_wils} shows the convergence history of the BiCGstab algorithm for $\DW u=b$.
The descent is ``wigglier'' than for the CG algorithm.
The sensitivity of the BiCGstab algorithm to numerical inaccuracy is visible; the six convergence histories in dp are not ``horizontally stretched carbon copies'' of each other.
As was true for the CG algorithm, with the Wilson kernel the convergence on the KNL chip is a bit faster than on the Skylake architecture.

\begin{figure}[p]
\vspace*{-8mm}
\includegraphics[width=0.99\textwidth]{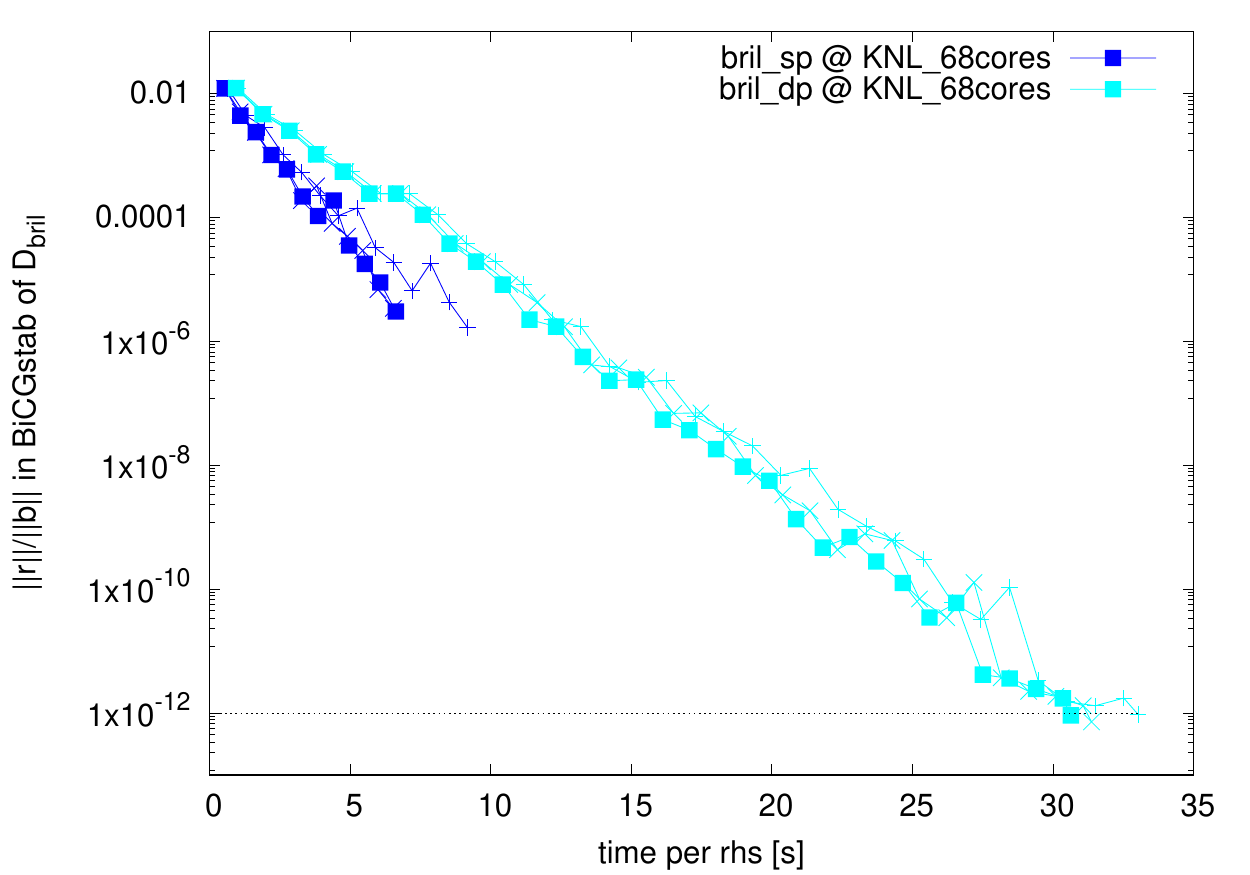}\\[-2mm]
\includegraphics[width=0.99\textwidth]{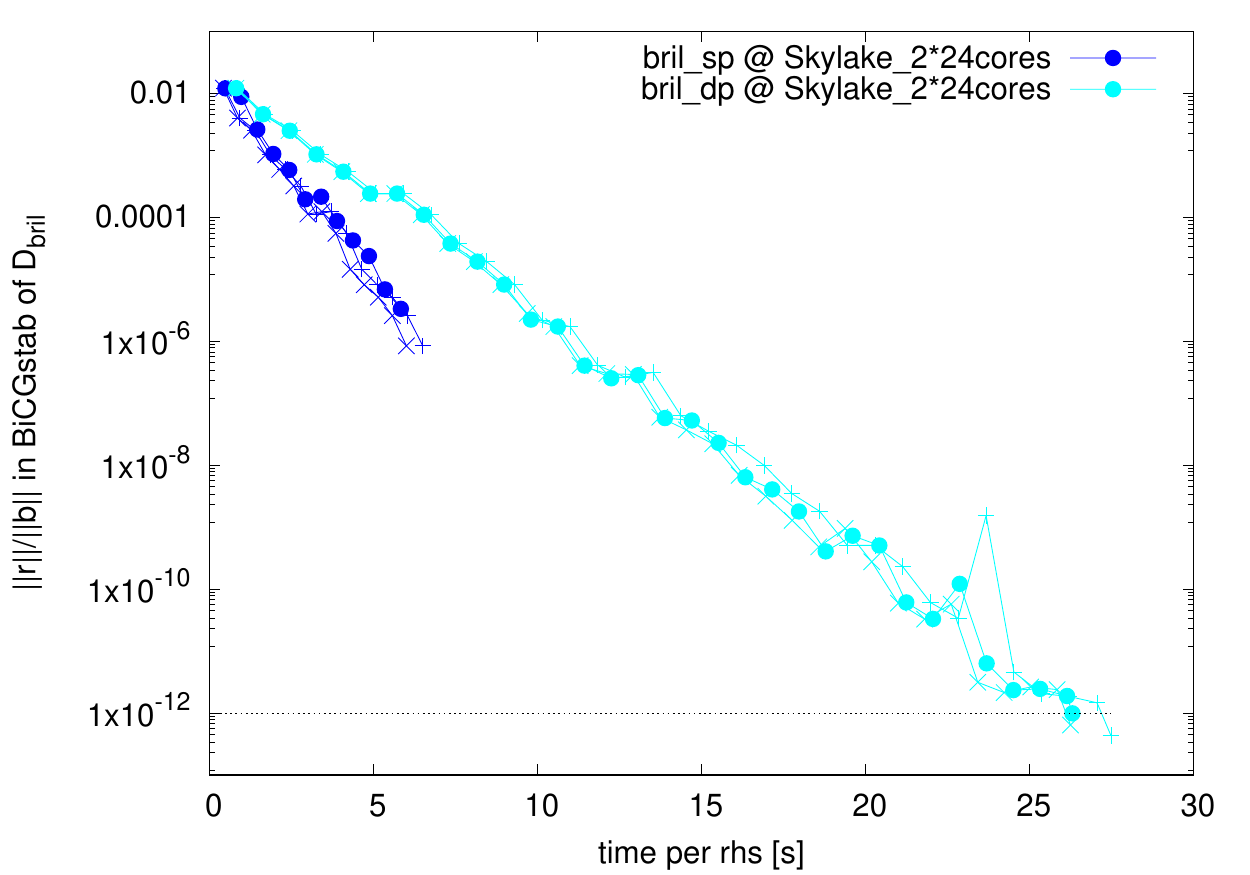}
\vspace*{-3mm}
\caption{\label{fig:conbicgstab_bril}%
Relative residual norm versus time of the BiCGstab solver for $\DB$ at $am_0=0.01$, $c_\mr{SW}=1$ on a $24^3\times48$ lattice in sp/dp (layout {\tt NvNsNc} filled, other $+,\times$) on the KNL and Skylake node.
In all cases the solver exits after 140 ($\sim322$) iterations in sp (dp).}
\vspace*{-3mm}
\end{figure}

Fig.\,\ref{fig:conbicgstab_bril} shows the convergence history of the BiCGstab algorithm for $\DB u=b$.
The overall characteristic is similar to the Wilson kernel, i.e.\ there are some wiggles, but they are not dramatic (at this heavy pion mass).
The main difference to the previous plot is the performance ratio between the two architectures.
As was true for the CG algorithm, with the Brillouin kernel the convergence on the Skylake architecture is significantly faster than on the KNL chip.
This is good news for the suitability of the Brillouin Dirac operator in phenomenological studies on architectures which are limited by memory bandwidth (cf.\ Refs.~\cite{Durr:2010ch,Durr:2012dw,Durr:2017wfi}).

Overall, explicit Krylov space solve-for operations show that the {\tt vec}-layout {\tt NvNsNc} usually fares best in terms of the total run-time.
And among the {\tt suv}-layouts {\tt NvNc} beats {\tt NcNv}.
These statements hold for to the KNL chip; on the Skylake node the time differences are marginal.
This matches the observations made in Secs.~\ref{sec:wils}, \ref{sec:bril}, \ref{sec:stag} for $\DW$, $\DB$, $\DS$, respectively.
The linear algebra rankings established in Sec.~\ref{sec:norms} are not crucial to the overall solver performance.
Accordingly, if one were to upgrade the code into a distribution with hybrid parallelization (MPI and OpenMP), one would restrict oneself to the {\tt NvNsNc} and {\tt NvNc} layouts, respectively.

As is well known \cite{Giusti:2002sm,Durr:2008rw,Baboulin:2009,Azzam:2020} the descent of $||r||$ in the sp-solver can be extended
beyond the sp-``limit'' $\ep\simeq10^{-6}$ by wrapping the sp-solver into a simple dp-updater (e.g.\ Richardson iteration).
Such ``mixed precision solvers'' are standard in the lattice community, and the ancillary code distribution contains a ``mixed precision'' version of each solver presented in this section.


\section{Summary\label{sec:summary}}


The goal of this article has been to explore whether a ``traditional'' strategy of implementing the Dirac-matrix times vector operation may yield acceptable performance figures
on many-core CPU architectures such as the KNL or more modern successors.
Here ``traditional'' means the implementation adopts a high-level language (e.g.\ Fortran\,2008), without assembly-tuning, and without cache-line optimization;
only OpenMP shared-memory parallelization and SIMD pragmas are used.
Furthermore, the freedom to choose the data layout is deliberately limited to the ordering of the internal degrees of freedom (color/spinor/RHS-index).
In this way, full portability of the code is ensured.

On the KNL processor the SIMD-index must be the fastest (in Fortran: first) one, so the layouts {\tt NvNcNs} and {\tt NvNsNc} are preferable for Wilson-type, and {\tt NvNc} for Susskind-type vectors.
For the three Dirac operators studied (Wilson, Brillouin and Susskind) acceptable performance figures are found.
In sp, they are of the order of 480\,Gflop/s, 880\,Gflop/s, and 780\,Gflop/s, respectively, on one 68-core KNL chip.
On the 24-core Skylake architecture the ordering of the internal degrees of freedom seems almost irrelevant.
This architecture has been tested in a dual-socket node, and with this configuration the sp performance figures read 350\,Gflop/s, 1320\,Gflop/s, and 550\,Gflop/s for the three operators, respectively.
In dp the performance figures are roughly halved, but in practice the sp performance is more relevant, since it determines the speed of a mixed-precision solver \cite{Giusti:2002sm,Durr:2008rw,Baboulin:2009,Azzam:2020}.
As an aside it was demonstrated that the relative residual norm $\ep=10^{-12}$ can be reached with dp vectors even if the underlying gauge field is in sp, i.e.\ if the matrix-vector operation is $v_\mr{dp} \leftarrow D_\mr{dp}[U_\mr{sp}]u_\mr{dp}$.

Comparing the time-to-solution of a typical CG or BiCGstab inverter on the KNL and on the Skylake node reveals an important difference between the Dirac operators involved in this study.
For the Wilson and the staggered Dirac operators (and the Wilson Laplacian) the KNL chip turns out to be faster, for the Brillouin Dirac operator (and the Brillouin Laplacian) the Skylake node is faster.
This difference is linked to the \emph{computational intensity} of these operators.
The Wilson operator uses between $0.89$ and $1.53$ flops/byte for $\Nc=3$ (depending on the number of RHS, see App.~\ref{app:E}), the staggered operator is a little higher.
For Brillouin fermions these landmarks are lifted to $2.21$ and $3.83$ flops/byte respectively (see App.~\ref{app:E}).
Hence, for operators with a large enough computational intensity the Skylake architecture can provide an advantage over KNL.
The KNL architecture is about to be phased out, and future architectures will come with a compute-to-bandwidth ratio comparable to that of the Skylake architecture (or higher).
As a result, the lattice community may be well advised to switch to Dirac operators with a high computational intensity, such as the Brillouin operator (\ref{def_bril}).

For the author the main conclusion is that the principles this implementation is based on (compiler only, multiple RHS,
SIMD-ization over the RHS-index rather than space-time indices, lack of large index tables) pass an important test on several architectures.
Hence a dedicated effort to use these routines in a multi-node hybrid implementation seems worth while.
Obviously this will be a difficult task, because the single-node kernel is reasonably efficient.
For the staggered Dirac operator Tab.~\ref{tab:stag_knl} documents a sp concurrency time of $0.024\,\mr{s}$ for $\Nv=12$, i.e.\ $0.002\,\mr{s}$ per RHS.
This happens with an unusually big lattice per node ($34^3\times68$, a $32^3\times64$ volume has $0.020\,\mr{s}$ for twelve RHS).
Scaling this to a local volume of $8^4$ (which is a common single-node volume in an MPI-OpenMP hybrid parallelization) leads to concurrency time $512$ times smaller, i.e.\ about $39\,\mu\mr{s}$.
To scale in the number of nodes, clever MPI calls must be arranged, with overlapping communication and computation.
Given the latency of a few microseconds in a standard MPI implementation, this is a non-trivial task.
Still, with an equally efficient staggered kernel which acts on a single RHS, the concurrency time would be down to $3\,\mu\mr{s}$;
in this case a good strong-scaling behavior in the number of MPI ranks would be a ``mission impossible''.
In short, this little \emph{Gedanken experiment} underlines the importance of design choices made early in the development process,
and it illustrates the ``curse of efficiency'' every HPC practitioner knows (and has somewhat mixed feelings about).

\subsection*{Acknowledgements}

Code development and computations were carried out on DEEP, as well as JURECA and JUWELS at IAS/JSC of the Forschungszentrum J\"ulich.
Access to DEEP was granted through the DEEP-EST Early Access Programme.
The author received partial support from the DFG via the collaborative research programme SFB/TRR-55.
Useful discussions with Eric Gregory from IAS/JSC of the Forschungszentrum J\"ulich are gratefully acknowledged.

\clearpage


\appendix


\section{Gamma matrices and Wilson projection trick\label{app:A}}


We employ the ``chiral'' or Weyl representation of the $\ga$-matrices
\bdm
\ga_1=
\begin{pmatrix}
  0 &  0 &  0 &-\ri\\
  0 &  0 &-\ri&  0 \\
  0 & \ri&  0 &  0 \\
 \ri&  0 &  0 &  0 
\end{pmatrix}
=\si_2\otimes\si_1
\;,\quad
\ga_2=
\begin{pmatrix}
  0 &  0 &  0 & -1 \\
  0 &  0 &  1 &  0 \\
  0 &  1 &  0 &  0 \\
 -1 &  0 &  0 &  0 
\end{pmatrix}
=\si_2\otimes\si_2
\;,\quad
\edm
\beq
\ga_3=
\begin{pmatrix}
  0 &  0 &-\ri&  0 \\
  0 &  0 &  0 & \ri\\
 \ri&  0 &  0 &  0 \\
  0 &-\ri&  0 &  0 
\end{pmatrix}
=\si_2\otimes\si_3
\;,\quad
\ga_4=
\begin{pmatrix}
  0 &  0 &  1 &  0 \\
  0 &  0 &  0 &  1 \\
  1 &  0 &  0 &  0 \\
  0 &  1 &  0 &  0 
\end{pmatrix}
=\si_1\otimes\si_0
\;,\quad
\label{def_gammamatrices}
\eeq
where the tensor-product notation uses the Pauli matrices
\beq
\si_0=
\begin{pmatrix}
  1 &  0 \\
  0 &  1 \\
\end{pmatrix}
\;,\quad
\si_1=
\begin{pmatrix}
  0 &  1 \\
  1 &  0 \\
\end{pmatrix}
\;,\quad
\si_2=
\begin{pmatrix}
  0 &-\ri\\
 \ri&  0 \\
\end{pmatrix}
\;,\quad
\si_3=
\begin{pmatrix}
  1 &  0 \\
  0 & -1 \\
\end{pmatrix}
\;,
\eeq
and the main feature of this representation is the diagonal form of
\beq
\ga_5=\ga_1\ga_2\ga_3\ga_4=(\si_2\si_2\si_2\si_1)\otimes(\si_1\si_2\si_3\si_0)=(-\ri\si_3)\otimes(\ri\si_0)=\mr{diag}(1,1,-1,-1)
\;.
\eeq

The Wilson Dirac operator (\ref{def_wils}) multiplies the contribution to the out-vector at site $n$ that comes from $n\pm\hat\mu$ with $V_{\pm\mu}(n)$ in color space and with $\frac{1}{2}(1\pm\ga_\mu)$ in spinor space.
A vector of length $4\Nc$ (with the color index moving fastest) is thus multiplied by $\frac{1}{2}(1\pm\ga_\mu)\otimes V_{\pm\mu}(n)$, and this is equivalent to the following procedure.
Reshape the vector into a $\Nc\times4$ matrix, multiply it with $V_{\pm\mu}(n)$ from the left and with $\frac{1}{2}(1\pm\ga_\mu)^\mr{trsp}$ from the right, and reshape the result back into a column vector.
In the right-multiplication we exploit that $\frac{1}{2}(1\pm\ga_\mu)$ is a projector, and that either associate eigenspace (to the eigenvalues $0$ and $1$, respectively) has dimension two.
This holds regardless of the representation used, and for each direction $\mu$.

Specifically, for the chiral representation the eigenvector decomposition takes the form
\bea
\frac{1}{2}(1+\ga_1)=PP\dag,\;
\frac{1}{2}(1-\ga_1)=QQ\dag
&\mbox{with}&
P=\frac{1}{\sqrt{2}}\begin{pmatrix} -\ri & 0 \\ 0 & 1 \\ 0 &  \ri \\ 1 & 0 \end{pmatrix}\!,\;
Q=\frac{1}{\sqrt{2}}\begin{pmatrix}  \ri & 0 \\ 0 & 1 \\ 0 & -\ri \\ 1 & 0 \end{pmatrix}\!,
\\
\frac{1}{2}(1+\ga_2)=PP\dag,\;
\frac{1}{2}(1-\ga_2)=QQ\dag
&\mbox{with}&
P=\frac{1}{\sqrt{2}}\begin{pmatrix} -1 & 0 \\ 0 & 1 \\ 0 &  1 \\ 1 & 0 \end{pmatrix}\!,\;
Q=\frac{1}{\sqrt{2}}\begin{pmatrix}  1 & 0 \\ 0 & 1 \\ 0 & -1 \\ 1 & 0 \end{pmatrix}\!,
\\
\frac{1}{2}(1+\ga_3)=PP\dag,\;
\frac{1}{2}(1-\ga_3)=QQ\dag
&\mbox{with}&
P=\frac{1}{\sqrt{2}}\begin{pmatrix} 0 & 1 \\  \ri & 0 \\ 0 &  \ri \\ 1 & 0 \end{pmatrix}\!,\;
Q=\frac{1}{\sqrt{2}}\begin{pmatrix} 0 & 1 \\ -\ri & 0 \\ 0 & -\ri \\ 1 & 0 \end{pmatrix}\!,
\\
\frac{1}{2}(1+\ga_4)=PP\dag,\;
\frac{1}{2}(1-\ga_4)=QQ\dag
&\mbox{with}&
P=\frac{1}{\sqrt{2}}\begin{pmatrix} 0 & -1 \\  1 & 0 \\ 0 & -1 \\ 1 & 0 \end{pmatrix}\!,\;
Q=\frac{1}{\sqrt{2}}\begin{pmatrix} 0 & -1 \\ -1 & 0 \\ 0 &  1 \\ 1 & 0 \end{pmatrix}\!,
\eea
where $P$ and $Q$ are unique up to arbitrary phases.
Hence, the following order of operations saves CPU time.
First act with $\sqrt{2}P^*$ from the right, next multiply the resulting $\Nc\times2$ matrix with $\frac{1}{2}V_\mu(n)$ from the left, and finally right-multiply with $\sqrt{2}P^\mr{trsp}$.
This holds for any $\mu>0$; for a negative $\mu$ simply replace $P$ by the respective $Q$.

Lattice practitioners refer to this procedure as the ``spin projection trick'' or ``shrink expand trick'', but it is sparsely documented in the literature
(notable exceptions include Refs.~\cite{Kamleh:2002is,Alexandru:2011ee}).
On a distributed memory machine, there is an obvious lemma.
In case the spinor is part of a halo component which is to be communicated to another MPI rank, the first right-multiplication (by a $4\times2$ matrix) is done prior to sending,
the second one (by a $2\times4$ matrix) after receipt.
This way the number of bits to be communicated is reduced by a factor two (applicable, again, for all directions $\pm\mu$).
Finally, the left-multiplication with the color-matrix $V_\mu(n)$ is done in the node where the latter resides (i.e.\ prior to sending for $\mu<0$, and after receipt for $\mu>0$).

Using the tensor notation (\ref{def_gammamatrices}) it is easy to write down the matrices $\si_{\mu\nu}=\frac{\ri}{2}[\ga_\mu,\ga_\nu]$, viz.
\bea
\si_{12}&=&\frac{\ri}{2}[\si_2\otimes\si_1,\si_2\otimes\si_2]=\frac{\ri}{2}\si_2^2\otimes[\si_1,\si_2]=\frac{\ri}{2}\si_0\otimes+2\ri\si_3=-\si_0\otimes\si_3\;,
\nonumber\\
\si_{13}&=&\frac{\ri}{2}[\si_2\otimes\si_1,\si_2\otimes\si_3]=\frac{\ri}{2}\si_2^2\otimes[\si_1,\si_3]=\frac{\ri}{2}\si_0\otimes-2\ri\si_2=+\si_0\otimes\si_2\;,
\nonumber\\
\si_{14}&=&\frac{\ri}{2}[\si_2\otimes\si_1,\si_1\otimes\si_0]=\frac{\ri}{2}[\si_2\si_1\otimes\si_1\si_0-\si_1\si_2\otimes\si_0\si_1]=+\si_3\otimes\si_1      \;,
\nonumber\\
\si_{23}&=&\frac{\ri}{2}[\si_2\otimes\si_2,\si_2\otimes\si_3]=\frac{\ri}{2}\si_2^2\otimes[\si_2,\si_3]=\frac{\ri}{2}\si_0\otimes+2\ri\si_1=-\si_0\otimes\si_1\;,
\nonumber\\
\si_{24}&=&\frac{\ri}{2}[\si_2\otimes\si_2,\si_1\otimes\si_0]=\frac{\ri}{2}[\si_2\si_1\otimes\si_2\si_0-\si_1\si_2\otimes\si_0\si_2]=+\si_3\otimes\si_2      \;,
\nonumber\\
\si_{34}&=&\frac{\ri}{2}[\si_2\otimes\si_3,\si_1\otimes\si_0]=\frac{\ri}{2}[\si_2\si_1\otimes\si_3\si_0-\si_1\si_2\otimes\si_0\si_3]=+\si_3\otimes\si_3      \;,
\eea
and this explicit form is used in the clover routine, see Sec.~\ref{sec:clov}.


\section{Details of the Wilson Laplace and Dirac routines\label{app:B}}


The structure of the Wilson Laplace routine (\ref{def_wlap}) for the vector layout {\tt NvNsNc} is
\small
\begin{verbatim}
!$OMP PARALLEL DO COLLAPSE(2) DEFAULT(private) FIRSTPRIVATE(mass) SHARED(old,new,V)
do t=1,Nt
do z=1,Nz; z_min=... ; z_plu=... ; t_min=modulo(t-2,Nt)+1; t_plu=modulo(t,Nt)+1
do y=1,Ny; y_min=... ; y_plu=...
do x=1,Nx; x_min=... ; x_plu=...
   !!! direction 0 gets mass term
   do concurrent(col=1:Nc,spi=1:4,rhs=1:Nv)
      site(rhs,spi,col)=(8.0+mass**2)*old(rhs,spi,col,x,y,z,t)
   end do
   !!! add contributions from -4 and +4 directions
   do concurrent(ccc=1:Nc,spi=1:4)
      !$OMP SIMD
      do rhs=1,Nv
         site(rhs,spi,:)-=conjg(V(ccc,:,4,x,y,z,t_min))*old(rhs,spi,ccc,x,y,z,t_min)
         site(rhs,spi,:)-=      V(:,ccc,4,x,y,z,t    ) *old(rhs,spi,ccc,x,y,z,t_plu)
      end do
   end do
   ...
   !!! plug site into new vector
   forall(col=1:Nc,spi=1:4,rhs=1:Nv) new(rhs,spi,col,x,y,z,t)=0.5*site(rhs,spi,col)
end do ! x=1,Nx
end do ! y=1,Ny
end do ! z=1,Nz
end do ! t=1,Nt
!$OMP END PARALLEL DO
\end{verbatim}
\normalsize
where we use a Fortran inspired notation.
The variables {\tt old,new,V} are shared among the threads, while {\tt mass} is copied from the master thread at the bifurcation point.
In a serial code the variables {\tt t\_min}, {\tt t\_plu} would be computed in the first line within the $t$-loop.
The clause {\tt COLLAPSE(2)} forces us to transfer these statements into the $z$-loop.
In the same line the clause {\tt SCHEDULE(static)} may be added to enforce compile-time thread scheduling.
In Fortran the notation {\tt site-=...} is wishful thinking; it is {\tt site=site-...} properly spelled out.
Within the SIMD loop the stride notation establishes an implicit {\tt forall(col=1:Nc) ...} construct, which is a one-line version of {\tt do concurrent(col=1:Nc); ...; end do}.
Taking everything together we thus have eight nested loops.
The line with dots only indicates that the other six directions are implemented analogously.
In Fortran this routine must be implemented separately for each layout {\tt NcNsNv}, {\tt NsNcNv}, {\tt NcNvNs}, {\tt NvNcNs}, {\tt NsNvNc}, {\tt NvNsNc},
and for {\tt old,new} being sp or dp.
Finally, a ``wrapper routine'' may be written to call them in a more convenient way.

The structure of the Wilson Dirac routine (\ref{def_wils}) for the vector layout {\tt NvNsNc} is the same, except minor differences.
The initialization of {\tt site} uses the factor {\tt (8.0+2.0*mass)}, since there is a factor $0.5$ in the end.
The most significant change is due to the ``Wilson projection trick'' discussed in App.~\ref{app:A}.
For the $\mp4$ directions the respective lines above are replaced by
\small
\begin{verbatim}
do ccc=1,Nc
   !$OMP SIMD PRIVATE(red,obj)
   do rhs=1,Nv
      red(1)=old(rhs,1,ccc,x,y,z,t_min)+     old(rhs,3,ccc,x,y,z,t_min)
      red(2)=old(rhs,2,ccc,x,y,z,t_min)+     old(rhs,4,ccc,x,y,z,t_min)
      red(3)=old(rhs,1,ccc,x,y,z,t_plu)-     old(rhs,3,ccc,x,y,z,t_plu)
      red(4)=old(rhs,2,ccc,x,y,z,t_plu)-     old(rhs,4,ccc,x,y,z,t_plu)
      forall(col=1:Nc,spi=1:2) obj(spi,col)=red(spi)*conjg(V(ccc,col,4,x,y,z,t_min))
      forall(col=1:Nc,spi=3:4) obj(spi,col)=red(spi)*      V(col,ccc,4,x,y,z,t    ) 
      site(rhs,1,:)=site(rhs,1,:)-     (obj(1,:)+obj(3,:))
      site(rhs,2,:)=site(rhs,2,:)-     (obj(2,:)+obj(4,:))
      site(rhs,3,:)=site(rhs,3,:)-     (obj(1,:)-obj(3,:))
      site(rhs,4,:)=site(rhs,4,:)-     (obj(2,:)-obj(4,:))
   end do
end do
\end{verbatim}
\normalsize
and similarly for the other directions.
The blank spaces mark positions where factors {\tt i\_sp*} would show up for some of the other directions
(with details depending on the choice of the Dirac matrices, see App.~\ref{app:A}).
Beyond this there is no change.

Overall, these implementations are compellingly simple, and it is perhaps a little surprising to the expert
that such simple routines can yield the timings reported in Sec.~\ref{sec:wils}.


\section{Details of the Brillouin Laplace and Dirac routines\label{app:C}}


The structure of the Brillouin Laplace routine (\ref{def_blap}) for the vector layout {\tt NvNsNc} is
\small
\begin{verbatim}
!$OMP PARALLEL DO COLLAPSE(3) DEFAULT(private) FIRSTPRIVATE(mass) SHARED(old,new,W)
do t=1,Nt
do z=1,Nz
do y=1,Ny
do x=1,Nx
   site(:,:,:)=cmplx(0.0,kind=sp)
   !!! add laplacian parts to site
   dir=0
   do go_t=-1,1; tsh=modulo(t+go_t-1,Nt)+1
   do go_z=-1,1; zsh=modulo(z+go_z-1,Nz)+1
   do go_y=-1,1; ysh=modulo(y+go_y-1,Ny)+1
   do go_x=-1,1; xsh=modulo(x+go_x-1,Nx)+1
      dir=dir+1 !!! note: dir=(go_t+1)*27+(go_z+1)*9+(go_y+1)*3+(go_x+2)
      if (dir<41) then
         tmp(:,:)=mask_lap(   dir)*W(:,:,dir,x,y,z,t)
      else if (dir>41) then
         tmp(:,:)=mask_lap(82-dir)*conjg(transpose(W(:,:,82-dir,xsh,ysh,zsh,tsh)))
      else
         tmp(:,:)=(1.875+0.5*mass**2)*eye(:,:)
      end if
      do ccc=1,Nc
         !$OMP SIMD PRIVATE(help)
         do rhs=1,Nv
            help(:)=old(rhs,:,ccc,xsh,ysh,zsh,tsh)
            do col=1,Nc
               site(rhs,:,col)=site(rhs,:,col)-help(:)*tmp(col,ccc)
            end do
         end do
      end do
   end do ! go_x=-1,1
   end do ! go_y=-1,1
   end do ! go_z=-1,1
   end do ! go_t=-1,1
   !!! plug site into new vector
   forall(col=1:Nc,spi=1:04,rhs=1:Nv) new(rhs,spi,col,x,y,z,t)=site(rhs,spi,col)
end do ! x=1,Nx
end do ! y=1,Ny
end do ! z=1,Nz
end do ! t=1,Nt
!$OMP END PARALLEL DO
\end{verbatim}
\normalsize
where the variables {\tt old,new,W} are shared among the threads, while {\tt mass} is copied from the master thread at the bifurcation point.
The clause {\tt SCHEDULE(dynamic)} may be added to this line to enforce run-time thread scheduling.
The four nested {\tt go} loops organize the harvesting within the $3^4$ hypercube around the point $n=(x,y,z,t)$ of $\ps_\mr{new}(n)$.
The factors $(\frac{1}{2}\la_1,\frac{1}{2}\la_2,\frac{1}{2}\la_3,\frac{1}{2}\la_4)$ are stored in {\tt mask\_lap(1:40)};
this is faster than evaluating {\tt count([go\_x,go\_y,go\_z,go\_t].ne.0)} and accessing a table with just four elements.
The factor $\frac{1}{2}\la_0+\frac{1}{2}m_0^2$ multiplies the $\Nc\times\Nc$ identity matrix stored in {\tt eye(1:Nc,1:Nc)}.
Within the SIMD loop there is an explicit color-loop, and the stride-notation establishes an implicit {\tt forall(spi=1:04)} construct.
Taking everything together we thus have twelve nested loops.

The structure of the Brillouin Dirac routine (\ref{def_bril}) for the vector layout {\tt NvNsNc} is the same, except minor differences.
The factor {\tt mask\_lap(min(dir,82-dir))} is not included in {\tt tmp} but stored in the variable {\tt lap}.
The variable {\tt absgo\_xyzt} is loaded from an array {\tt mask\_absgo(1:40)}, since this is faster than evaluating {\tt count([go\_x,go\_y,go\_z,go\_t].ne.0)}.
In fact, with this variable in hand, one can take {\tt lap} from an array {\tt mask\_lap(0:4)}.
Similarly, the derivatives are put together from an array {\tt mask\_der(-1:1,0:4)}.
Hence, the {\tt ccc} loop above is replaced by
\small
\begin{verbatim}
lap=mask_lap(absgo_xyzt)
der_tz=cmplx(mask_der(go_t,absgo_xyzt),mask_der(go_z,absgo_xyzt),kind=sp)
der_yx=cmplx(mask_der(go_y,absgo_xyzt),mask_der(go_x,absgo_xyzt),kind=sp)
do ccc=1,Nc
   !$OMP SIMD PRIVATE(myold,help)
   do rhs=1,Nv
      myold(:)=old(rhs,:,ccc,xsh,ysh,zsh,tsh)
      help(1)=-lap*myold(1)+conjg(der_tz)*myold(3)-      der_yx *myold(4)
      help(2)=-lap*myold(2)+conjg(der_yx)*myold(3)+      der_tz *myold(4)
      help(3)=-lap*myold(3)+      der_tz *myold(1)+      der_yx *myold(2)
      help(4)=-lap*myold(4)-conjg(der_yx)*myold(1)+conjg(der_tz)*myold(2)
      do col=1,Nc
         site(rhs,:,col)=site(rhs,:,col)+help(:)*tmp(col,ccc)
      end do
   end do
end do
\end{verbatim}
\normalsize
and taking everything together, we end up with a set of twelve nested loops.

Overall, these implementations are fairly straightforward.
The timings of these routines, for $\Nc=3$ and $\Nv=12$ on a $34^3\times68$ lattice, are discussed in Sec.~\ref{sec:bril}.


\section{Details of the Susskind ``staggered'' Dirac routine\label{app:D}}


The structure of the staggered Dirac routine for the layout {\tt NvNc} is
\small
\begin{verbatim}
!$OMP PARALLEL DO DEFAULT(private) FIRSTPRIVATE(mass) SHARED(old,new,V)
do t=1,Nt; t_min=...; t_plu=...; eta4=1.000
do z=1,Nz; z_min=...; z_plu=...; eta4=-eta4; eta3=1.000
do y=1,Ny; y_min=...; y_plu=...; eta4=-eta4; eta3=-eta3; eta2=1.000
do x=1,Nx; x_min=...; x_plu=...; eta4=-eta4; eta3=-eta3; eta2=-eta2
   !!! direction 0 gets mass term
   forall(col=1:Nc,rhs=1:Nv) site(rhs,col)=(2.0*mass)*old(rhs,col,x,y,z,t)
   !!! add non-trivial directions
   do ccc=1,Nc
      !$OMP SIMD
      do rhs=1,Nv
         site(rhs,:)-=eta4*conjg(V(ccc,:,4,x,y,z,t_min))*old(rhs,ccc,x,y,z,t_min)
         site(rhs,:)+=eta4*      V(:,ccc,4,x,y,z,t    ) *old(rhs,ccc,x,y,z,t_plu)
      end do
   end do
   ...
   !!! plug site into new vector
   forall(col=1:Nc,rhs=1:Nv) new(rhs,col,x,y,z,t)=0.5_sp*site(rhs,col)
end do ! x=1,Nx
end do ! y=1,Ny
end do ! z=1,Nz
end do ! t=1,Nt
!$OMP END PARALLEL DO
\end{verbatim}
\normalsize
where the variables {\tt old,new,V} are shared among the threads, while {\tt mass} is copied from the master thread at the bifurcation point.
In the actual code there is the clause {\tt COLLAPSE(3)} in the {\tt PARALLEL DO} construct, and this delays the initialization of {\tt eta3,eta4} by one or two loops, respectively.
In the same line the clause {\tt SCHEDULE(static)} is added to enforce compile-time thread scheduling.
The line with dots only indicates that the other six directions are implemented analogously.
In the end, the variable {\tt site(1:Nv,1:Nc)} is written, with a factor $\frac{1}{2}$, into the respective field of {\tt new}.
Within the SIMD loop the stride notation establishes an implicit {\tt forall(col=1:Nc)} construct.
In total there are thus seven nested loops.


\section{Flop count and memory traffic details\label{app:E}}


It is common practice to count the number of additions and multiplications, coined ``floating-point operations'' (flop).
In view of the capabilities of modern processors it would make sense to count fused-multiply-add operations, but for backward compatibility everyone adopts the traditional counting rule.
Hence a complex-plus-complex addition takes $2$ flops, a real-times-complex multiplication takes $2$ flops, and a complex-times-complex multiplication takes $6$ flops.
Accordingly, a $SU(\Nc)$ left-multiplication of a $\Nc\times4$ Dirac spinor takes $\Nc^2\times4$ multiplications and $\Nc(\Nc-1)\times4$ additions, which amounts to $[8\Nc^2-2\Nc]\times4$ flops.
Based on this we arrive at the following flop counts for the operators considered in this article.


\subsection{Wilson Laplace operator\label{app:wlap}}

A matrix-vector operation with the hpd operator $-\frac{1}{2}\lap^\mr{std}+\frac{1}{2}m_0^2$ proceeds as follows:
\begin{enumerate}
\itemsep-1pt
\item[(i)]
$SU(\Nc)$-multiply the $\Nc\times4$ block for each direction.
In this step we assume the prefactor $-\frac{1}{2}$ is included into the local copy of $V_\mu(n)$ and thus consider it for free.
Overall, this takes $[8\Nc^2-2\Nc]\cdot4\cdot8$ flops.
\item[(ii)]
Accumulate these 8 terms, as well as the $0$-hop contribution which is multiplied with the precomputed factor $(4+\frac{1}{2}m_0^2)$.
Overall, this takes $2\Nc\cdot4\cdot9$ flops.
\end{enumerate}
All together we have a grand total of $256\Nc^2+8\Nc$ flops per site, or $2328$ flops for $SU(3)$.

Here we assume the vector has the same spinor$\otimes$color structure per lattice site as for the Wilson Dirac operator (see~\ref{app:wils}).
In the event the vector has no spinor degree of freedom, the flop count is four-fold reduced, i.e.\ $64\Nc^2+2\Nc$ flops per site, or $582$ for $SU(3)$.

In addition, the following memory access operations are due (per site, for one RHS):
\begin{enumerate}
\itemsep-1pt
\item[(i)]
Read one color-spinor object for each point of the $9$-point stencil from the ``in'' vector; this requires $2\Nc\cdot4\cdot9$ floats (doubles) in sp (dp).
\item[(ii)]
Read one $SU(\Nc)$ matrix%
\footnote{Based on unitarity, one could omit the last row or column, and reconstruct it ``on the fly''; this reduces the load to $2\Nc(\Nc-1)\cdot8$ floats.
This does usually pay off for $\Nc=3$ but barely so for higher $\Nc$, and this is why no gauge compression is used in this code.
Using the matrix ``exp'' and ``log'' functions, one could even reduce this number to $(\Nc^2-1)\cdot8$ floats, but the latter function is tricky to implement (for arbitrary $\Nc$).}
(in this code suite always in sp, cf.\ the discussion in Sec.~\ref{sec:guidelines}) per direction; this requires $2\Nc^2\cdot8$ floats.
\item[(iii)]
Write one color-spinor object to the ``out'' vector; this requires $2\Nc\cdot4$ floats (doubles) in sp (dp).
\end{enumerate}
With several RHS, the memory footprint in (i) and (iii) is multiplied by $\Nv$, while (ii) stays unchanged.
All together we thus arrive at $80\Nc\Nv+16\Nc^2$ floats of traffic per site in sp (the number in dp follows by replacing $\Nv\to2\Nv$).
For $\Nc=3,\Nv=1$ the $2328$ flops and $384$ floats per site amount to a computational intensity of $1.52$ flops/byte in sp.
For $\Nc=3,\Nv=12$ the $27936$ flops and $3024$ floats per site yield $2.31$ flops/byte in sp.
Increasing $\Nv$ increases the computational intensity, but there is an asymptotic bound of $(256\Nc^2+8\Nc)/(320\Nc)$ flops/byte in sp for $\Nv\to\infty$, which evaluates to $2.42$ flops/byte for $\Nc=3$.


\subsection{Wilson Dirac operator\label{app:wils}}

A matrix-vector operation with the Wilson Dirac operator $\DW$ proceeds as follows:
\begin{enumerate}
\itemsep-1pt
\item[(i)]
Spin project (from 4 to 2 components) the $\Nc\times4$ matrix for each direction.
Overall, this takes $\Nc\cdot4\cdot8$ flops.
\item[(ii)]
$SU(\Nc)$-multiply the $\Nc\times2$ block for each direction, and expand it back to the $\Nc\times 4$ format (for free).
Overall, this takes $[8\Nc^2-2\Nc]\cdot2\cdot8$ flops.
\item[(iii)]
Accumulate these 8 terms, as well as the $0$-hop contribution which uses the precomputed factor $(4+m_0)$.
Overall, this takes $2\Nc\cdot4\cdot9$ flops.
\end{enumerate}
All together we have a grand total of $128\Nc^2+72\Nc$ flops per site, or $1368$ flops for $SU(3)$.

In the older literature a different normalization of the Dirac operator was used, where the factor $(4+m_0)$ is absent.
In this case only $2\Nc\cdot4\cdot8$ flops are spent in step (iii), and the grand total amounts to $128\Nc^2+64\Nc$ flops, or $1344$ flops for $SU(3)$.
Sometimes the Dirac operator \emph{without} the $0$-hop contribution is considered.
In this case step (iii) uses only $2\Nc\cdot4\cdot7$ flops, and the grand total amounts to $128\Nc^2+56\Nc$ flops, or $1320$ flops for $SU(3)$.

In addition, for a matrix-vector operation we must perform the same memory operations as for the Wilson Laplace operator, see~\ref{app:wlap}.
This was $80\Nc\Nv+16\Nc^2$ floats of traffic per site in sp (in dp the first term doubles).
For $\Nc=3,\Nv=1$ the $1368$ flops and $384$ floats per site amount to a computational intensity of $0.89$ flops/byte in sp.
For $\Nc=3,\Nv=12$ the $16416$ flops and $3024$ floats per site yield $1.36$ flops/byte in sp.
Again increasing $\Nv$ increases the computational intensity, but there is an asymptotic bound of $(128\Nc^2+72\Nc)/(320\Nc)$ flops/byte in sp for $\Nv\to\infty$, which evaluates to $1.53$ flops/byte for $\Nc=3$.


\subsection{Brillouin Laplace operator\label{app:blap}}

A matrix-vector operation with the hpd operator $-\frac{1}{2}\lap^\mr{bri}+\frac{1}{2}m_0^2$ proceeds as follows:
\begin{enumerate}
\itemsep-1pt
\item[(i)]
$SU(\Nc)$-multiply the $\Nc\times4$ block for each of the $80$ non-trivial directions.
Overall, this takes $[8\Nc^2-2\Nc]\cdot4\cdot80$ flops.
\item[(ii)]
Multiply the resulting $\Nc\times4$ matrix for $81$ directions with the weight factor as given by the Brillouin Laplacian.
This weight factor is real, and the mass term may be incorporated into the $0$-hop contribution of the Laplacian.
Overall, this takes $\Nc\cdot4\cdot2\cdot81$ flops.
\item[(iii)]
Accumulate the 81 contributions to the out-spinor.
Overall, this takes $2\Nc\cdot4\cdot80$ flops.
\end{enumerate}
All together we have a grand total of $2560\Nc^2+648\Nc$ flops per site, or $24984$ flops for $SU(3)$.

Here we assume the vector has the same spinor$\otimes$color structure per lattice site as for the Brillouin Dirac operator (see~\ref{app:bril}).
In the event the vector has no spinor degree of freedom, the flop count is four-fold reduced, i.e.\ $640\Nc^2+162\Nc$ flops per site, or $6246$ for $SU(3)$.

In addition, for a matrix-vector operation we must (per site, for one RHS):
\begin{enumerate}
\itemsep-1pt
\item[(i)]
Read one color-spinor object for each point of the $81$-point stencil from the ``in'' vector; this requires $2\Nc\cdot4\cdot81$ floats (doubles) in sp (dp).
\item[(ii)]
Read one complex $\Nc\times\Nc$ matrix%
\footnote{For $36$ of the $40$ stored directions this matrix is \emph{not unitary}; compare the caption of Tab.~\ref{tab:off-axis}.}
(in this code suite always in sp, cf.\ the discussion in Sec.~\ref{sec:guidelines}) per non-trivial direction; this requires $2\cdot\Nc^2\cdot80$ floats.
\item[(iii)]
Write one color-spinor object to the ``out'' vector; this requires $2\Nc\cdot4$ floats (doubles) in sp (dp).
\end{enumerate}
With several RHS, the memory footprint in (i) and (iii) is multiplied by $\Nv$, while (ii) stays unchanged.
All together we thus arrive at $656\Nc\Nv+160\Nc^2$ floats of traffic per site in sp (the number in dp follows by replacing $\Nv\to2\Nv$).
For $\Nc=3,\Nv=1$ the $24984$ flops and $3408$ floats per site amount to a computational intensity of $1.83$ flops/byte in sp.
For $\Nc=3,\Nv=12$ the $299808$ flops and $25056$ floats per site yield $2.99$ flops/byte in sp.
Again increasing $\Nv$ increases the computational intensity, but there is an asymptotic bound of $(2560\Nc^2+648\Nc)/(2624\Nc)$ flops/byte in sp for $\Nv\to\infty$, which evaluates to $3.17$ flops/byte for $\Nc=3$.


\subsection{Brillouin Dirac operator\label{app:bril}}

A matrix-vector operation with the Brillouin Dirac operator $\DB$ proceeds as follows:
\begin{enumerate}
\itemsep-1pt
\item[(i)]
$SU(\Nc)$-multiply the $\Nc\times4$ block for each of the $80$ non-trivial directions.
Overall, this takes $[8\Nc^2-2\Nc]\cdot4\cdot80$ flops.
\item[(ii)]
Multiply the resulting $\Nc\times4$ matrix with the weight factors as given by the isotropic derivatives and the Brillouin Laplacian.
These weight factors are either real or purely imaginary, and for each $\nab_\mu^\mr{iso}$ non-zero only for $54$ out of the $81$ directions.
The mass term may be incorporated into the $0$-hop contribution of the Laplacian.
Overall, this takes $2\Nc\cdot4\cdot(4\cdot54+81)$ flops.
\item[(iii)]
Accumulate the 81 contributions to the out-spinor.
Overall, this takes $2\Nc\cdot4\cdot80$ flops.
\end{enumerate}
All together we have a grand total of $2560\Nc^2+2376\Nc$ flops per site, or $30168$ flops for $SU(3)$.

In addition, for a matrix-vector operation we must perform the same memory operations as for the Brillouin Laplace operator, see~\ref{app:blap}.
This was $656\Nc\Nv+160\Nc^2$ floats of traffic per site in sp (in dp the first term doubles).
For $\Nc=3,\Nv=1$ the $30168$ flops and $3408$ floats per site amount to a computational intensity of $2.21$ flops/byte in sp.
For $\Nc=3,\Nv=12$ the $362016$ flops and $25056$ floats per site yield $3.61$ flops/byte in sp.
Again increasing $\Nv$ increases the computational intensity, but there is an asymptotic bound of $(2560\Nc^2+2376\Nc)/(2624\Nc)$ flops/byte in sp for $\Nv\to\infty$, which evaluates to $3.83$ flops/byte for $\Nc=3$.


\subsection{Susskind ``staggered'' Dirac operator\label{app:stag}}

In terms of the flop count a matrix-vector operation with the ``staggered'' Dirac operator $\DS$ mimics an application of the Wilson Laplacian on a scalar argument (see~\ref{app:wlap}).
The staggering brings factors of $-1$, which are considered for free, and the center-point of the stencil is multiplied by $m_0$ rather than by $4+\frac{1}{2}m_0^2$, but this does not affect the flop count either.
The final result is thus still $64\Nc^2+2\Nc$ flops per site, or $582$ flops for $SU(3)$.

Also the memory requirement is a quarter of the number given in~\ref{app:wlap}, i.e.\ $20\Nc\Nv+4\Nc^2$ floats of traffic per site in sp (the first term doubles in dp) with $\Nv$ RHS.
The computational intensity is thus $(64\Nc^2\Nv+2\Nc\Nv)/(80\Nc\Nv+16\Nc^2)$ flops/byte in sp.
For $\Nc=3,\Nv=1$ this yields $(64\Nc+2)/(80+16\Nc)\simeq1.52$ flops/byte in sp.
For $\Nc=3,\Nv=12$ this yields $(768\Nc+24)/(960+16\Nc)\simeq2.31$ flops/byte in sp.
The asymptotic bound for $\Nv\to\infty$ is again $(64\Nc^2+2\Nc)/(80\Nc)$ flops/byte in sp, which evaluates to $2.42$ flops/byte for $\Nc=3$.


\subsection{Clover improvement operator\label{app:clov}}

A matrix-vector operation with (\ref{def_clov}, \ref{options_clov}), using the precomputed $F_{\mu\nu}(n)$, proceeds as follows:
\begin{enumerate}
\itemsep-1pt
\item[(i)]
$SU(\Nc)$-multiply the $\Nc\times4$ block for any of the $6$ clover orientations (specified by $1\leq\mu<\nu\leq4$) at the given space-time position.
Overall, this takes $[8\Nc^2-2\Nc]\cdot4\cdot6$ flops.
\item[(ii)]
Add the $6$ contributions per color and spinor index to the out-vector
(here we use that each $\si_{\mu\nu}$ has one non-zero entry in spinor space per row or column; factors of $\ri$ are considered for free).
Overall, this takes $2\Nc\cdot4\cdot6$ flops.
\end{enumerate}
All together we have a grand total of $192\Nc^2$ flops per site, or $1728$ flops for $SU(3)$.

In the interest of speed the special properties of the $\si_{\mu\nu}$ matrices can be exploited (see App.~\ref{app:A}).
Ignoring the numerical effort to form the linear combinations of the $F_{\mu\nu}$ matrices (justified for large enough $\Nc$), the effective number of $SU(\Nc)$-multiplications is reduced to two.
In this case one ends up with $[8\Nc^2-2\Nc]\cdot4\cdot2$ flops under (i), and $2\Nc\cdot4\cdot2$ flops under (ii).
All together this yields a grand total of $64\Nc^2$ flops per site, or $576$ flops for $SU(3)$.
The ancillary code distribution uses this trick, and hence the smaller number when a timing information is converted to a Gflop/s figure for the clover routine or for $\DW,\DB$ at $c_\mr{SW}>0$.


\section{Overview of ancillary Fortran\hspace{3pt}2008 code distribution\label{app:F}}


The ancillary code distribution contains a makefile, as well as the source files {\tt testknl\_$\{$main, globals,util,cond,vec,suv,clov,wlap,wils,blap,bril,wsuv,bsuv,stag$\}$.f90}.

The source files are written in compliance with the Fortran\,2008 standard, with preprocessor macros (with a hash \# in the first column).
The makefile is designed to convert each source file into an object file by issuing either the command
\begin{verbatim}
ifort -implicitnone -stand f08 -fpp -warn all -nogen-interfaces -qopenmp
      -O2 -xmic-avx512  -c testknl_xxx.f90
\end{verbatim}
or (the choice is made by commenting/uncommenting the respective lines in the makefile)
\begin{verbatim}
gfortran -fimplicit-none -std=f2008 -cpp -ffree-line-length-none -Wall -Wextra
         -fall-intrinsics -fopenmp -Ofast -mcmodel=medium -c testknl_xxx.f90
\end{verbatim}
where {\tt testknl\_xxx.f90} refers to any of the files above, and to link them into a single executable named {\tt testknl\_main}.
The architecture flag {\tt-xmic-avx512} can be replaced by another one via a simple comment/uncomment operation in the makefile.

To obtain good performance figures it is important to set the right environment variables.
On KNL the shell is requested via {\tt srun --cpu-bind=threads}.
On either architecture the variables {\tt OMP\_PLACES=threads OMP\_PROC\_BIND=true OMP\_DYNAMIC=false} are exported;
it is advisable to \emph{not} export {\tt KMP\_AFFINITY}, as this would overwrite these settings.

The modules listed below include functions, subroutines and constants from other modules via the {\tt use} command,
and state (in their headers) which of its routines are ``public'' (i.e.\ callable form other modules) or ``private'' (not callable).
For simplicity the main program includes all publicly available routines from all modules.

\begin{table}[tb]
\centering
\begin{tabular}{|l|llll|}
\hline
testknl\_util.o:   & testknl\_globals.o &                 &                 &                 \\
testknl\_cond.o:   & testknl\_globals.o &                 &                 &                 \\
testknl\_vec.o:    & testknl\_globals.o & testknl\_util.o &                 &                 \\
testknl\_suv.o:    & testknl\_globals.o & testknl\_util.o &                 &                 \\
testknl\_clov.o:   & testknl\_globals.o & testknl\_vec.o  & testknl\_util.o &                 \\
testknl\_wlap.o:   & testknl\_globals.o & testknl\_vec.o  & testknl\_cond.o &                 \\
testknl\_wils.o:   & testknl\_globals.o & testknl\_vec.o  & testknl\_cond.o & testknl\_clov.o \\
testknl\_blap.o:   & testknl\_globals.o & testknl\_vec.o  & testknl\_cond.o &                 \\
testknl\_bril.o:   & testknl\_globals.o & testknl\_vec.o  & testknl\_cond.o & testknl\_clov.o \\
testknl\_wsuv.o:   & testknl\_globals.o & testknl\_suv.o  & testknl\_cond.o &                 \\
testknl\_bsuv.o:   & testknl\_globals.o & testknl\_suv.o  & testknl\_cond.o &                 \\
testknl\_stag.o:   & testknl\_globals.o & testknl\_suv.o  & testknl\_cond.o &                 \\
testknl\_main.o:   & \multicolumn{4}{l|}{testknl\_globals.o testknl\_util.o testknl\_cond.o testknl\_vec.o testknl\_suv.o} \\
                   & \multicolumn{4}{l|}{testknl\_clov.o testknl\_wlap.o testknl\_wils.o testknl\_blap.o testknl\_bril.o} \\
                   & \multicolumn{4}{l|}{testknl\_wsuv.o testknl\_bsuv.o testknl\_stag.o} \\
\hline
\end{tabular}
\caption{\label{tab:dependencies}
Dependencies of the modules as stated by the {\tt makefile}; the file {\tt testknl\_globals} does not import anything, the file {\tt testknl\_main} does not export anything.}
\end{table}

The main purpose of the makefile is to shorten compilation times during the code development process
by properly reflecting the tree of dependencies as listed in Tab.~\ref{tab:dependencies}.
We finish with a brief description of the services provided by each file.

\subsubsection*{$\bullet$ testknl\_globals}

Defines the publicly accessible \emph{parameters} (Fortran variables which cannot be changed in subsequent code portions) $\Nc,\Nv$,
the box dimensions $N_x,N_y,N_z,N_t$, the kind parameters {\tt sp}, {\tt dp} (4\,bytes, 8\,bytes for {\tt integer}, {\tt real}, or Re/Im of {\tt complex} variables),
and the imaginary unit {\tt i\_sp=cmplx(0.0\_sp,1.0\_sp,kind=sp)}, {\tt i\_dp=cmplx(0.0\_dp,1.0\_dp,kind=dp)} in single/double precision, respectively.

The routine {\tt xneighbors} provides information about the box geometry in the $x$-direction.
It returns, for a given {\tt x}, the values {\tt [x\_min,x\_plu]=modulo([x-2,x],Nx)+1}.
Alternatively, {\tt integer,parameter,dimension(0:Nx+1)\,::\,list\_x=[Nx,(m,m=1,Nx),1]} defines an array of length $N_x+2$ whose $m$-th slot contains the
coordinate $m\in\{1,\ldots,N_x\}$, while the $0$-th slot carries $N_x$, and the $N_x+1$-th (last) slot carries $1$.
The $y,z,t$ directions are handled analogously.

Finally, it contains the two \emph{global} variables {\tt N\_cpu}, {\tt N\_thr} which will carry the number of physical cores per socket and the number of threads
to be used in a {\tt PARALLEL DO} construct.
Though representing a slight lapse in style, this is incredibly convenient for handling the number of active OpenMP threads from within the code.

\subsubsection*{$\bullet$ testknl\_util}

Contains various service functions for the multiplication of up to four $SU(\Nc)$ matrices;
for instance {\tt mult\_oodd(A,B,C,D)} returns the product $ABC\dag D\dag$.
In addition, {\tt color\_linsolve(A,B)} returns the matrix $X$ which solves $AX=B$,
{\tt color\_eye()} the identity, and
{\tt color\_expm(A)} the matrix exponential of $A$ (all for arbitrary $\Nc$).
The functions {\tt color\_ncminrealtrace(M)} and {\tt color\_imagtrace(M)} return $\Nc-\mr{Re\,Tr}(M)$ and $\mr{Im\,Tr}(M)$,
respectively, where $\mr{Tr}$ takes the trace in color space.
Finally, {\tt color\_swil(V)} returns the average Wilson action $1-\mr{Re\,Tr}(V)/\Nc$ for a gauge field $V$ (separately for spatial and temporal plaquettes).
All functions mentioned so far are declared {\tt pure}, i.e.\ they can be called from within a thread-parallel region.

The subroutine {\tt color\_stoutsmear(V,rho\_stout,n\_smear)} applies $n$ steps of stout-smear\-ing with parameter $\rh$ \cite{Morningstar:2003gk}
to the possibly smeared gauge field $V$, and returns the result in the same variable.
The subroutine {\tt color\_vtow(V,W)} constructs the gauge field $W_\mr{dir}(x,y,z,t)$ from the possibly smeared field $V_\mu(x,y,z,t)$ as defined in Tab.~\ref{tab:off-axis}.
The $81$ directions are numbered via the set of nested loops {\tt do concurrent(go\_t=-1:1,go\_z=-1:1,go\_y=-1:1,go\_x=-1:1); ... ; end do},
but only the first $40$ of them are stored in $W_\mr{dir}(x,y,z,t)$, since $W_{41}(x,y,z,t)$ is the identity,
and the remaining ones can be reconstructed by daggering the entry in a nearby site (see the discussion around Tab.\,\ref{tab:off-axis}).
The reverse operation {\tt color\_wtov(W,V)} just selects the four on-axis directions.
The routines {\tt color\_vtof(V,F)} and {\tt color\_wtof(W,F)} construct (from $V$ or $W$) the field strength {\tt F(:,:,munu,x,y,z,t)},
as used in the definition of the clover term, where {\tt munu} ranges from one to six; see the discussion after Eq.~(\ref{options_clov}).

There are service routines to initialize random variables in color/spinor/RHS space with a ``memory function''.
The purpose of the latter is to make sure that a vector can be initialized, for several orderings of the color/spinor/RHS slots, with the same ``physics content''.
And there are routines to initialize gauge fields%
\footnote{$W$ is always defined via $V$, and the latter may or may not coincide with $U$, depending on $(\rh,n)$ passed to the stout-smearing routine.
The user may allow the code to check whether there is, in the directory {\tt confs}, a file with a name matching the current values of $\Nc,N_x,N_y,N_z,N_t$ and $\beta$.
In such an event its content is read in and used, assuming that the $2\cdot\Nc^2\cdot4\cdot N_x N_y N_z N_t$ little-endian floats in the file are written
in exactly the order the complex Fortran variable {\tt U(1:Nc,1:Nc,1:4,1:Nx,1:Ny,1:Nz,1:Nt)} would expect.},
Dirac vectors {\tt vec\_sp, vec\_dp}, and Susskind vectors {\tt suv\_sp, suv\_dp}.
Finally, the routines {\tt gflops\_$\{$clov,wlap,wils,blap,bril,stag$\}$} convert the best measured time, $t_\mr{min}$, into a performance in Gflop/s,
according to the flop counts listed in App.~\ref{app:E}.

\subsubsection*{$\bullet$ testknl\_cond}

Contains subroutines to determine $\la_\mr{max},\la_\mr{min}$ and $\mr{cond}(A)=\la_\mr{max}/\la_\mr{min}$ of a HPD operator $A$.
For each RHS index, the expected input is the diagonal $(a_1,\ldots,a_n)$ and the first super/sub-diagonal $(b_1,\ldots,b_{n-1})$ of the pertinent (symmetric) Lanczos matrix.
Also, a routine is provided to convert the $(\al_1,\ldots,\al_n)$ and $(\be_1,\ldots,\be_{n-1})$ that occur in the CG algorithm into the Lanczos $(a_1,\ldots,a_n)$ and $(b_1,\ldots,b_{n-1})$.
After the job is completed for each $i\in\{1,\ldots,\Nv\}$ the spread among the $\Nv$ values $\la_\mr{max}^{(i)}$ provides an estimate for the uncertainty of $\la_\mr{max}$.
A similar statement holds for $\la_\mr{min}$.

\subsubsection*{$\bullet$ testknl\_vec}

Contains service routines to perform basic operations with the $\Nv$-RHS Dirac vectors (with $\Nc$ color and $4$ spinor degrees of freedom per lattice site),
for the six possible orderings of the ``internal'' degrees of freedom.
Accordingly, the suffix {\tt str\_layout} may take the values {\tt NcNsNv}, {\tt NsNcNv}, {\tt NcNvNs}, {\tt NvNcNs}, {\tt NsNvNc}, {\tt NvNsNc}.
The subroutine {\tt vec\_shapecheck\_$\{$sp,dp$\}$} verifies the object {\tt vec} (in sp or dp) has dimensions consistent%
\footnote{Note it is not always possible to let Fortran ``detect'' the layout from the vector shape alone,
since some of the values in the set $\{\Nc,N_s\equiv4,\Nv\}$ may numerically coincide.}
with {\tt str\_layout}.

The subroutine {\tt vec\_zero\_$\{$sp,dp$\}$} sets the vector to zero, {\tt vec\_scale\_r$\{$sp,dp$\}$} multiplies it with a real number which depends on the RHS index.
The subroutines {\tt vec\_copy\_$\{$sp,dp$\}$}, {\tt vec\_cast\_sp2dp}, {\tt vec\_cast\_dp2sp} copy one argument to the other one with any of the four possible precision options.
The subroutine {\tt vec\_normsqu\_$\{$sp,dp$\}$} calculates the squared two-norm $||v||_2^2$ (set up as a driver routine which calls a dedicated implementation for each layout).
Similarly {\tt vec\_dotdiag\_$\{$sp,vp,dp$\}$} calculates the $\Nv$ dot-products $u^{(i)\dagger}v^{(i)}$ with $i$ the RHS index (specialty: {\tt vp} takes one argument in {\tt dp} and one in {\tt sp}).
The subroutine {\tt vec\_incr\_r$\{$sp,vp,dp$\}$} performs the operation (\ref{def_incr}) for real $\al$ and various sp and dp options of $u$ and $v$.
Similarly {\tt vec\_anti\_r$\{$sp,vp,dp$\}$} performs the operation (\ref{def_anti}) for real $\al$ and various sp and dp options of $u$ and $v$.
For both ``incr'' and ``anti'' there is a sibling routine where the ``r'' is replaced by ``c'' and $\al$ is complex rather than real.
The subroutine {\tt app\_gamma5\_$\{$sp,dp$\}$} applies the matrix $\ga_5$ in spinor space to the argument $v$ which is supposed to be in sp or dp, respectively.

The subroutine {\tt vec\_pointsource\_$\{$sp,dp$\}$} initializes a multi-RHS vector with non-zero entries in $\Nv/(4\Nc)$ randomly chosen source positions.
For instance with $\Nv=48,\Nc=3$ four positions are chosen, and in each case twelve vectors are initialized (to cover all directions in color-spinor space).
The subroutine {\tt vec\_pseudoscalar\_$\{$sp,dp$\}$} performs spectroscopy with two degenerate quarks in the pseudoscalar channel.
It needs to be fed with an argument $u$ that solves $Du=v$, with $v$ initialized by the pointsource routine.
Finally, {\tt vec\_residualnormadjust\_$\{$sp,dp$\}$} is a service routine for the Krylov space solvers CG and BiCGstab which operator on Dirac vectors.

\subsubsection*{$\bullet$ testknl\_suv}

Contains service routines to perform basic operations with the $\Nv$-RHS Susskind vectors (with $\Nc$ color degrees of freedom per lattice site),
for the two possible orderings of the ``internal'' degrees of freedom.
Accordingly, the suffix {\tt str\_layout} may take the values {\tt \_NcNv\_}, {\tt \_NvNc\_} (to stay with a string of length six).
The subroutine {\tt suv\_shapecheck\_$\{$sp,dp$\}$} verifies the object {\tt suv} (in sp or dp) has dimensions consistent with {\tt str\_layout}.

The remaining subroutines in this module provide exactly the same functionality for {\tt suv}-arrays as the sibling routine described in the previous bullet point for {\tt vec}-arrays.
Only in the case of the routine {\tt app\_ga5xi5\_$\{$sp,dp$\}$}, which applies the diagonal site-parity%
\footnote{Note the difference to ordinary parity which maps $n=(x,y,z,t)$ into $(-x,-y,-z,t)$.}
matrix $\ep=(-1)^{x+y+z+t}$ to the input argument, this correspondence%
\footnote{It yields a left-multiplication with $\ga_5\otimes\xi_5$ in the spinor$\otimes$taste interpretation of the staggered formalism.}
is not-so-obvious.

\subsubsection*{$\bullet$ testknl\_clov}

The driver routine {\tt add\_clov\_$\{$sp,dp$\}$(old,new,F,csw,str\_layout)} applies the site-diagonal operator $C(m,n)$ as defined in (\ref{def_clov}) to the array {\tt old}, and increments the array {\tt new} by the result.
Which option in (\ref{options_clov}) is used depends on the content of the six-letter string {\tt str\_layout}, and it is checked that the layout of the vectors {\tt old} and {\tt new} complies with it.
The field-strength $F_{\mu\nu}(n)$, precomputed by the respective routine in \textbf{testknl\_util}, needs to be provided in {\tt F(Nc,Nc,6,Nx,Ny,Nz,Nt)} along with the real variable $c_\mr{SW}$.

For the layout {\tt NvNsNc} the routine {\tt add\_clov\_V\_$\{$sp,dp$\}$(old,new,V,csw)} does the same thing, but $F_{\mu\nu}(n)$ it built on the fly from the (possibly smeared) gauge field $V_\mu(n)$.
This option is not competitive on the architectures considered in this paper, but this may change soon.

\subsubsection*{$\bullet$ testknl\_wlap}

The driver routine {\tt app\_wlap\_$\{$sp,dp$\}$} applies the operator $A\equiv-\frac{1}{2}\lap^\mr{std}+\frac{1}{2}m_0^2$ to the argument vector, which is supplied along with the six-letter string {\tt str\_layout}.
Depending on the content of the latter, it calls one of the internal routines which implement the operation in an efficient manner for the respective layout.
The routine {\tt resultcheck\_wlap\_$\{$sp,dp$\}$} checks its input vector, which is supposed to be the output of {\tt app\_wlap\_$\{$sp,dp$\}$}, against a predefined%
\footnote{This is useful when optimizing a routine; as soon as one introduces an error, the mishap is recognized.\label{foot:resultcheck}}
vector.

The solver {\tt cg\_wlap\_sp(b,x,V,mass,str\_layout,tol,maxit,do\_precond)} solves the equation $Ax=b$ for $x$,
for a given multi-RHS vector $b$, with {\tt V} being the possibly smeared gauge field $V_\mu(n)$, and {\tt mass} being $m_0$.
The {\tt vec}-objects {\tt b,x} must be in {\tt sp} and in the layout as described in {\tt str\_layout}.
In addition, {\tt tol} specifies the relative norm $\ep=||Ax-b||/||b||$, {\tt maxit} the maximum number of iterations, and the boolean argument {\tt do\_precond} decides whether the operator is to be preconditioned%
\footnote{Currently, polynomial preconditioning is used, but the user should replace it by a dedicated preconditioner.}
or not.
The ``double precision'' solver {\tt cg\_wlap\_dp} solves the analogous task for vectors {\tt b,x} in {\tt dp}.
The ``mixed precision'' solver {\tt cg\_wlap\_mp}, again for {\tt b,x} in {\tt dp}, does the same job by internally calling {\tt cg\_wlap\_sp} several times.

\subsubsection*{$\bullet$ testknl\_wils}

The driver routine {\tt app\_wils\_$\{$sp,dp$\}$} applies the operator $\DW$ as specified in (\ref{def_wils}) to the argument vector, which is supplied along with a six-letter string {\tt str\_layout}.
Depending on its content an internal routine is called which implements the operation in an efficient way for the respective layout.
The routine {\tt resultcheck\_wils\_$\{$sp,dp$\}$} is useful in optimization work, see footnote \ref{foot:resultcheck}.

The solver {\tt cg\_wilsdagwils\_sp(b,x,V,F,mass,csw,str\_layout,tol,maxit,do\_precond)} solves the equation $Ax=b$ with $A\equiv \DW\dag \DW$ for $x$, with a given multi-RHS vector $b$.
The argument {\tt V} is the possibly smeared gauge field $V_\mu(x)$, and {\tt mass} is $m_0$ in $\DW$.
The remaining arguments have been described in the previous bullet point, except for {\tt F} which describes the field strength $F_{\mu\nu}(n)$ and {csw} which is $c_\mr{SW}$.
Similarly, {\tt cg\_wilsdagwils\_dp} and the ``mixed precision'' solver {\tt cg\_wilsdagwils\_mp} do the same job for {\tt b,x} in {\tt dp}.
Internally, there is a routine {\tt app\_wilsdagwils\_$\{$sp,dp$\}$} which applies $\DW\dag \DW=\ga_5 \DW \ga_5 \DW$ to the argument.
In {\tt cg\_wilsdagwils\_$\{$sp,dp$\}$} it is used in much the same way as {\tt app\_wlap\_$\{$sp,dp$\}$} in {\tt cg\_wlap\_$\{$sp,dp$\}$}, except for a technical difference.
In {\tt app\_wlap\_$\{$sp,dp$\}$} no internal vector is needed, while in {\tt app\_wilsdagwils\_$\{$sp,dp$\}$} an internal vector {\tt tmp\_$\{$sp,dp$\}$} is needed, if the in-vector is not to be overwritten.
In principle {\tt app\_wilsdagwils\_$\{$sp,dp$\}$} could allocate/deallocate it in every step.
One some architectures this is time-consuming, and this is why {\tt cg\_wilsdagwils\_$\{$sp,dp$\}$} allocates it once and provides it as a ``scratch area'' to {\tt app\_wilsdagwils\_$\{$sp,dp$\}$}.

The solver {\tt bicgstab\_wils\_sp(b,x,V,F,csw,mass,str\_layout,tol,maxit,do\_precond)} is designed to solve the equation $\DW x=b$ for $x$ without acting on both sides with $\DW\dag$,
with the multi-RHS vector $b$ and all other arguments as described above.
Similarly, {\tt bicgstab\_wils\_dp} and the ``mixed precision'' solver {\tt bicgstab\_wils\_mp} do the same job for {\tt b,x} in {\tt dp}.

\subsubsection*{$\bullet$ testknl\_blap}

The concepts are similar to the file {\tt testknl\_wlap}, except that the operator $A\equiv-\frac{1}{2}\lap^\mr{bri}+\frac{1}{2}m_0^2$ is applied.
Again {\tt app\_blap\_$\{$sp,dp$\}$} is a driver routine, i.e.\ the content of {\tt str\_layout} decides which of the layout-specific implementations is actually called.

The solver {\tt cg\_blap\_sp(b,x,W,mass,str\_layout,tol,maxit,do\_precond)} solves the equation $Ax=b$ for $x$, with a given multi-RHS vector $b$.
The argument {\tt W} contains the hypercubic links, which are linear combinations of chains of the possibly smeared gauge field $V_\mu(x)$, and {\tt mass} is $m_0$.
The {\tt vec}-objects {\tt b,x} must be in {\tt sp} and in the layout as described in {\tt str\_layout}.
Again, {\tt tol} specifies the relative norm $\ep=||Ax-b||/||b||$, while {\tt maxit} is the maximum number of iterations, and the boolean argument {\tt do\_precond} decides whether the operator is to be preconditioned or not.
The ``double precision'' solver {\tt cg\_blap\_dp} and the ``mixed precision'' solver {\tt cg\_blap\_mp} perform the same job for vectors {\tt b,x} in {\tt dp}.

\subsubsection*{$\bullet$ testknl\_bril}

The concepts are similar to the file {\tt testknl\_wils}, except that the Brillouin operator $\DB$ replaces the Wilson operator $\DW$.
The driver routine {\tt app\_bril\_$\{$sp,dp$\}$} applies $\DB$ as specified in (\ref{def_bril}) to the argument vector.
The six-letter string {\tt str\_layout} decides which of the layout-specific implementations of the matrix-times-vector operation is actually called.

The solver {\tt cg\_brildagbril\_sp(b,x,W,F,mass,csw,str\_layout,tol,maxit,do\_precond)} solves the equation $Ax=b$ with $A\equiv \DB\dag \DB$ for $x$, with a given multi-RHS vector $b$.
The argument {\tt W} contains the hypercubic links, {\tt F} the precomputed field-strength tensor, and the parameters $m_0,c_\mr{SW}$ follow.
The remaining arguments have been described in previous bullet points.
Similarly, {\tt cg\_brildagbril\_dp} and the ``mixed precision'' solver {\tt cg\_brildagbril\_mp} do the same job for {\tt b,x} in {\tt dp}.
Again, for reasons of expedience, an internal vector {\tt tmp\_$\{$sp,dp$\}$} is allocated and passed as ``scratch area'' to {\tt app\_brildagbril\_$\{$sp,dp$\}$},
to prevent it from allocating/deallocating such a vector in every iteration of the solver.

The solver {\tt bicgstab\_bril\_sp(b,x,V,F,csw,mass,str\_layout,tol,maxit,do\_precond)} is designed to solve the equation $\DB x=b$ for $x$ without acting on both sides with $\DB\dag$,
with the multi-RHS vector $b$ and all other arguments as described above.
Similarly, {\tt bicgstab\_bril\_dp} and the ``mixed precision'' solver {\tt bicgstab\_bril\_mp} do the same job for {\tt b,x} in {\tt dp}.

\subsubsection*{$\bullet$ testknl\_wsuv}

This module applies the operator $A\equiv-\frac{1}{2}\lap^\mr{std}+\frac{1}{2}m_0^2$ to Susskind vectors (color structure only, i.e.\ \emph{without} spinor structure), dubbed {\tt suv}.
Hence everything is in complete analogy to the description under the bullet point {\tt testknl\_wlap}, except that the vectors are a factor four shorter,
and {\tt str\_layout} must take one of the values {\tt \_NcNv\_} or {\tt \_NvNc\_}.

\subsubsection*{$\bullet$ testknl\_bsuv}

This module applies the operator $A\equiv-\frac{1}{2}\lap^\mr{bri}+\frac{1}{2}m_0^2$ to Susskind vectors (color structure only, i.e.\ \emph{without} spinor structure), dubbed {\tt suv}.
Hence everything is in complete analogy to the description under the bullet point {\tt testknl\_blap}, except that the vectors are a factor four shorter,
and {\tt str\_layout} must take one of the values {\tt \_NcNv\_} or {\tt \_NvNc\_}.

\subsubsection*{$\bullet$ testknl\_stag}

The driver routine {\tt app\_stag\_$\{$sp,dp$\}$} applies the operator $\DS$ as specified in (\ref{def_stag}) to the argument vector (of Susskind type), which is supplied along with the string {\tt str\_layout}.
Depending on the latter being {\tt \_NcNv\_} or {\tt \_NvNc\_}, an internal routine is called which implements the operation in an efficient way for the respective layout.
The routine {\tt resultcheck\_stag\_$\{$sp,dp$\}$} is useful in optimization work, see footnote \ref{foot:resultcheck}.

The solver {\tt cg\_stagdagstag\_sp(b,x,V,mass,str\_layout,tol,maxit,do\_precond)} solves the equation $Ax=b$ with $A\equiv \DS\dag \DS$ for $x$, with a given multi-RHS vector $b$.
The argument {\tt V} is the possibly smeared gauge field $V_\mu(x)$, and {\tt mass} is $m_0$.
Similarly, the ``double precision'' solver {\tt cg\_stagdagstag\_dp} and the ``mixed precision'' solver {\tt cg\_stagdagstag\_mp} do the same job for {\tt b,x} in {\tt dp}.
Internally, there is a routine {\tt app\_stagdagstag\_$\{$sp,dp$\}$} which applies $\DS\dag \DS=\ep \DS \ep \DS$ to the argument.
In {\tt cg\_stagdagstag\_$\{$sp,dp$\}$} it is used in much the same way as {\tt app\_wsuv\_$\{$sp,dp$\}$} in {\tt cg\_wsuv\_$\{$sp,dp$\}$} or {\tt app\_bsuv\_$\{$sp,dp$\}$} in {\tt cg\_bsuv\_$\{$sp,dp$\}$}.
Similar to the CG solvers for $\DW\dag\DW$ and $\DB\dag\DB$ a temporary vector {\tt tmp\_$\{$sp,dp$\}$} is allocated in the calling solver, to avoid an allocation/deallocation operation in 
{\tt app\_stagdagstag\_$\{$sp,dp$\}$} for every iteration of the solver.

\subsubsection*{$\bullet$ testknl\_main}

The main program generates (as a subset of its capabilities) the data presented in this paper.
It uses preprocessor macros like {\tt \#define fpp\_vec} or {\tt \#define fpp\_suv} which, if set, make the program call the routines discussed in Sec.~\ref{sec:norms}, and print the timing information.
If this information is not desired, one simply comments these two lines.
The parameters {\tt Nc,Nv}, as well as {\tt Nx,Ny,Nz,Nt} are to be set in the module {\tt testknl\_globals}.

Setting the preprocessor macro {\tt \#define fpp\_clov} makes the code call the clover routine discussed in Sec.~\ref{sec:clov}, looping over sp and dp, as well as the six options for the vector layout.
Setting the preprocessor macro {\tt \#define fpp\_wlap} and/or {\tt \#define fpp\_wils} results in the code calling the Wilson Laplace and/or Dirac matrix-times-vector routines discussed in Sec.~\ref{sec:wils}.
Setting the preprocessor macro {\tt \#define fpp\_blap} and/or {\tt \#define fpp\_bril} results in the code calling the Brillouin Laplace and/or Dirac matrix-times-vector routines discussed in Sec.~\ref{sec:bril}.
Setting the preprocessor macro {\tt \#define fpp\_wsuv} and/or {\tt \#define fpp\_bsuv} results in the code applying the Wilson Laplace and/or Brillouin Laplace operator to a Susskind vector.
Though not explicitly discussed in the body of the paper, these routines are useful to gauge the performance of the staggered routine, which is steered by the preprocessor macro {\tt \#define fpp\_stag} and discussed in Sec.~\ref{sec:stag}.

Any one of these preprocessor macros can be combined with {\tt \#define fpp\_scaling}; in this case the code will provide the data for the scaling of the respective operator in the number of OpenMP threads.
Alternatively, a given operator macro, e.g.\ {\tt \#define fpp\_wils}, may be combined with {\tt \#define fpp\_inverse}.
In this case the inversion of the Wilson Dirac operator by means of the CG and BiCGstab algorithms is performed, in sp and dp, with a loop over the layout options which can be restricted to the option {\tt NvNsNc}.
Another alternative is to combine a given operator macro, e.g.\ {\tt \#define fpp\_blap}, with {\tt \#define fpp\_chiral}.
In this case the Brillouin Laplacian is inverted for a series of mass values by means of the CG algorithm, in sp and dp, with a (restrictable) loop over the layout options.
Finally, an operator macro like {\tt \#define fpp\_stag} may be combined with {\tt \#define fpp\_spectro}.
In this case some rudimentary staggered spectroscopy (Goldstone pion only) is performed.
In fact, any one of the operator macros {\tt fpp\_$\{$wlap,wils,blap,bril,wsuv,bsuv,stag$\}$} can be combined with any one of the task macros {\tt fpp\_$\{$scaling,inverse,chiral,spectro$\}$}.







\clearpage


\end{document}